\documentclass[preprint,12pt,number,square,sort&compress]{elsarticle}

\usepackage[paperwidth=8.5in,paperheight=11.0in,left=1.0in,right=1.0in,
top=1.0in,bottom=1.0in,heightrounded]{geometry}
\usepackage{blindtext}
\usepackage[separate-uncertainty=true,multi-part-units=single]{siunitx}
\usepackage{graphicx}
\usepackage[export]{adjustbox}
\graphicspath{ {Figures/} }
\usepackage{fnpct}
\usepackage[colorlinks]{hyperref}
\usepackage{amsmath}
\usepackage{amsfonts}
\usepackage{fixmath}
\usepackage{accents}
\usepackage{amsxtra}
\usepackage{amstext}
\usepackage{amssymb}
\usepackage{latexsym}
\usepackage{dsfont}
\usepackage{textcomp}
\usepackage[normalem]{ulem}
\usepackage{multirow}
\usepackage{color}
\usepackage{epstopdf}
\usepackage{caption, subcaption}
\usepackage{bm}
\usepackage{soul}
\usepackage{natbib}
\usepackage{hyperref}
\hypersetup{
	colorlinks=true,
	linkcolor=cyan,
	filecolor=magenta,      
	urlcolor=blue,
}
\usepackage{tabstackengine}
\usepackage{booktabs}
\stackMath

\journal{Wave Motion}

\DeclareUnicodeCharacter{0B}{~}
\begin{document}
\let\Gamma\varGamma

\newcommand{\vect}[1]{\bm{#1}}
\newcommand{\TM}{\mathsf{T}}
\newcommand{\SM}{\mathsf{S}}
\newcommand{\TSM}{\tilde{\SM}}
\newcommand{\PM}{\mathsf{M}}
\newcommand{\NM}{\mathsf{N}}
\newcommand{\xbar}[1]{\mkern 1.7mu\overline{\mkern-1.7mu#1\mkern-1.7mu}\mkern 1.7mu}
\newcommand{\rood}[1]{\textcolor{red}{[#1]}} 
\newcommand{\gre}[1]{{#1}} 
\newcommand{\bluu}[1]{\textcolor{blue}{#1}} 
\newcommand{\ut}[1]{\underaccent{\tilde}{#1}}
\renewcommand{\vec}[1]{\ut{#1}}
\newcommand{\pd}[2]{\partial_{#1}#2}
\newcommand{\pdx}[1][]{\pd{x}#1}
\newcommand{\pdt}[1][]{\pd{t}#1}
\newcommand{\eps}{\varepsilon}
\newcommand{\sig}{\sigma}
\newcommand{\kap}{\kappa}
\newcommand{\kapt}{\tilde{\kappa}}
\newcommand{\gam}{\gamma}
\newcommand{\om}{\omega}
\newcommand{\bbet}{\bm{\beta}}
\newcommand{\bpsi}{\bm{\psi}}
\newcommand{\bkap}{\bm{\kap}}
\newcommand{\zp}{z^{+}}
\newcommand{\zm}{z^{-}}
\newcommand{\kp}{k^{+}}
\newcommand{\kpj}{\kp_{j}}
\newcommand{\kmj}{\km_{j}}
\newcommand{\cp}{c^{+}}
\newcommand{\Vol}{\mathcal{V}}
\newcommand{\Sur}{\mathcal{S}}
\newcommand{\Mur}{\mathcal{M}}
\newcommand{\Rur}{\mathcal{R}}
\newcommand{\Rsf}{\mathsf{R}}
\newcommand{\Tsf}{\mathsf{T}}
\newcommand{\Oh}{\mathcal{O}}
\newcommand{\spr}{s_2}
\newcommand{\snpr}{S}
\newcommand{\ii}{\mathrm{i}}
\newcommand{\beginsupplement}{%
	\setcounter{table}{0}
	\setcounter{figure}{0}
	\setcounter{section}{0}
	\setcounter{subsection}{0}
	\setcounter{equation}{0}
	\setcounter{paragraph}{0}
	\setcounter{page}{1}
	\renewcommand{\thetable}{S\arabic{table}}%
	\renewcommand{\thefigure}{S\arabic{figure}}%
	\renewcommand{\thesection}{S\arabic{section}}%
	\renewcommand{\thesubsection}{S\arabic{section}.\arabic{subsection}}%
	\renewcommand{\theequation}{S\arabic{equation}}%
	\renewcommand{\theparagraph}{S\arabic{paragraph}}%
	\renewcommand{\thepage}{S\arabic{page}}%
	\renewcommand{\theHtable}{S\the\value{table}}%
	\renewcommand{\theHfigure}{S\the\value{figure}}%
	\renewcommand{\theHsection}{S\the\value{section}}%
	\renewcommand{\theHsubsection}{S\the\value{subsection}}%
	\renewcommand{\theHequation}{S\the\value{equation}}%
	\renewcommand{\theHparagraph}{S\the\value{paragraph}}%
}
\begin{frontmatter}
	
\title{In-Plane Linear Stress Waves in Layered Media: I. Non-Hermitian Degeneracies and Modal Chirality}

\author{Vahidreza Alizadeh and Alireza V. Amirkhizi*}

\address{Department of Mechanical Engineering\\
University of Massachusetts, Lowell\\
One University Avenue, Lowell, MA 01854\\
*Corresponding author: \href{mailto:alireza\_amirkhizi@uml.edu}{alireza\_amirkhizi@uml.edu}}

\date{\today}

\begin{abstract}
We study the band structure and scattering of in-plane coupled longitudinal and shear stress waves in linear layered media and observe that exceptional points (EP) appear for elastic (lossless) media, when parameterized with real-valued frequency and tangential wave vector component. The occurrence of these EP pairs is not limited to the original stop bands. They could also appear in all mode pass bands, leading to the formation of new stop bands. The scattered energy near these locations is studied along with the associated polarization patterns. The broken phase symmetry is observed inside the frequency bands book-ended by these EP pairs. This is especially manifested by the chirality of the trajectory of the particle velocity, which gets selected by a ``direction'' of the wave, e.g. the imaginary part of normal component of the wave vector, or the energy flux direction just outside the band. Additionally, EP pairs also appear in the spectrum of the (modified) scattering matrix when mechanical gain is theoretically included to balance the loss in a parity-time symmetric finite structure. These EP pairs lead to amplification of transmission to above 1 and single-sided reflectivity, both phenomena associated with broken phase symmetry, with intriguing potential applications. 
\end{abstract}


\end{frontmatter}

\section{Introduction}

Tracing the existence of exceptional points (EPs) in a non-Hermitian system is an important aspect of characterizing such a system. 
EPs are associated with points in a parameter space at which two eigenstates of a parameterized non-Hermitian operator coalesce. Unlike Hermitian operators, for which this could be a simple algebraic degeneracy and one can still find a complete basis of eigenvectors, the EPs represent geometric degeneracies, where the set of linearly independent eigenvectors of the non-Hermitian operator is not large enough to constitute a complete basis. The EPs were first studied in the context of quantum mechanics \cite{Heiss2012}. In recent years, they have been connected to a variety of exotic behaviors in electromagnetism and photonic crystals \cite{Dembowski2001, Ding2015, Hodaei2017,Miri2019}, phononic crystals \cite{Christensen2016, Lu2018, Lustig2019, wang2020exceptional}, and acoustics \cite{Zhu2014, Achilleos2017, Ding2016, Merkel2018}. They have been associated with avoided crossing of branches in the band structure of a periodic system \cite{Lu2018}.
In contrast, in regular crossings, while eigenvalues coalesce, the eigenvectors remain linearly independent (diabolic points, DPs). A method was proposed in \cite{Seyranian2005} to quantitatively study the behavior of eigenvalues in the vicinity of mode coalescence for both EPs and DPs of an operator. The topological structure of exceptional points were also observed experimentally in various studies \cite{Dembowski2001, Lee2009, Gao2015}. Metamaterials can be engineered in such a way to bring EPs of various operators within a range that makes them accessible for experimental evaluation and to exploit them for applications that correspond to such a singularity. EPs have also been discussed and studied in the systems that lack parity and time symmetry (e.g. due to loss) but are $\mathcal{PT}$-symmetric. $\mathcal{PT}$-symmetric micro-structured materials, sometimes referred to as ``synthetic materials'' \cite{Feng2013}, can be achieved in acoustics and optics via introducing and carefully tailoring loss and gain constituents \cite{Merkel2018, wang2020exceptional}. A thorough study of the symmetry breaking conditions, focused on pseudo-Hermitian operators, is recently given by Melkani~\cite{melkani_degeneracies_2023}. 
In such cases, the exceptional points also mark the transition between exact and broken $\mathcal{PT}$-symmetry phases \cite{Ding2015, El-Ganainy2018}. Fleury et al. \cite{Fleury2015} introduced an acoustic $\mathcal{PT}$-symmetric metameterial in which loud speakers combined with non-Foster electrical circuits bring the scattering matrix to its exceptional point so that ``unidirectional cloaking'' can be observed. Other examples of tuning a $\mathcal{PT}$-symmetric system for unidirectional invisibility as a result of EP degeneracy of the non-Hermitian scattering matrix was reported in \cite{Lin2011, Ge2012, Christensen2016, Shi2016, Auregan2017}. Constructing true $\mathcal{PT}$-symmetric systems involving gain is not a straight-forward experimental task \cite{Guo2009, Ruter2010, Kang2013}. It has been suggested that fully passive non-Hermitian systems with no actual gain (or unbalanced gain/loss) can demonstrate similar asymmetric behavior \cite{Feng2013,  Wu2014, Sun2014, Liu2018}. While in some of these efforts a potential gain from environment is suggested and in others a rescaling of the two modes recast the system in somewhat similar structure as one with balanced gain, in most cases the behavior is essentially significant asymmetry of reflection and sometimes a counter-intuitive increase in transmission with increasing loss. The occurrence of non-Hermitian degeneracies is not limited to the scattering matrix. In fact, they have also been widely observed in the band structure when eigenstates coalesce in photonics \cite{Zhen2015, Othman2017} and phononics \cite{Lu2018, Lustig2019, Yang2019}. Special note must be made of the recent work of Gupta and Thevamaran~\cite{gupta_requisites_2023}, where they note the emergence of EPs in lossy mechanical systems under certain requirements associated with nature of the loss spectrum. Focusing on coupled wave propagation in layered media, Lustig et al. \cite{Lustig2019} studied longitudinal and in-plane shear waves in media with elastic constituents. They observed that exceptional points could be achieved at a specific real-valued $k_2$ (parallel to the layers interface, or tangential component of the wave vector), which has to be constant everywhere to satisfy continuity along all the interfaces (Snell's law). While their approach using hybrid matrix method has successfully been used to calculate the band structure, the transfer matrix method (TMM) is also a suitable and easy to implement method for calculating not only the dispersion curves, but the scattering coefficients \cite{Li2003, Nemat-Nasser2015, Nanda2018, Psiachos2019, Amirkhizi2018, wang2020exceptional}, as well as the mode shapes of the system. Furthermore, with TMM the inclusion of loss (or gain, if desired) in frequency domain requires very little additional effort. 

In this work we discuss the nature of modes and scattering using 4th order TMM analysis for elastic as well as lossy layered structures, with particular attention to the polarizations in the proximity of EPs. Of particular interest is the broken phase behavior of the modes signified by their polarization chirality. While chirlaity affected through encircling of the EPs via material modulation has been shown (e.g. spatial in \cite{elbaz_encircling_2022} or temporal in \cite{geng_topological_2021}), to the best of our knowledge, this ``polarization chirality'' has not been studied in detail, especially as it ties in with energy flux. We expect that this chirality selection, once properly tuned, might lead to sensing modalities that have not been utilized so far. Such instances can even occur in the middle of pass bands. Additionally, similar behavior has been suggested in optical systems for directional lensing \cite{peng_chiral_2016}.The occurrence of EPs in the interior of stop bands has been observed before and studied here as well with attention to polarizations. The TMM approach also enables easy calculation of scattering in the presence of lossy constituents and is not limited to lossless media. Of course the EPs are no longer located in the real domain when lossy layers are introduced. It is not unlikely to be able to bring the EPs back into the real domain (of parameters) when a balanced gain (of enough magnitude) is also added (e.g. similar to \cite{wang2020exceptional}), and to also study the EPs of the modified scattering matrix in such cases, though such an exploration is left for future work. Additionally, these results can easily be leveraged for field integration, as the basis for homogenization and extraction of effective overall parameters \cite{Amirkhizi2017,Amirkhizi2018}, which is deferred to a follow-up study as well.

The structure of this paper is as follows. In Section \ref{sec:TM}, the transfer matrix of layered solids for coupled longitudinal and in-plane shear waves is derived. Then, the oblique scattering calculations will be presented for a typical 3-phase layered system. with and without loss. The band structure and mode shapes are then studied for the same 3-phase system and an even simpler 2-phase system, and the modal coalescence and broken phase symmetry, modal shape and chirality just outside and inside the broken phase region, and energy flux calculations are presented near such coalescence pairs. Of course introduction of loss will eliminate such regions. The conclusions are summarized at the end and suggestions for further study, particularly derivation of effective parameters and study of balanced gain ($\mathcal{PT}$-symmetric systems) complete the paper. 

\section{Transfer matrix of layered structures for oblique coupled P-SV waves} \label{sec:TM}

Here we will study the propagation of coupled in-plane shear (SV) and longitudinal (P) stress waves in finite or infinite (periodic) layered linear solids. Consider a slab consisting of multiple homogeneous and isotropic layers normal to the $x_1$-axis. Each layer, $j$, has thickness, $d^j = x_1^j - x_1^{j-1}$, where $x_1^{j-1}$ and $x_1^j$ indicate the boundary coordinates of layer $j$, and is unbounded in the two normal directions. Due to the assumed isotropy of each layer and without loss of generality, waves traveling in $x_1 x_2$ plane are studied here.

\subsection{Oblique propagation of coupled P-SV waves}

The continuous independent field variables over the entirety of system are the $x_1$ and $x_2$ components of particle velocities, $v_1$ and $v_2$, as well as the normal and shear stress components, $\tau_1 = \sigma_{11}$, and $\tau _6 = \sigma_{12}$ (Voigt notation for stress components is used in the remainder). Thus the point-wise state vector, with $e^{\ii \omega t}$ phasor time dependence factored out, can be written for $x_1^{j-1} \leq x_1 \leq x_1^j$ as,
\begin{equation}
    \mathbold{\psi}(x_1,\omega, s_2) =
	\begin{pmatrix}
		v_1\\
		\tau_1\\
		v_2\\
		\tau_6
	\end{pmatrix}
	= \vect{\zeta}^j(s_2)\vect{\delta}^j(x_1, \omega, s_2)\vect{A}^j e^{-\ii \omega s_2 x_2},
\end{equation}
where,
\begin{equation}
	\begin{aligned}
		\setstacktabbedgap{5ex}
		\def\stackalignment{r}
		\stackanchor[10pt]{%
			\vect{\zeta}^j(s_2) = \left(
			\tabbedCenterstack[c]{%
				s_1^{Pj}/s^{Pj} & -s_1^{Pj}/s^{Pj} \\ 
				-(Z_1^{Pj}s_1^{Pj} + Z_{12}^{j}s_2)/s^{Pj} & -(Z_1^{Pj}s_1^{Pj} + Z_{12}^{j}s_2^{Pj})/s^{Pj} \\ 
				s_2/s^{Pj} & s_2/s^{Pj} \\ 
				-(Z_6^{Pj}s_2 + Z_{62}^{j} s_1^{Pj})/s^{Pj} & (Z_6^{Pj}s_2 + Z_{62}^{j} s_1^{Pj})/s^{Pj}
			}%
			\right.\kern-1ex
		}{
			\left.
			\tabbedCenterstack[c]{%
				-s_2/s^{Sj} & s_2/s^{Sj} \\ 
				(Z_1^{Sj}s_2 - Z_{12}^{j}s_1^{Sj})/s^{Sj} &      (Z_1^{Sj}s_2 - Z_{12}^{j}s_1^{Sj})/s^{Sj} \\ 
				 s_1^{Sj}/s^{Sj} & s_1^{Sj}/s^{Sj} \\ 
				 -(Z_6^{Sj}s_1^{Sj} - Z_{62}^{j}s_2)/s^{Sj} & (Z_6^{Sj}s_1^{Sj} - Z_{62}^{j} s_2)/s^{Sj}
			}%
			\right),%
		}\\
	\end{aligned}
\end{equation}
\begin{equation}
	\begin{aligned}
		\vect{\delta}^j(x_1, \omega, s_2) =
		\begingroup
		\setlength\arraycolsep{-3pt}
		\begin{pmatrix}
			e^{-\ii \omega s_{1}^{Pj}(x_1 - x_1^{j-1})} & 0 & 0 & 0\\
			0 & e^{\ii \omega s_{1}^{Pj}(x_1 - x_1^{j-1})} & 0 & 0\\
			0 & 0 & e^{-\ii \omega s_{1}^{Sj}(x_1 - x_1^{j-1})} & 0\\
			0 & 0 &0 & e^{\ii \omega s_{1}^{Sj}(x_1 - x_1^{j-1})}
		\end{pmatrix},
		\endgroup
	\end{aligned}
\end{equation}
with $s_1^{Pj} = k_{1}^{Pj} / \omega = \sqrt{\left(s^{Pj}\right)^2 - s_2^2}$ and $s_1^{Sj} = k_{1}^{Sj} / \omega = \sqrt{\left(s^{Sj}\right)^2 - s_2^2}$, as the normal components of the slowness vector (wave vector divided by the angular frequency) in which $s^{Pj} = 1 / c_L^j$ and $s^{Sj} = 1 / c_T^j$ are the longitudinal and shear slowness in layer $j$, respectively\footnote{The wave speeds notations $c_L$ and $c_T$ are consistent with literature, while slowness $s^{Pj}$ and $s^{Sj}$ are used for uniformity in this work, especially as in the laminates the waves are not pure modes. Slownesses are also less used in the literature.}, $s_2$ is the tangential component of the slowness vector and is proportional to the wave vector component $k_2 = \omega s_2$ also parallel to the layers (or the non-dimensional phase $Q_2=k_2 d$, where $d$ is the total thickness of a repeating unit cell). Both are required to be continuous throughout the system  due to the Snell's law. However, formulating based on $s_2$ leads to clearer results, especially as $\omega$ approaches $0$\footnote{Prescribing finite values of $k_2$ (or $Q_2$), would lead to the appearance of stop bands at zero frequency as the total size of the wave vector $|\vect{k}|$ should also approach zero. In contrast, when finite $s_2$ is prescribed, this issue no longer occurs as $k_2$ is proportional to the frequency and also vanishes. Therefore while some examples with prescribed $k_2$ (or $Q_2$) are shown below, for the analysis and major results, $s_2$ and $\omega$ are used as the main parameters. In the following formulation the parameterization is dropped for sake of brevity, unless needed to avoid confusion.}.  
The vector $\vect{A}^j$ collects the complex amplitude of left and right traveling P- and SV-waves in layer $j$ (first two components are P-wave amplitudes). The definitions of characteristic impedances are as follows: $Z_1^{Pj} =  C_{11}^j s_{1}^{Pj}$, $Z_{12}^{j} = C_{12}^j s_2$, $Z_1^{Sj} = C_{11}^j s_{1}^{Sj}$, $Z_6^{Pj} = C_{66}^j s_{1}^{Pj}$, $Z_{62}^{j} =  C_{66}^j s_2$, and $Z_6^{Sj} = C_{66}^j s_{1}^{Sj}$.

\subsection{Transfer matrix and scattering}

The transfer matrix for each layer can be defined as,
\begin{equation}
	\vect{\TM}^j = \vect{\zeta}^j\vect{\delta}^j(d^j)({\vect{\zeta}^j})^{-1},
\end{equation}
where, $\vect{\delta}^j$ is evaluated at $x_1^j$ and therefore only a function of $d^j = x_1^j - x_1^{j-1}$ (as well as $k_2 = \omega s_2$). The transfer matrix of a multilayer medium ($\Omega = \{x_1: x_1^0<x_1<x_1^n = x_1^0 + d\}$) can be calculated as,
\begin{equation}\label{eq:tm}
	\vect{\TM} = \vect{\TM}^n \vect{\TM}^{n-1} \cdots \vect{\TM}^j \cdots \vect{\TM}^1,
\end{equation}
where, $n$ is the number of layers. If this multilayer medium is the unit cell of an infinitely periodic domain, the Bloch-Floquet periodicity will require that the transfer matrix would be effectively a phase advance operation with (at most) four possible eigensolutions. In other words,
\begin{equation}
    \vect{\psi}(x_1^n) = \vect{\TM} \vect{\psi}(x_1^0) = e^{-\ii Q_1^\alpha} \vect{\psi}(x_1^0),
\end{equation}
where $\alpha = (\text{P}/\text{S})^{(\pm)}$ represents the four solutions of the P-SV case. The phase advance $Q_1^\alpha$ can also define the effective wave vector component and slowness components along $x_1$ through $Q_1^\alpha = K_1^\alpha d = \omega S_1^\alpha d$. Of course due to the reciprocity, the four solutions are simply two pairs of opposite signs, representing the left and right ``traveling'' solutions, though they can be potentially evanescent.
To remove the phase ambiguity of the solution if possible, an integer multiple of $2\pi$ is added whenever needed to force continuity of the wave vector. 

For a finite multi-layer slab, the wave amplitudes of the right and left hand side of the medium can be related using the propagation matrix,
\begin{equation}
	\vect{A}^b = \vect{\PM}\vect{A}^a,
\end{equation}
where, superscripts $a$ and $b$ represent the wave amplitudes in the left and right half-spaces, respectively, and $\vect{\PM}$ is the propagation matrix (also referred to as transfer matrix in literature, e.g. \cite{markos2008wave}) which can be obtained from the transfer matrix: 
\begin{equation}
	\vect{\PM} = (\vect{\delta}^b (x_1^b))^{-1}(\vect{\zeta}^b)^{-1}\vect{\TM}\vect{\zeta}^a\vect{\delta}^a(x_1^a).
\end{equation}
The scattering parameters with incident P- or SV-wave from left ($a$) side of the finite slab are calculated as
\begin{equation}\label{eq:scatt_P}
	\begin{aligned}
		\SM^{PP}_{aa} &= (\PM_{24}\PM_{41} - \PM_{21}\PM_{44})/\Delta, &
		\SM^{PP}_{ba} &= \PM_{11} + \PM_{12}\SM^{PP}_{aa} + \PM_{14}	\SM^{SP}_{aa}, &\\
		\SM^{SP}_{aa} &= (\PM_{21}\PM_{42} - \PM_{22}\PM_{41})/\Delta, &
		\SM^{SP}_{ba} &=  \PM_{31} + \PM_{32}\SM^{PP}_{aa} + \PM_{34}	\SM^{SP}_{aa},
	\end{aligned}
\end{equation}
\begin{equation}\label{eq:scatt_S}
	\begin{aligned}
		\SM^{PS}_{aa} &= (\PM_{24}\PM_{43} - \PM_{23}\PM_{44})/\Delta, &
		\SM^{PS}_{ba} &= \PM_{13} + \PM_{12}\SM^{PS}_{aa} + \PM_{14}	\SM^{SS}_{aa}, &\\
		\SM^{SS}_{aa} &=(\PM_{23}\PM_{42} - \PM_{22}\PM_{43})/\Delta, &
		\SM^{SS}_{ba} &=  \PM_{33} + \PM_{32}\SM^{PS}_{aa} + \PM_{34}	\SM^{SS}_{aa},
	\end{aligned}
\end{equation}
where first and second subscripts represent the outgoing (scattered) and incoming (incident) domains, respectively, and the first and second superscript represent the polarization of the outgoing and incoming waves. For example, $\SM^{SP}_{ba}$ represents the SV-wave ($S$) transmission into domain $b$ due to the P-wave ($P$) incidence from domain $a$.
Additionally, $\Delta = \PM_{22}\PM_{44} - \PM_{24}\PM_{42}$. In order to calculate the scattering parameters for incidence from right (i.e. with $b$ as the second superscript), one only needs to use the inverse propagation matrix $\mathsf{N} = \PM^{-1}$, and swap the indices as $1 \leftrightarrow 2$ and $3 \leftrightarrow 4$ in the formulas above.

The energy conservation for a lossless slab under P-wave incidence from left, ($a$) side, requires
\begin{equation}
	\begin{aligned}
		\Re(Z^P_1)\left(1 - \left|\SM^{PP}_{aa}\right|^2 - \left|\SM^{PP}_{ba}\right|^2\right) - \Re(Z^S_6) \left(\left|\SM^{SP}_{aa}\right|^2 + \left|\SM^{SP}_{ba}\right|^2\right) &= 0,\\
        \Re(Z^S_6)\left(1 - \left|\SM^{SS}_{aa}\right|^2 - \left|\SM^{SS}_{ba}\right|^2\right) - \Re(Z^P_1) \left(\left|\SM^{PS}_{aa}\right|^2 + \left|\SM^{PS}_{ba}\right|^2\right) &= 0.
    \end{aligned}
\end{equation}
Here it is assumed that the properties of both left ($a$) and right ($b$) side are the same, as they show up on the impedance pre-factors. Similar necessary conditions may be written for incidences from right side, which may lead to slightly different conditions for asymmetric laminates. 

\section{Mode shapes and scattering near band structure degeneracies} 
\begin{figure}[!ht]
\centering\includegraphics[height=230pt,center]{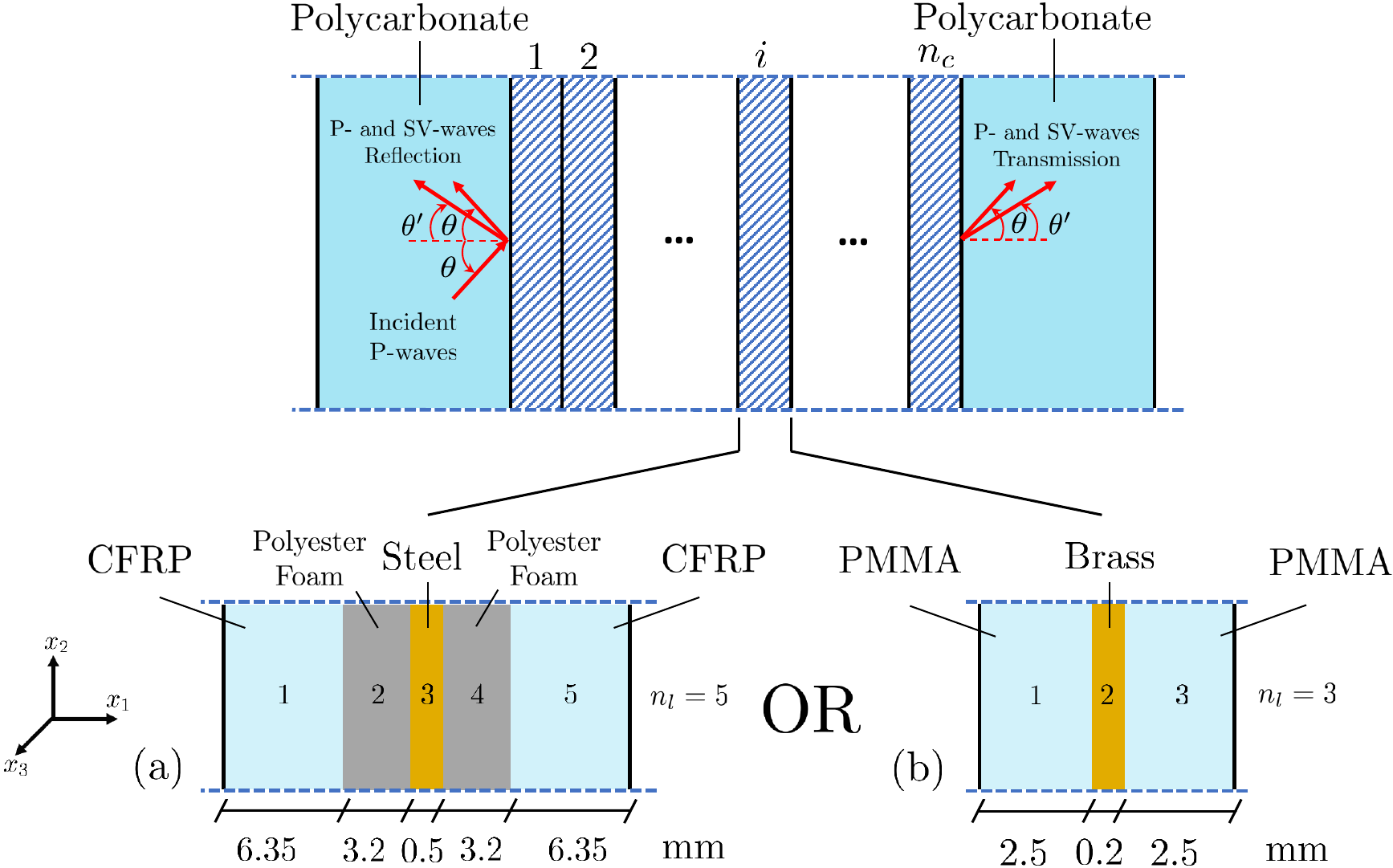}
    \caption{\label{fig:mat} Scattering problem setup and example unit cell geometries. The 5-layer, 3-phase unit cell (bottom left) was studied earlier in \cite{Amirkhizi2018}, while the 3-layer, 2-phase system was studied in \cite{alizadeh_overall_2021}. Polycarbonate is selected as both left and right scattering media. For an incident P-wave (shown) he reflected and transmitted P-waves are propagating with the same angle while reflected and transmitted SV-waves are propagating with $\theta'$ such that $\sin(\theta')/c_T = \sin(\theta)/c_L$. For incident SV-waves (not shown), similar calculation may be made. The relevant base material properties can be found in Table~\ref{tab:prop}.} 
\end{figure}

To study the coupled stress wave propagation in linear layered media further and identify occurrence and behavior of various degeneracies in such systems, two example unit cells are introduced in in Figure~(\ref{fig:mat}). The base material constants and geometries for each case are shown in Table~\ref{tab:prop}. The unit system [mm, $\mu$s, mg] for length, time, and mass are used throughout, unless explicitly stated other wise. This leads to the units of [km/s, GPa, g/cm$^{3}$, MHz, MRayl] units for wave velocity, elastic moduli, density, frequency, and impedance, respectively. For better presentation, frequency units of kHz will be used. 

\begin{table}[!ht]
\centering
\caption{\label{tab:prop} Material properties for various components and domains in this study.}
\begin{tabular}{cccc}
\cline{2-4}
 & 5-layer RUC & 3-layer RUC & Polycarbonate \\ \cline{2-4} 
\begin{tabular}[c]{@{}c@{}}Thickness \\ ($d$) [mm]\end{tabular} & (6.35, 3.20, 0.50, 3.20, 6.35) & (2.50, 0.20, 2.50) & - \\ \hline
\begin{tabular}[c]{@{}c@{}}Density \\ ($\rho$) [g/cm$^3$]\end{tabular} & (1.53, 0.36, 7.82, 0.36, 1.53) & (1.18, 8.64, 1.18) & 1.20 \\ \hline
\begin{tabular}[c]{@{}c@{}}Long. Wave Speed \\ ($c_L$) [mm/$\mu$s]\end{tabular} & (1.98, 0.23, 5.90, 0.23, 1.98) & (2.75, 4.70, 2.75) & 2.13 \\ \hline
\begin{tabular}[c]{@{}c@{}}Shear Wave Speed \\ ($c_T$) [mm/$\mu$s]\end{tabular} & (1.04, 0.13, 3.20, 0.13, 1.04) & (1.39, 2.10, 1.39) & 0.88 \\ \hline
\end{tabular}
\end{table}

\subsection{Degeneracies in band structure and mode shapes}

\begin{figure}[!ht]
	\centering\includegraphics[height=200pt,center]{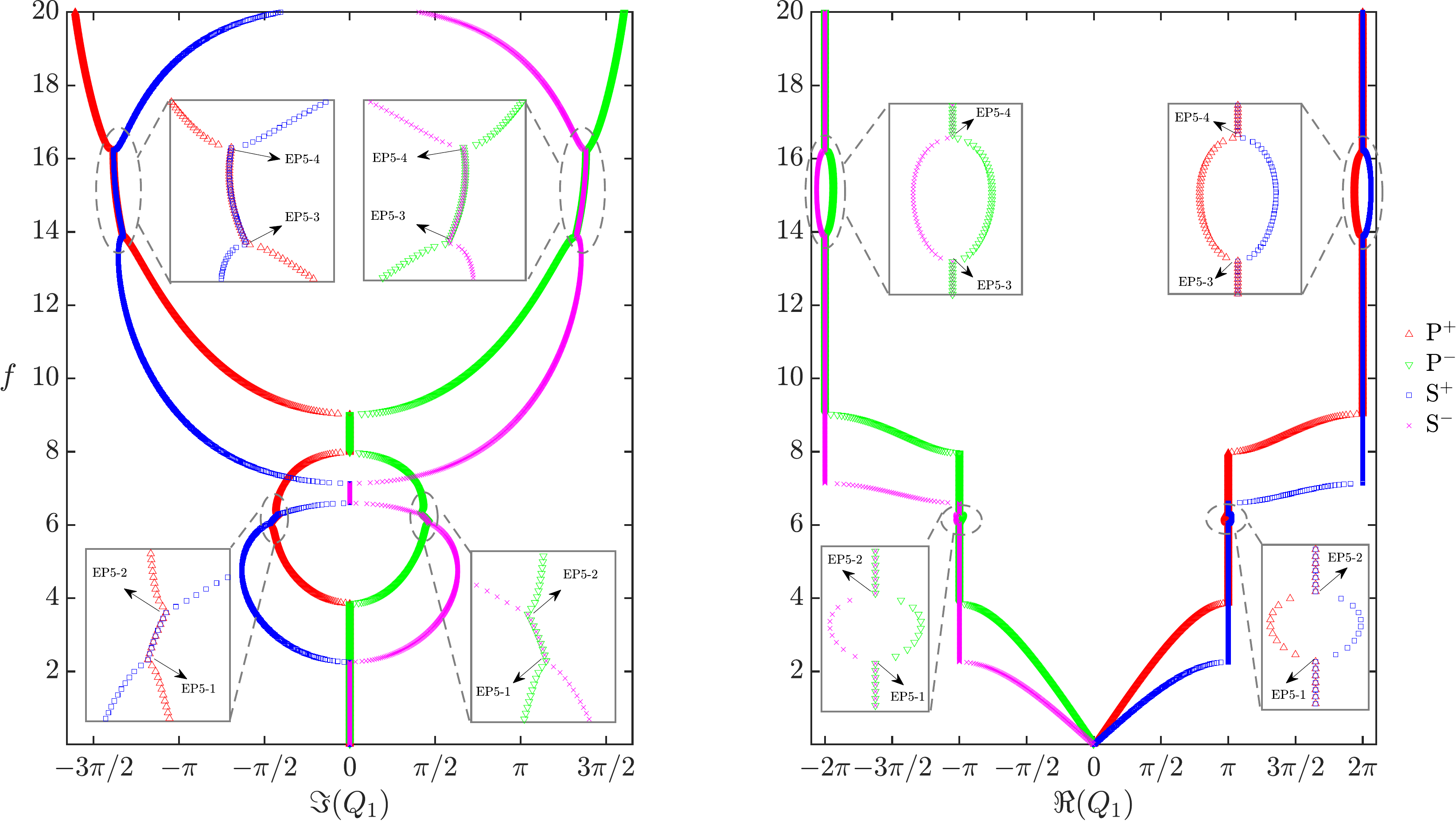}
	\caption{\label{fig:bnd_strc_s2_Q} The real and imaginary parts of the phase advance, $Q_1^{\alpha} = K_1^{\alpha}d$ at $s_2 = k_2/\omega = 0.15$. The superscript $^{\alpha}$ represent $\text{P}^+, \text{P}^-, \text{S}^+$, and $\text{S}^-$, in which ``P'' and ``S'' denote the dominant longitudinal and shear deformation, respectively. The real part of the phase advance is unwrapped by adding an integer multiple of $2\pi$ when needed to maintain continuity. Note that none of the 4 branches are completely longitudinal or shear as evidenced by extracting the mode shapes. Due to the symmetry of the unit cell, the phase advance for ``$^-$" branches are exactly negative of the ``$^+$" branches. Both real and imaginary parts of the phase advance coalesce at 4 frequencies, $f \approx 6, 6.3, 13.9, 16.2$ kHz, identified as EP5-1 to 4 in the insets. The fact that they are exceptional points of the band structure is confirmed by observing the mode shape collapse; See~Figure~(\ref{fig:mode_s2}).}
\end{figure}

\begin{figure}[!ht]
	\begin{subfigure}[b]{0.5\linewidth}
		\centering\includegraphics[height=120pt,left]{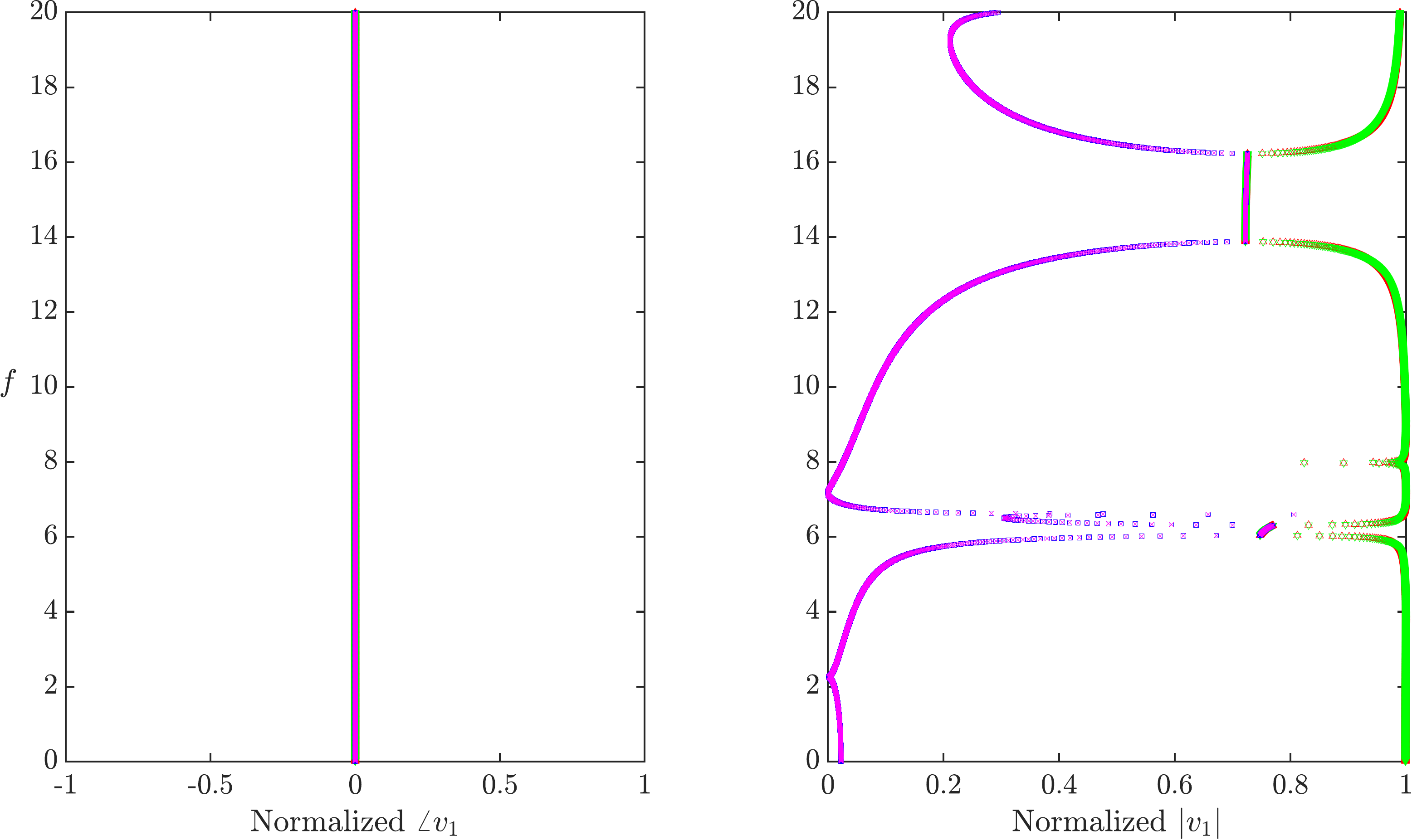}
		\caption{\label{fig:v1_s2}}
	\end{subfigure}%
	\begin{subfigure}[b]{0.5\linewidth}
		\centering\includegraphics[height=120pt,left]{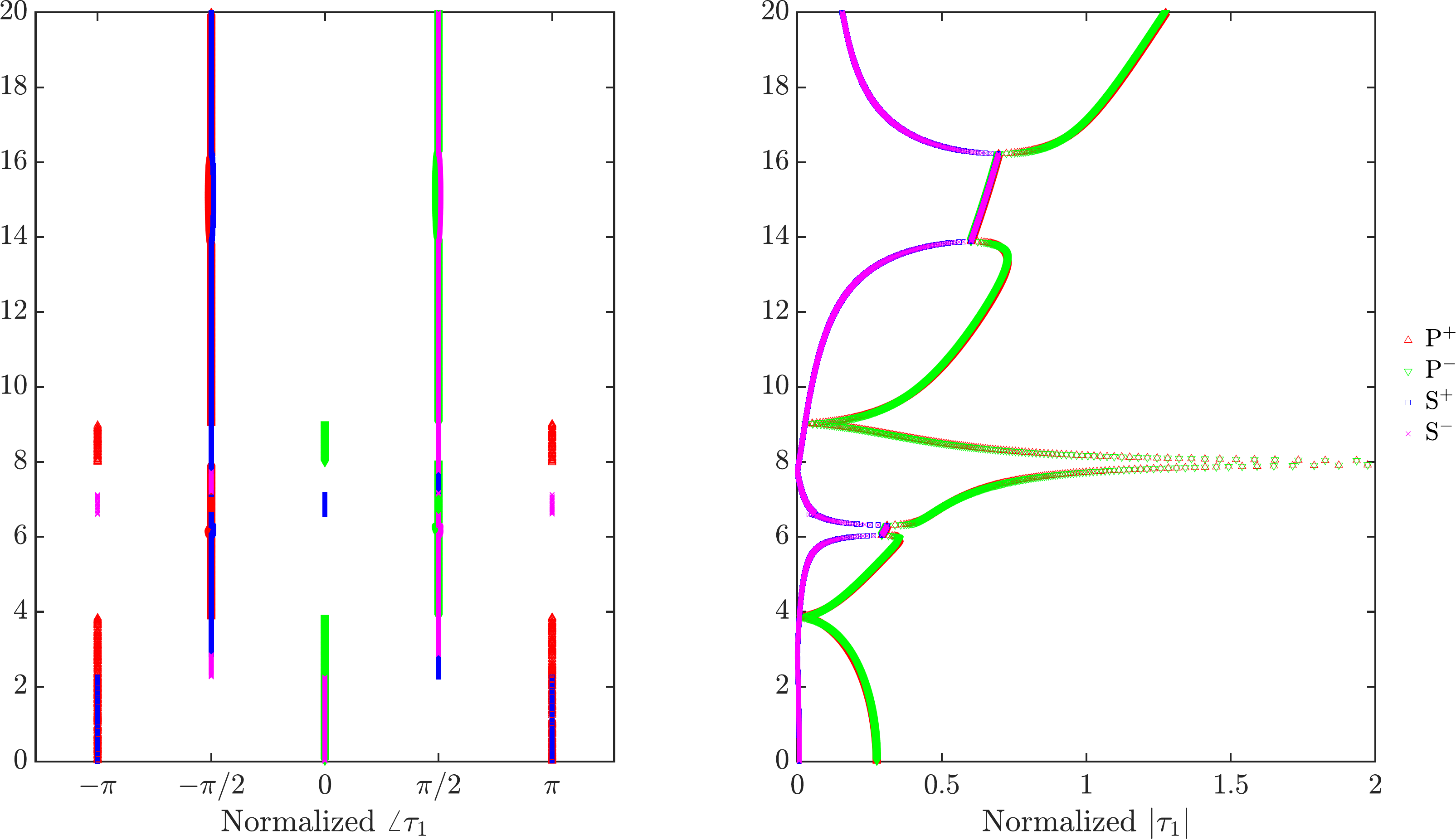}
		\caption{\label{fig:t1_s2}}
	\end{subfigure}
	\begin{subfigure}[b]{0.5\linewidth}
		\centering\includegraphics[height=120pt,left]{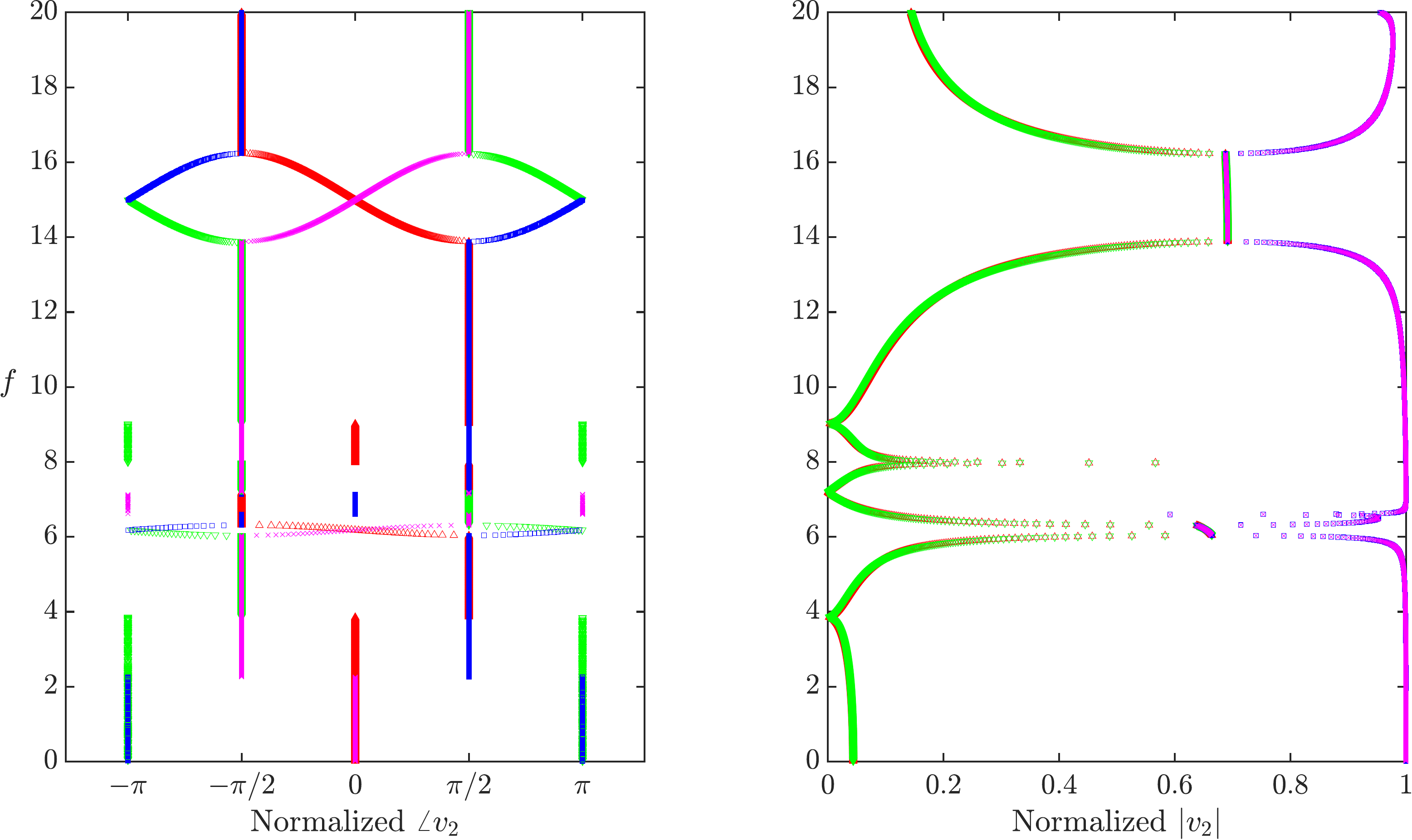}
		\caption{\label{fig:v2_s2}}
	\end{subfigure}%
	\begin{subfigure}[b]{0.5\linewidth}
		\centering\includegraphics[height=120pt,left]{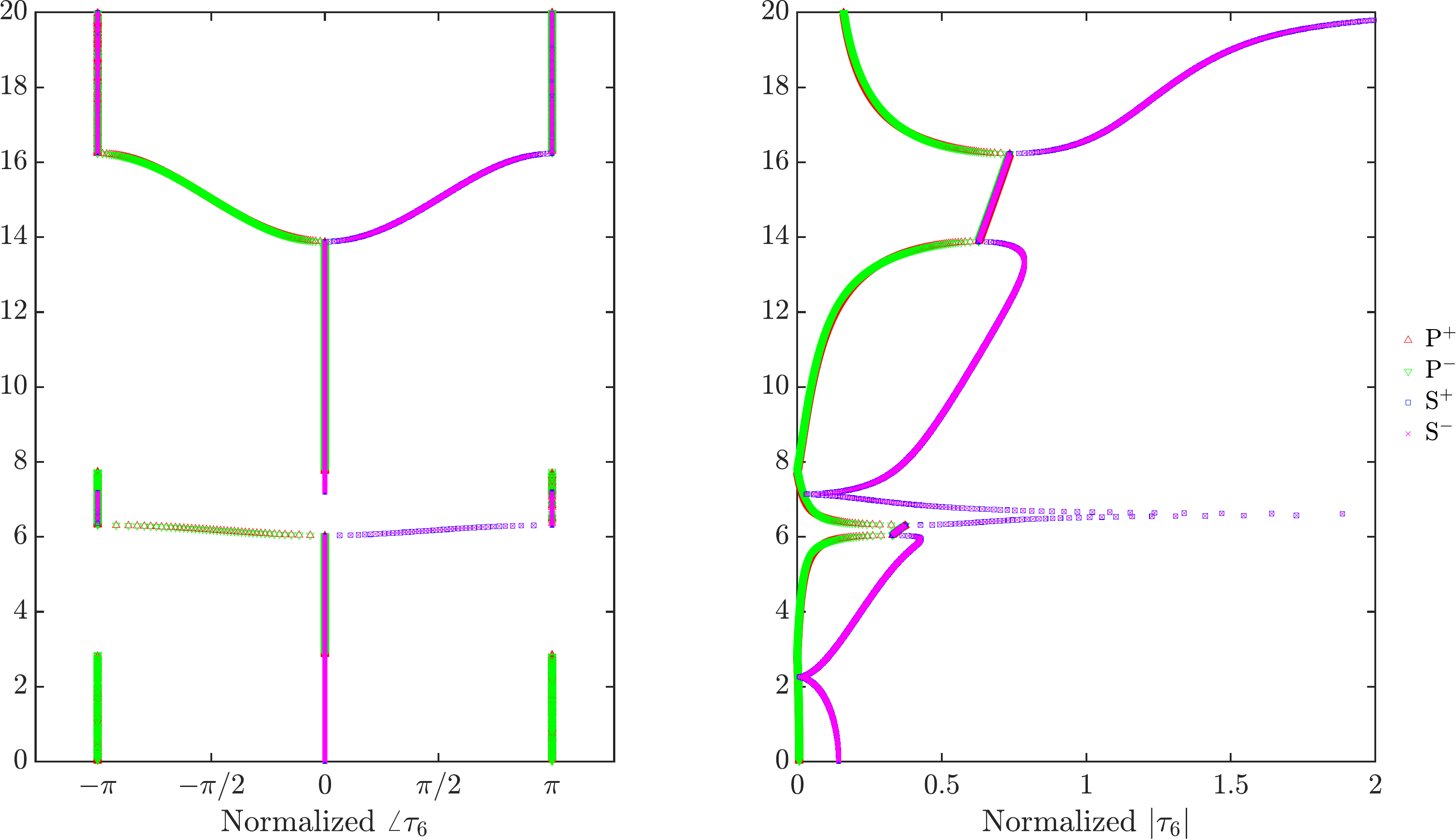}
		\caption{\label{fig:t6_s2}}
	\end{subfigure}
	\caption{\label{fig:mode_s2} 
     The normalized phase and amplitude of particle velocity components, $v_1$, $v_2$, and stress components, $\tau_1$, $\tau_6$, at $s_2 = 0.15$. The phase normalization is done based on the phase of particle velocity, $\angle v_1$. Amplitudes are normalized according to the particle velocity vector amplitude, i.e. $|v| = \sqrt{|v_1|^2+|v_2|^2}$, thus the unit of normalized stress components is MRayl. Both amplitude and phase coalesce for P$^+$ and S$^+$ at the 4 exceptional points with frequencies, $f \approx 6, 6.3, 13.9, 16.2$ kHz. 
    }
\end{figure}

The band structure of a layered medium is obtained by extracting the eigenvalues of the transfer matrix of a unit cell. They can be arranged in such a way that wave speed and phase advance are continuous over the frequency range of interest (with potential need for unwrapping the phase as frequency is increased). For lossy materials, this process is relatively straightforward, while for lossless materials a limiting approach may be needed~\cite{Amirkhizi2017,abedi_use_2020}. For oblique P- and SV-waves there would be 4 distinguishable branches which explain the behavior of left and right traveling P- and SV-waves within the layered structure. Figure~(\ref{fig:bnd_strc_s2_Q}) shows the real and imaginary parts of phase advance across a unit cell, $Q_1^\alpha = K_1^\alpha d$, for the lossless 3-phase, 5 layer, unit cell shown in Figure~(\ref{fig:mat}) with a prescribed slowness vector component $s_2 = 0.15$. At these four points both real and imaginary parts of $Q_1^{\alpha}$ coalesce demonstrating a potential for exceptional points, for which degeneracy (or defectiveness) of the mode shapes need to be also established. Figure~(\ref{fig:mode_s2}) shows that the mode shapes of the corresponding eigenvalues coalesce at these four frequencies. Of course and as seen in Figure~(\ref{fig:mode_s2}), the polarization of each branch is neither fully shear nor longitudinal, but it is generally mixed. Therefore, they are named based on the dominance of wave mode over the frequency range. For normal P- or SV-waves, i.e. at $s_2 = 0$, four modes are decoupled and they are either fully longitudinal or shear in nature. 

\begin{figure}[!ht]
	\begin{subfigure}[b]{0.33\linewidth}
		\centering\includegraphics[height=90pt,center]{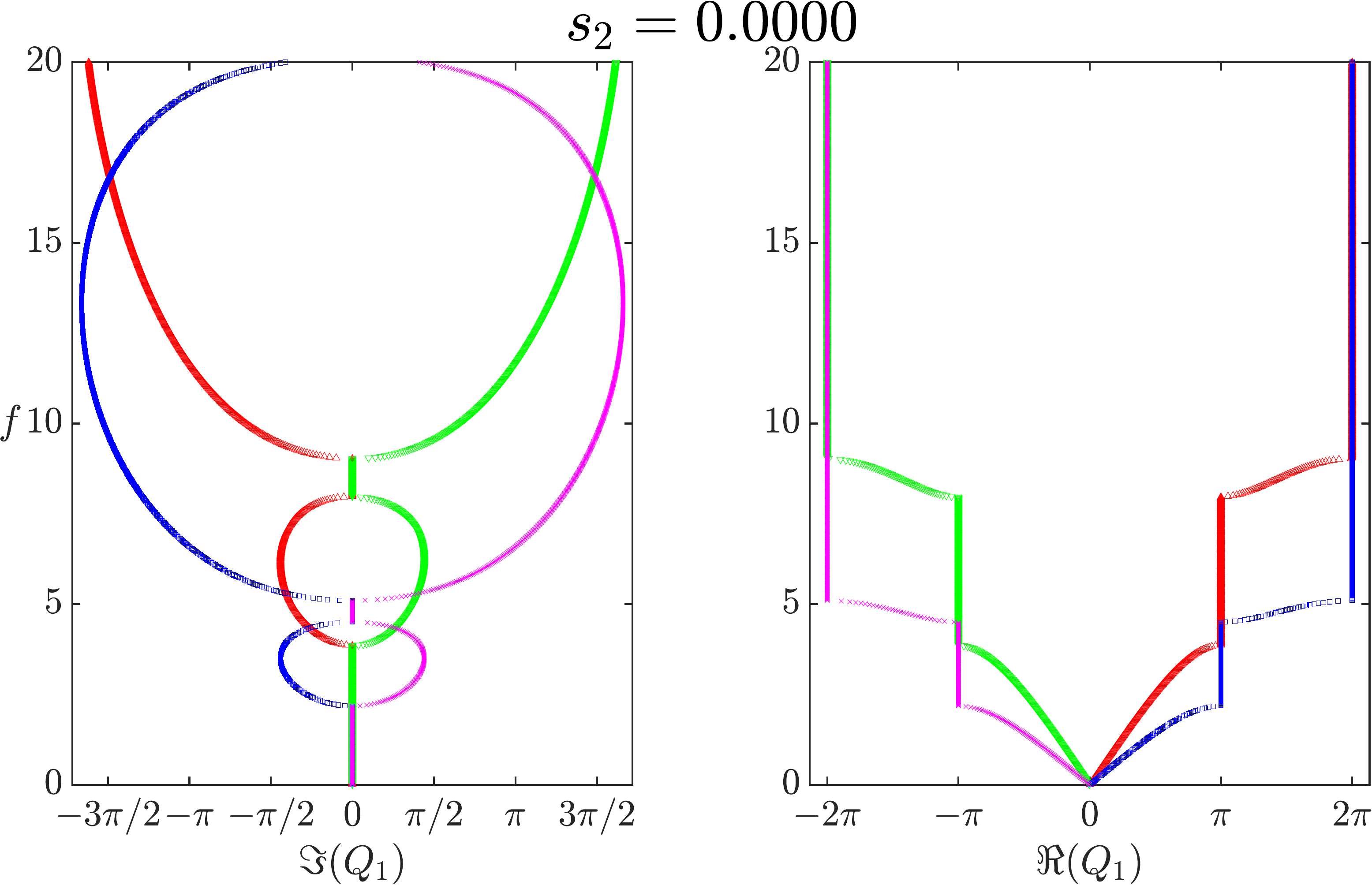}
		\caption{\label{fig:Q2_0}}
	\end{subfigure}%
	\begin{subfigure}[b]{0.33\linewidth}
		\centering\includegraphics[height=90pt,center]{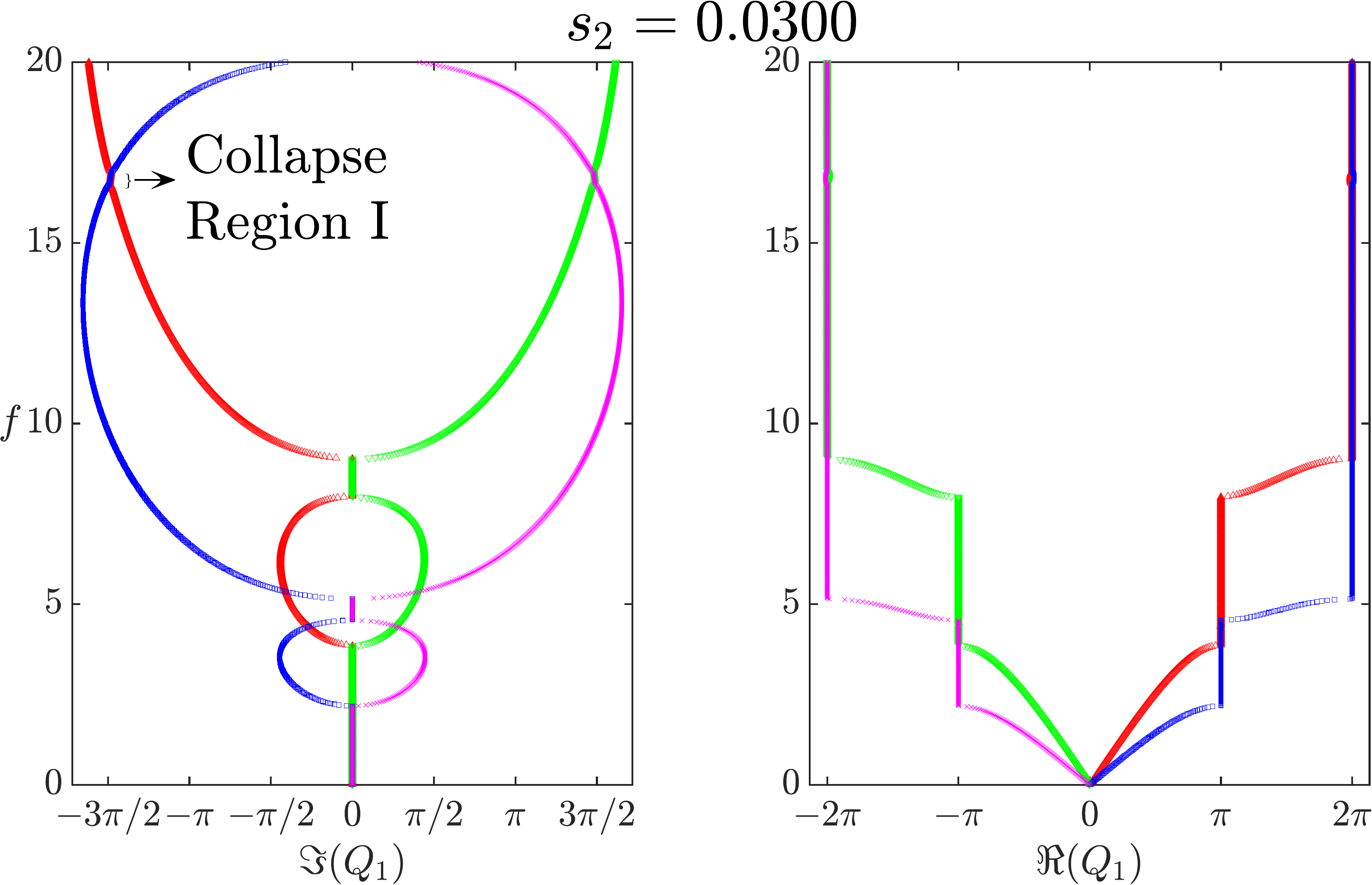}
		\caption{\label{fig:Q2_015}}
	\end{subfigure}%
	\begin{subfigure}[b]{0.33\linewidth}
		\centering\includegraphics[height=90pt,center]{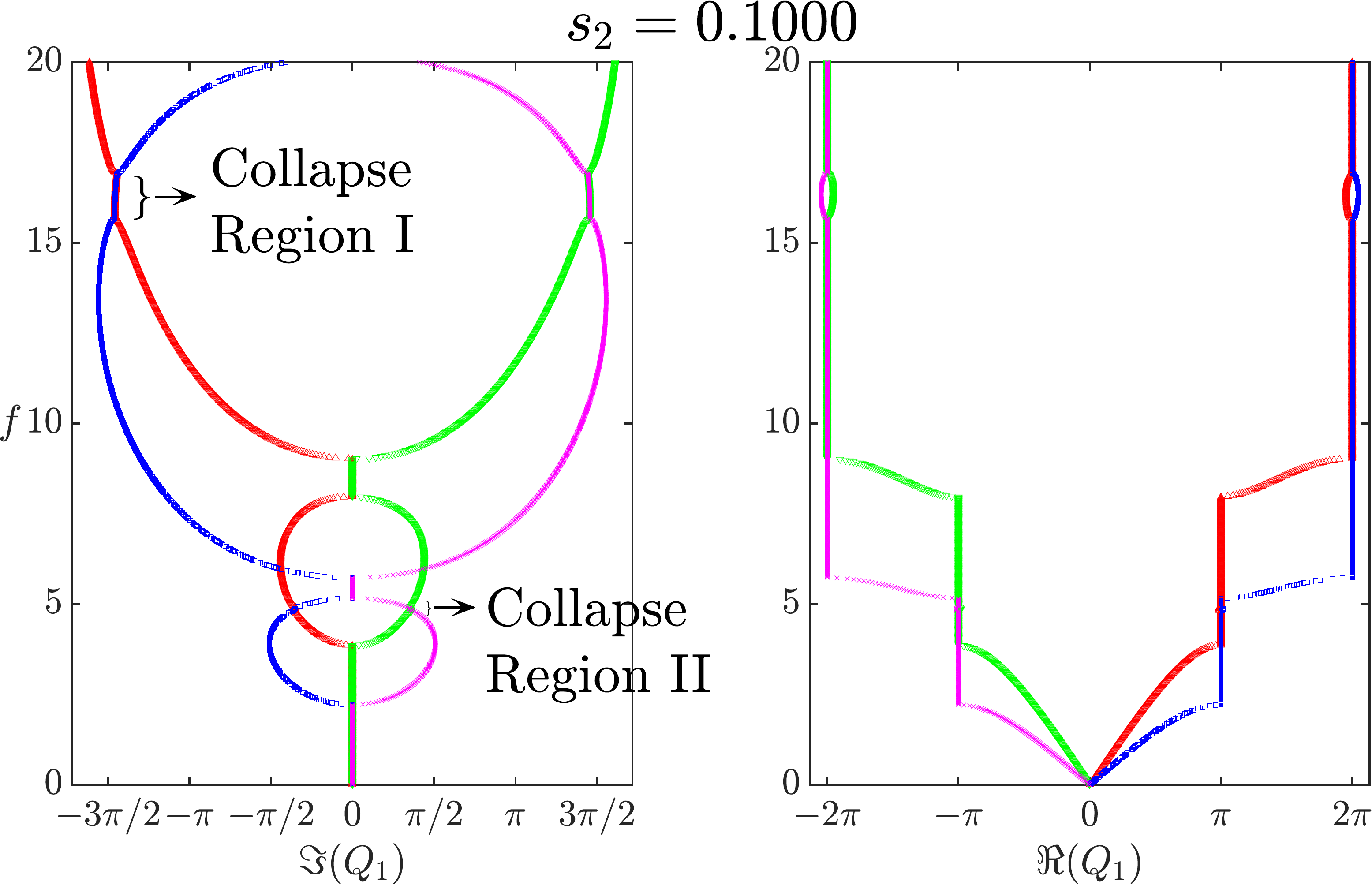}
		\caption{\label{fig:Q2_030}}
	\end{subfigure}
	\begin{subfigure}[b]{0.33\linewidth}
		\centering\includegraphics[height=90pt,center]{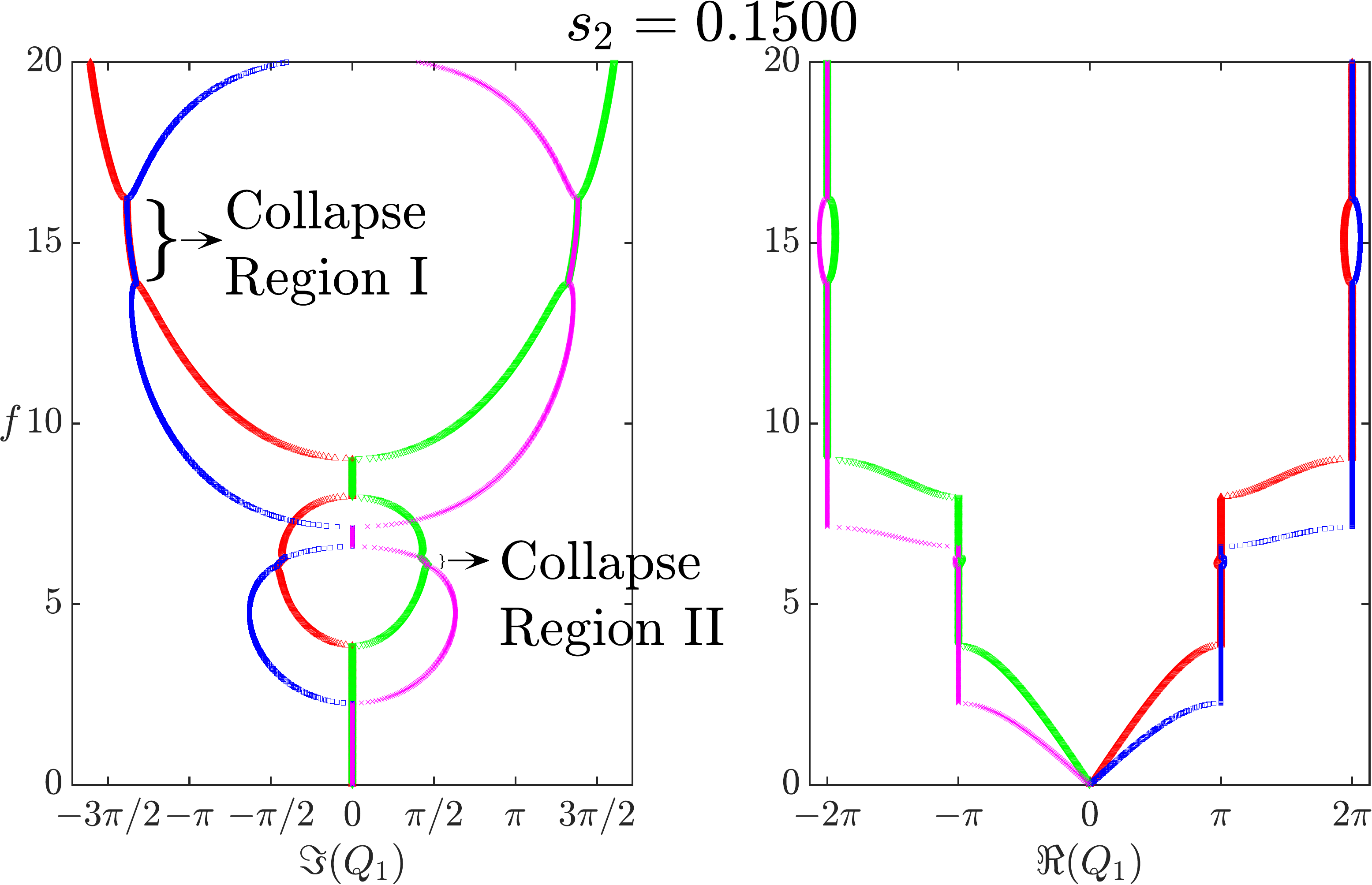}
		\caption{\label{fig:Q2_005}}
	\end{subfigure}%
	\begin{subfigure}[b]{0.33\linewidth}
		\centering\includegraphics[height=90pt,center]{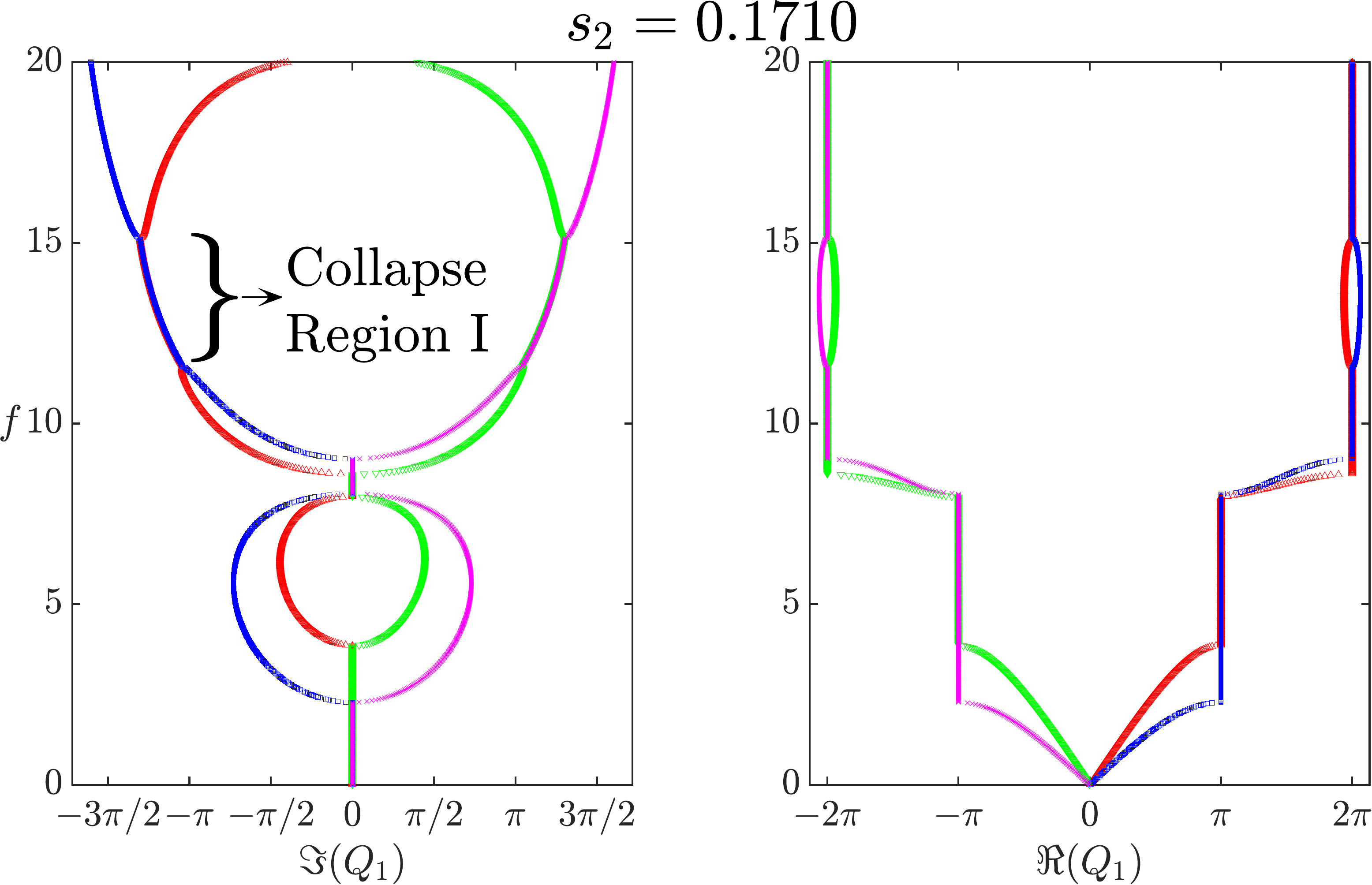}
		\caption{\label{fig:Q2_020}}
	\end{subfigure}%
	\begin{subfigure}[b]{0.33\linewidth}
		\centering\includegraphics[height=90pt,center]{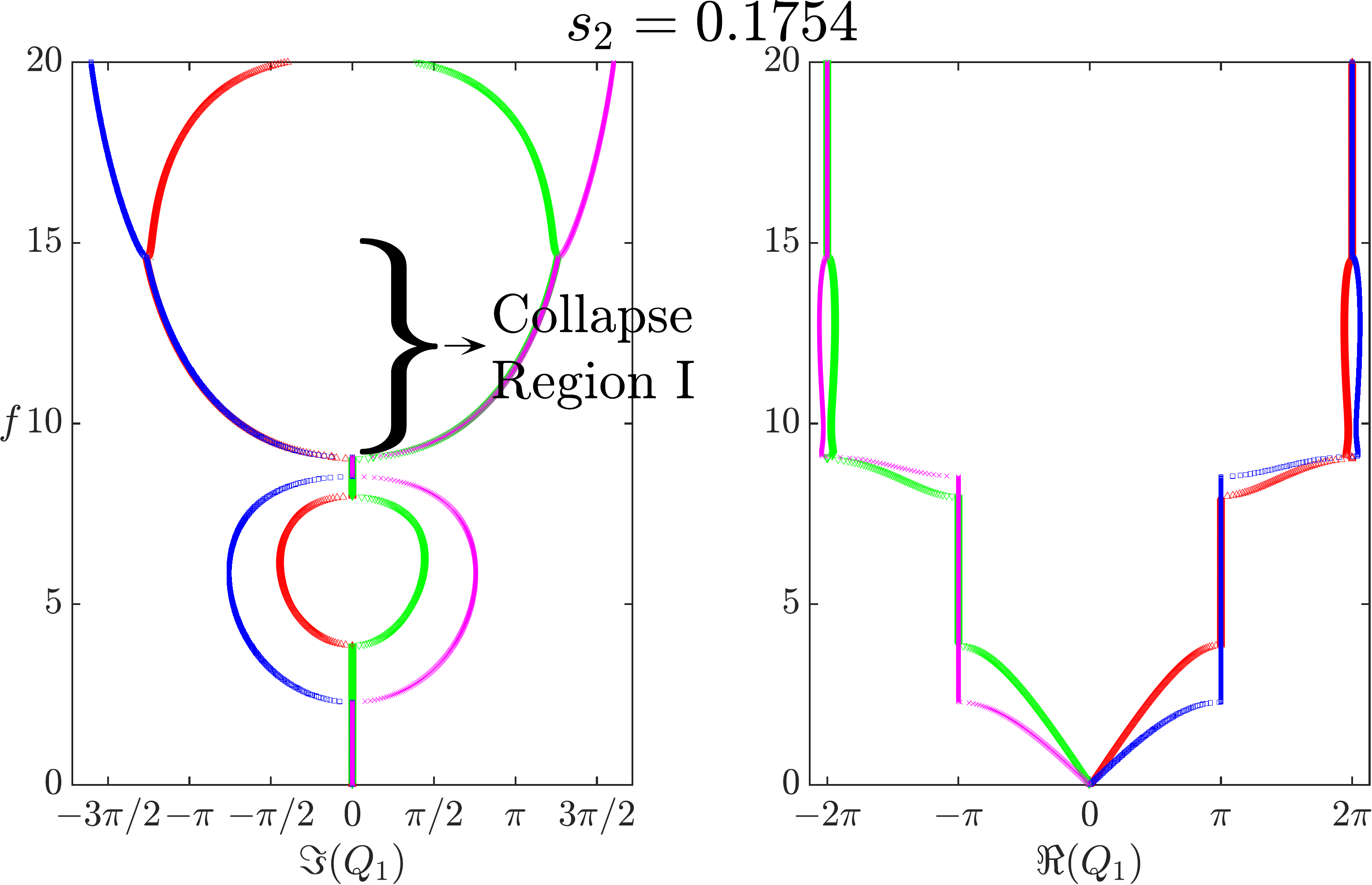}
		\caption{\label{fig:Q2_035}}
	\end{subfigure}
	\begin{subfigure}[b]{0.33\linewidth}
		\centering\includegraphics[height=90pt,center]{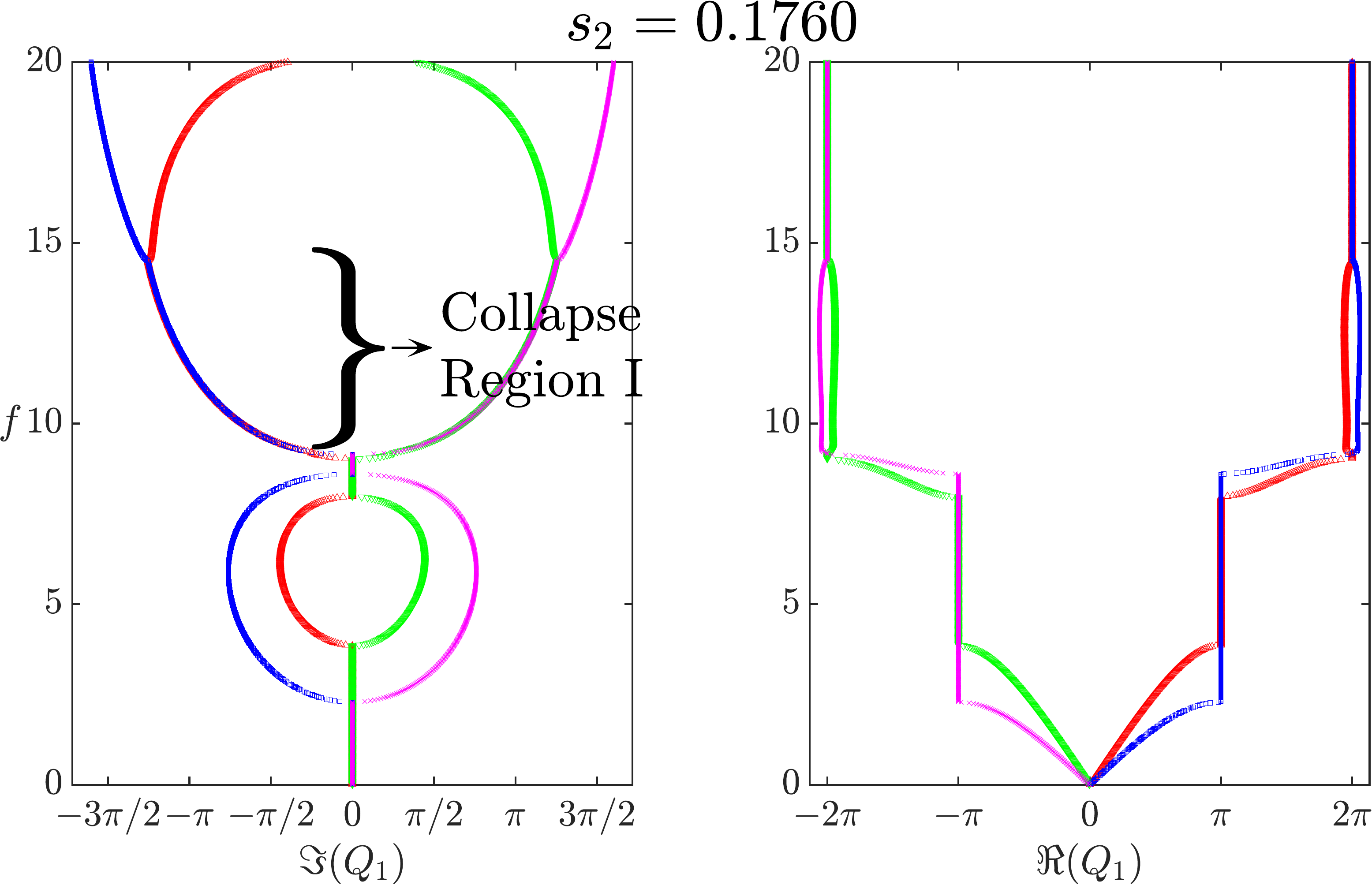}
		\caption{\label{fig:Q2_010}}
	\end{subfigure}%
	\begin{subfigure}[b]{0.33\linewidth}
		\centering\includegraphics[height=90pt,center]{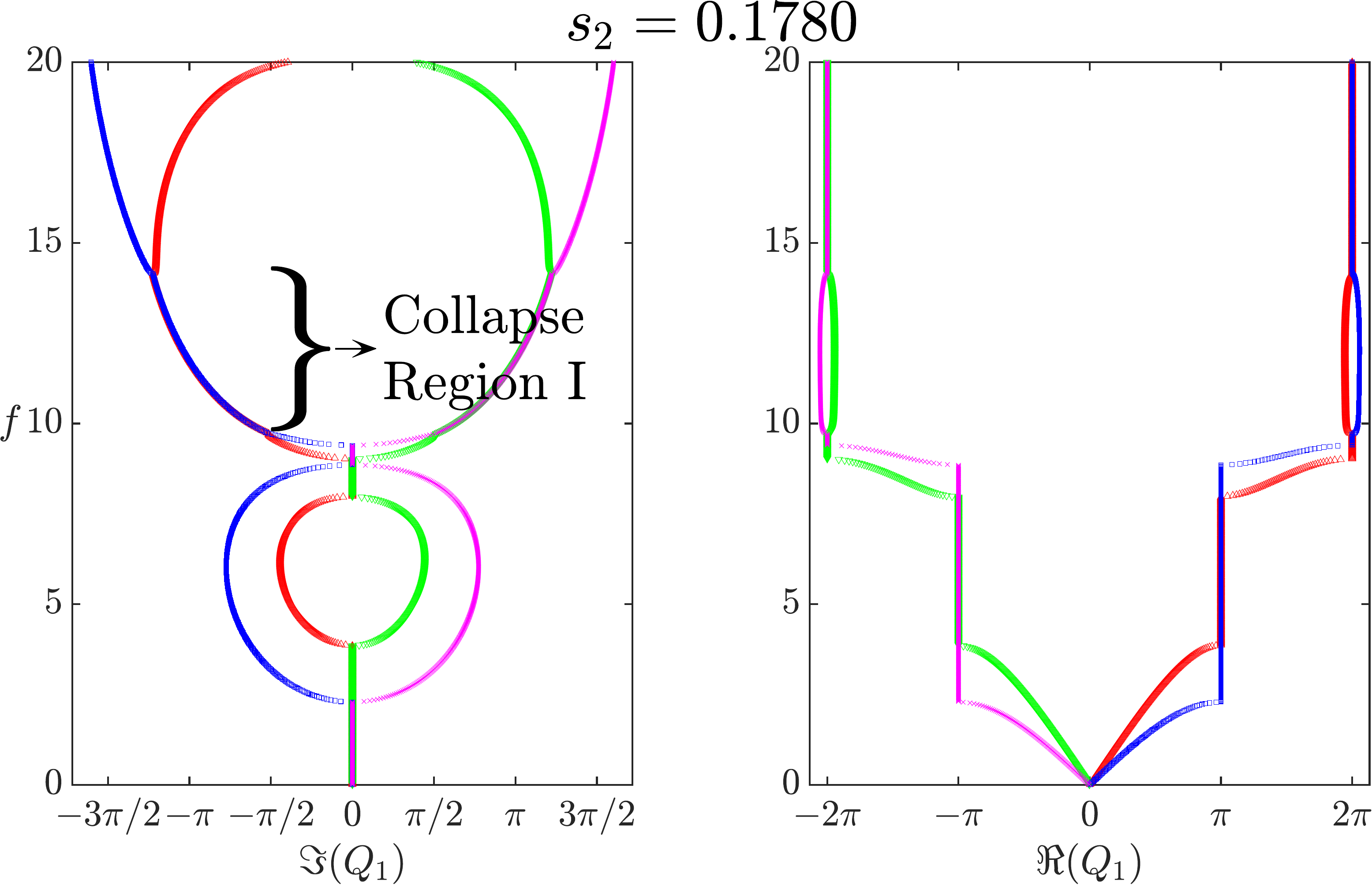}
		\caption{\label{fig:Q2_025}}
	\end{subfigure}%
	\begin{subfigure}[b]{0.33\linewidth}
		\centering\includegraphics[height=90pt,center]{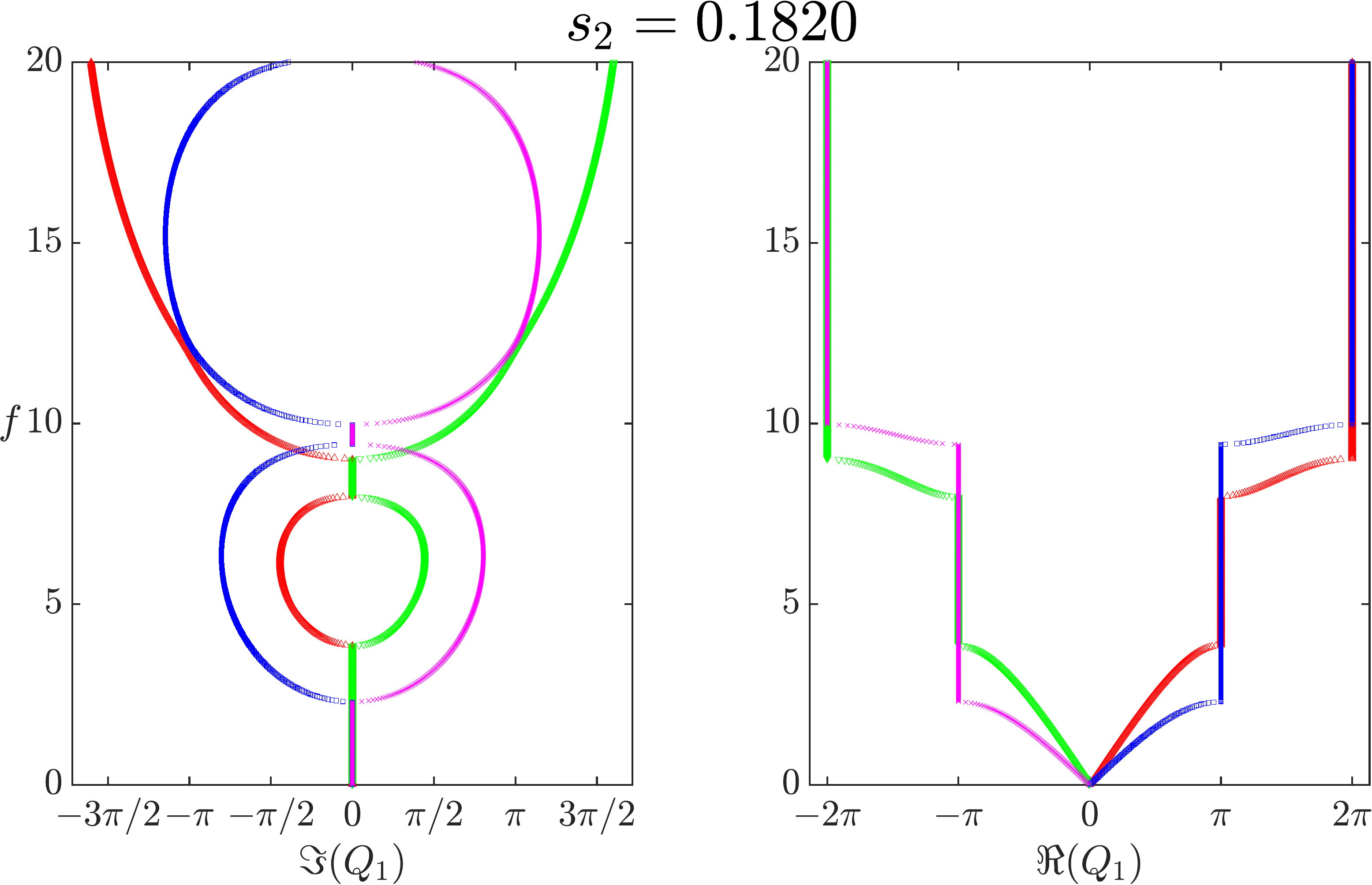}
		\caption{\label{fig:Q2_040}}
	\end{subfigure}
	\caption{\label{fig:bnd_strc_Q2} The real and imaginary parts of the phase advance at different values of $s_2$. $s_2$ changes from 0 to 0.1820 at 9 levels shown in (\subref{fig:Q2_0}) to (\subref{fig:Q2_040}). The first longitudinal-dominant stop band remain relatively stationary while the shear-dominant one shifts up as $s_2$ increases and the simple crossing (at $s_2 = 0$) turn into mode coalescence regions (around \SI{6}{kHz}). The imaginary parts of the phase advance are equal over the entire regions, while real parts match only at the end point, i.e. the locations of the exceptional points. The Collapse region I, at around \SI{15}{kHz} for $s_2=0.15$, shows up sooner and takes longer to evolve, and in the process flips what appears to be longitudinal- and shear-dominant branches twice. 
    }
\end{figure}

Figure~(\ref{fig:bnd_strc_Q2}) depicts the real and imaginary parts of the phase advance for increasing values of $s_2$. Note that it appears that the first stop band of the longitudinal-dominant mode is almost stationary while the shear-dominant one seems to shift up with increasing $s_2$. 
Over the one or two collapse regions, the stop bands of the shear dominant mode are colliding with the longitudinal ones. In the lower frequency one, the evolution seems to leave the first longitudinal-dominant stop band mostly intact. However, in the higher frequency collapse region evolution, the stop band dominant feature appears to flip twice. 

\begin{figure}[!ht]
   \centering\includegraphics[height=200pt,center]{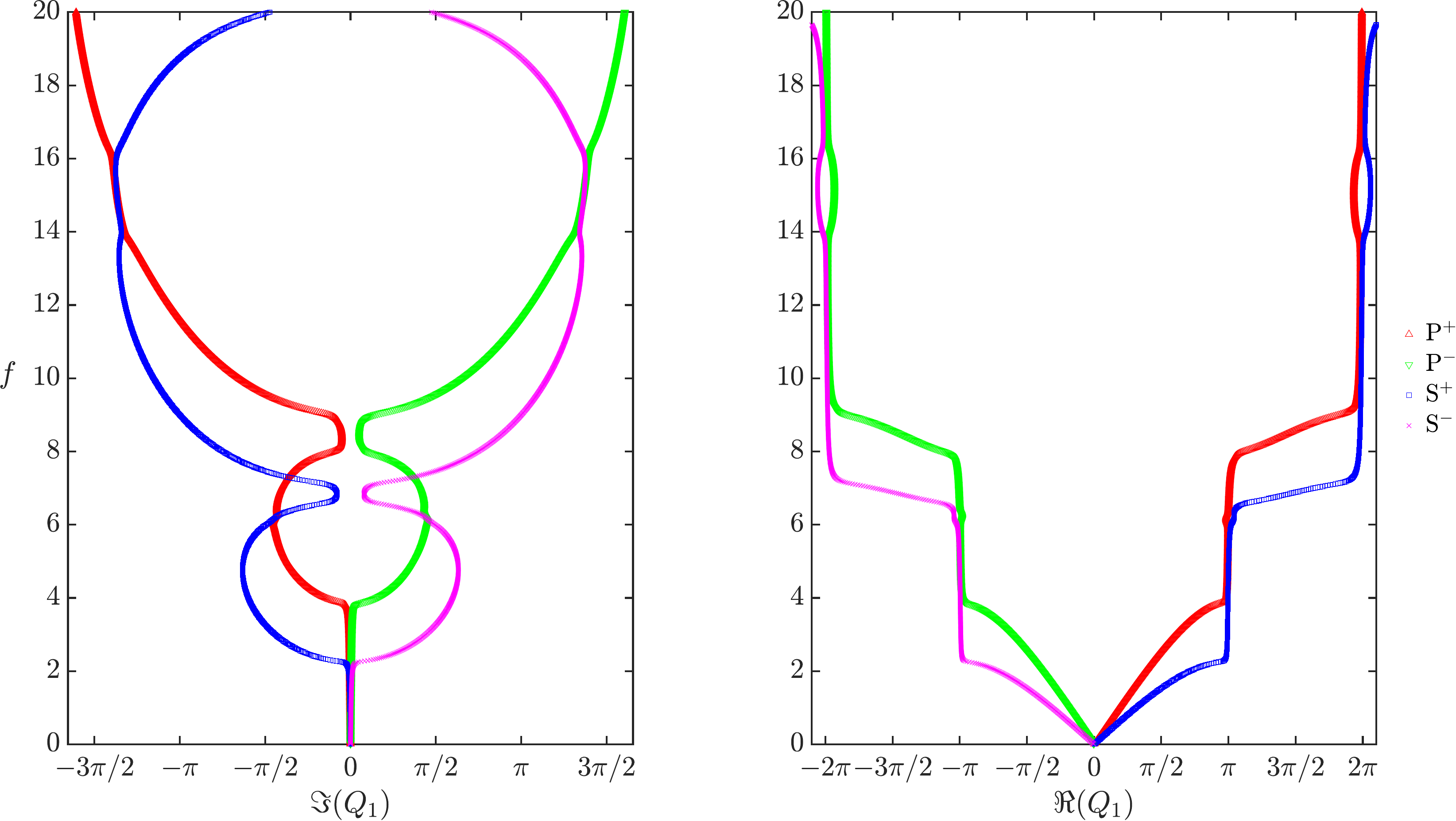}
   \caption{\label{fig:bnd_strc_s2_Q_lossy} The real and imaginary parts of the phase advance, $Q_1^{\alpha} = K_1^{\alpha}d$ at $s_2 = k_2/\omega = 0.15$ considering 1\% loss in the shear wave speed for layers 2 and 4 ($c_T''/c_T' = 0.01$). \gre{Frequency unit is kHz.}
   }
\end{figure}

Figure~(\ref{fig:bnd_strc_s2_Q_lossy}) shows the same for a lossy case (1\% added loss to both shear and longitudinal wave speeds in layers 2 and 4). The coalescence in $Q_1^\alpha$ occurs at four different frequencies below \SI{20}{kHz}. Note that while the band structure looks similar in the lossy case, one can no longer observe coalescence of complex eigenvalues, and the imaginary part of the eigenvalues simply cross at a single point (where the real parts are different) rather than collapsing for a finite frequency band.

\begin{figure}[!ht]
	\centering\includegraphics[height=230pt,center]{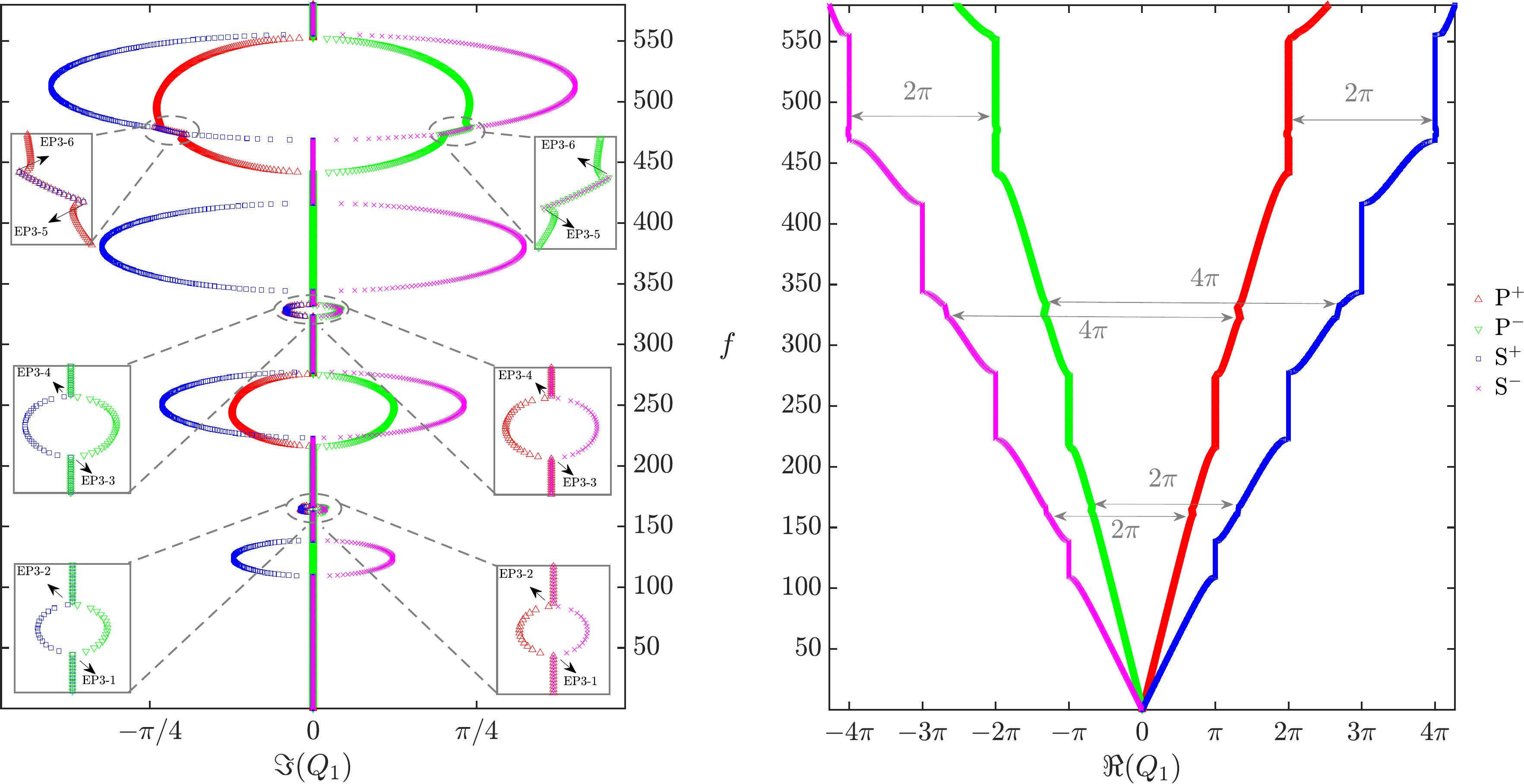}
	\caption{\label{fig:3L_Q2_020_ri} The real and imaginary part of phase advance at $s_2 = 0.04$ for the 2-phase, 3-layer structure studied in \cite{Amirkhizi2018}. Mode coalescence happens at the end points of the bands marked with number (1) to (3), the first two of which occur in the middle of pass bands, and are associated with pairs of opposite flux. \gre{Frequency unit is kHz.}
    }
\end{figure}

All of the four exceptional points studied for the 3-phase, 5-layer unit cell are located at the first and the second stop bands. In contrast, Figure~(\ref{fig:3L_Q2_020_ri}) shows the phase advance at $s_2 = 0.04$ for the 2-phase, 3-layer unit cell studied in \cite{Amirkhizi2018} with exceptional points located in the second and third pass bands (EP3-1 to -4)\footnote{A third pair, EP3-5 and -6 also appears in the fourth stop band, which due to its similarity to the previous case is not discussed further.}. 
Unlike the previous case in which imaginary parts of the phase advance were equal for the width of the ``collapse bands'', in this case the real parts match while the imaginary parts only do so at the two limits. Additionally, while the signs of the real parts are the same, the signs of the imaginary part are opposite, indicating appearance of evanescent modes with decaying and increasing amplitudes. This is indicated by the $2\pi$ or $4\pi$ differences in the real part of the phase advance occurring between P and S pairs with opposite flux ($\pm$). To the best of our knowledge, such a broken phase symmetry phenomenon has not been observed before, particularly in coupled stress waves in linear layered media, 

\subsection{Modal chirality and flux}

\begin{figure}[!ht]
    \begin{subfigure}[b]{0.20\linewidth}
        \centering\includegraphics[height=90pt,center]{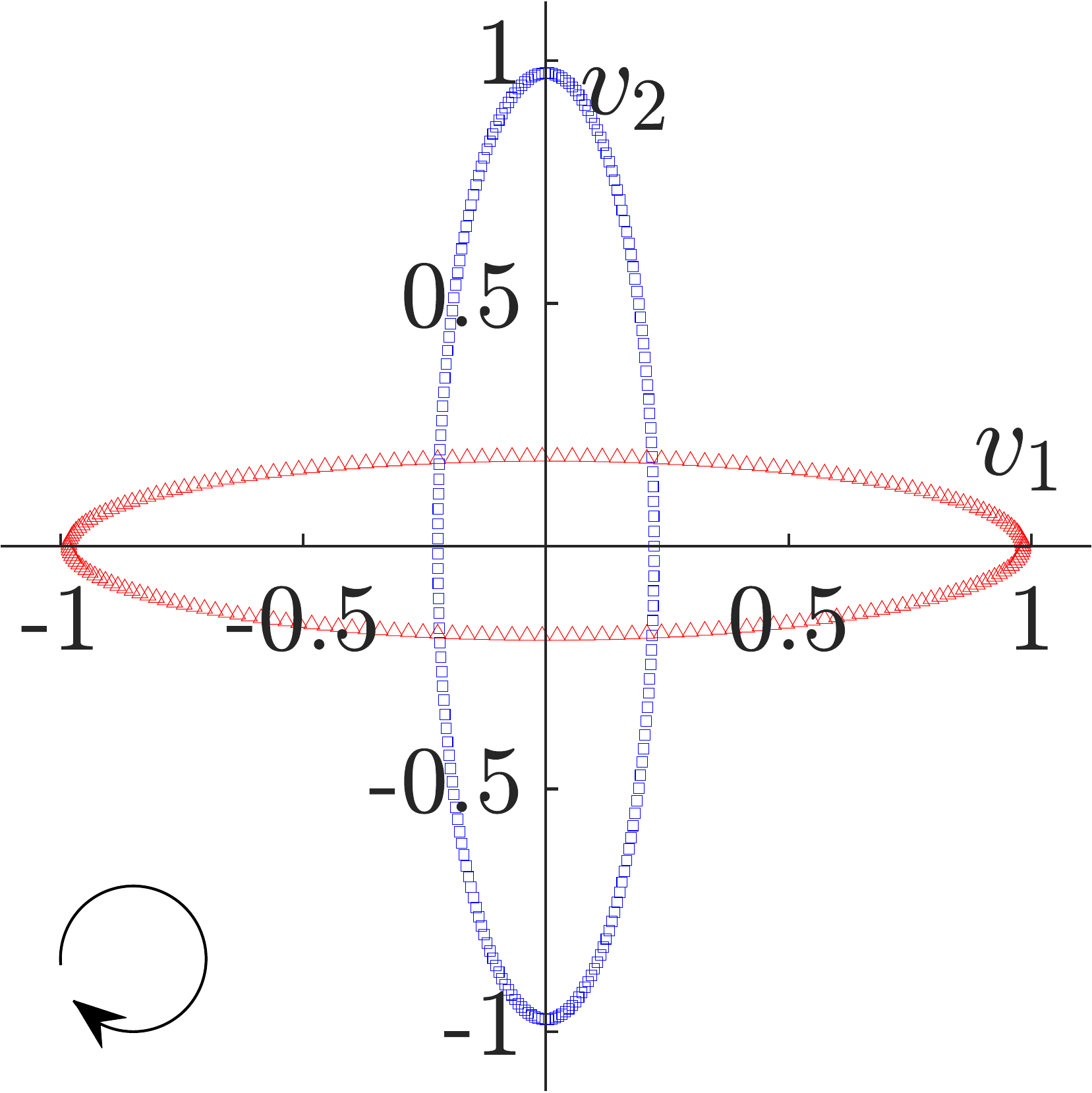}
        \caption{\label{fig:EP512_1}}
    \end{subfigure}%
    \begin{subfigure}[b]{0.20\linewidth}
        \centering\includegraphics[height=90pt,center]{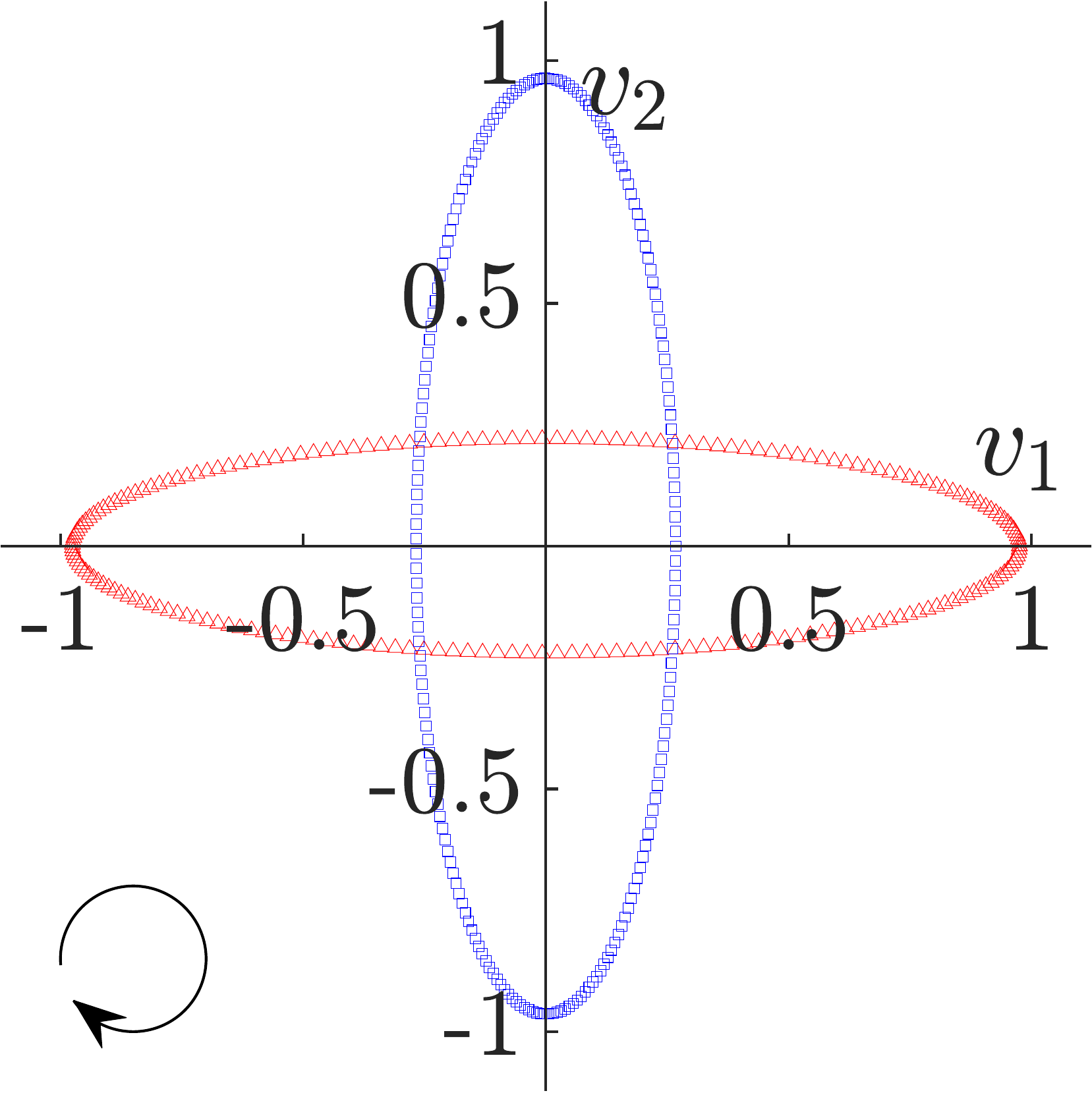}
        \caption{\label{fig:EP512_2}}
    \end{subfigure}%
    \begin{subfigure}[b]{0.20\linewidth}
        \centering\includegraphics[height=90pt,center]{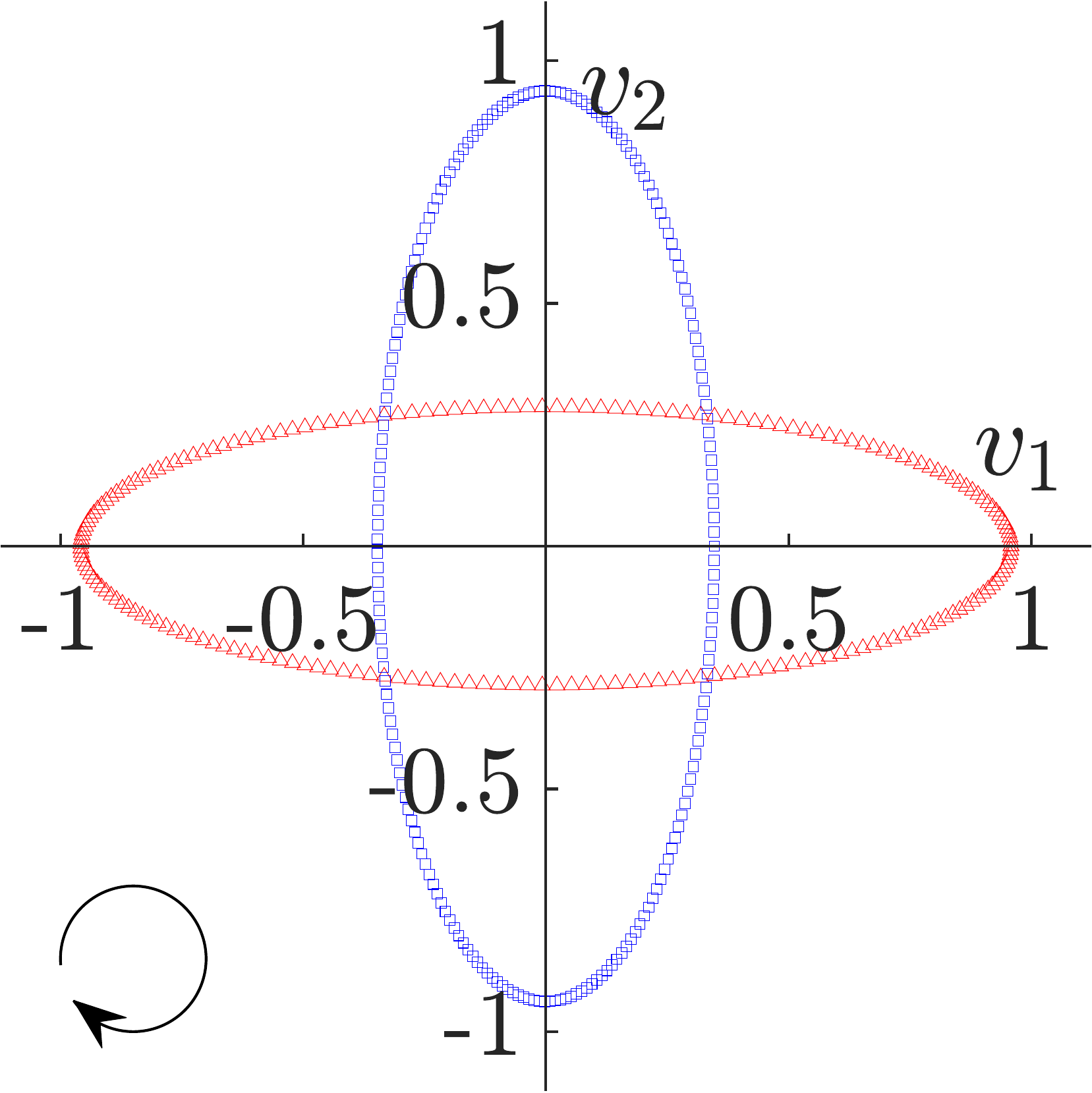}
        \caption{\label{fig:EP512_3}}
    \end{subfigure}%
    \begin{subfigure}[b]{0.20\linewidth}
        \centering\includegraphics[height=90pt,center]{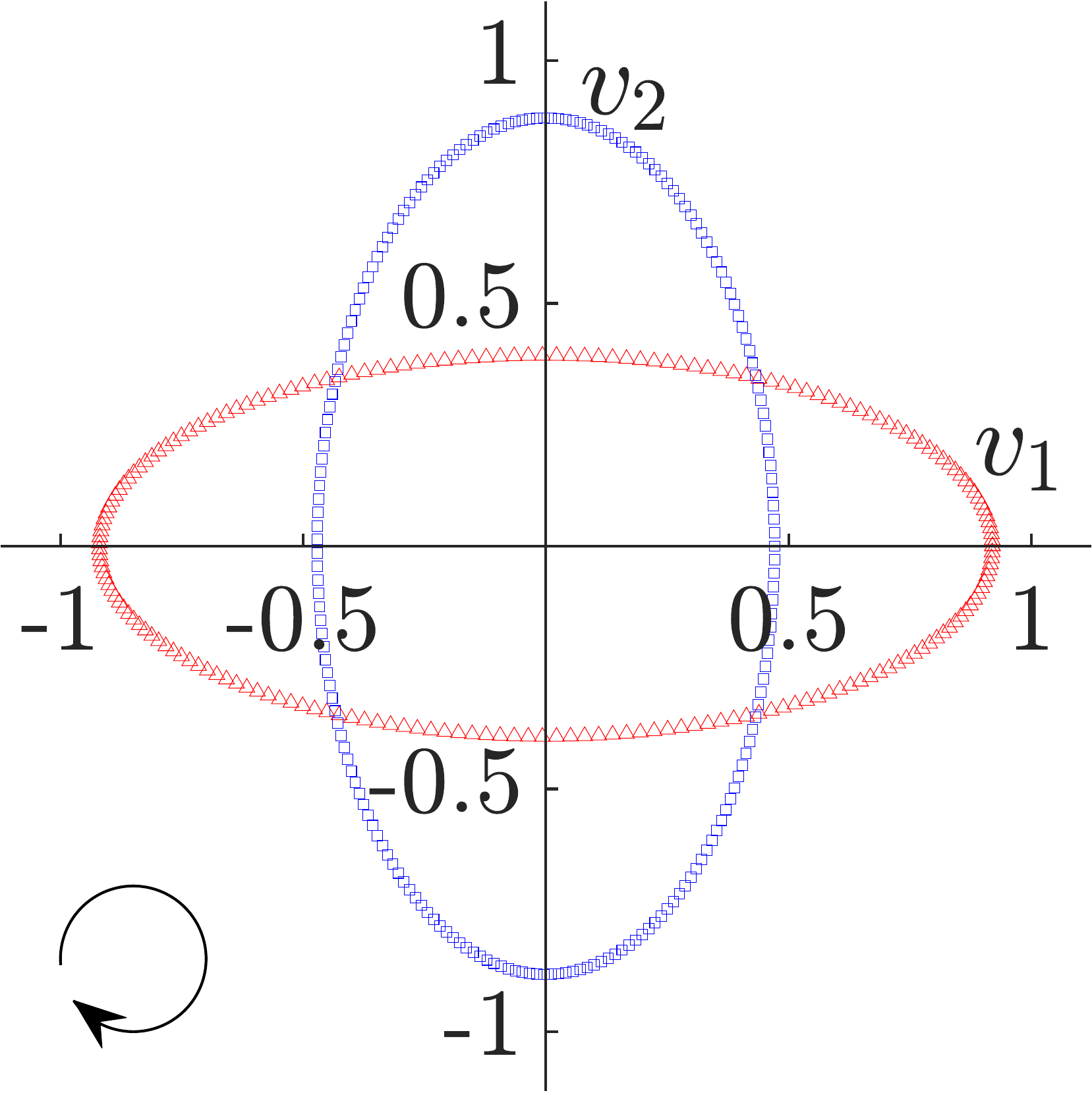}
        \caption{\label{fig:EP512_4}}
    \end{subfigure}%
    \begin{subfigure}[b]{0.20\linewidth}
        \centering\includegraphics[height=90pt,center]{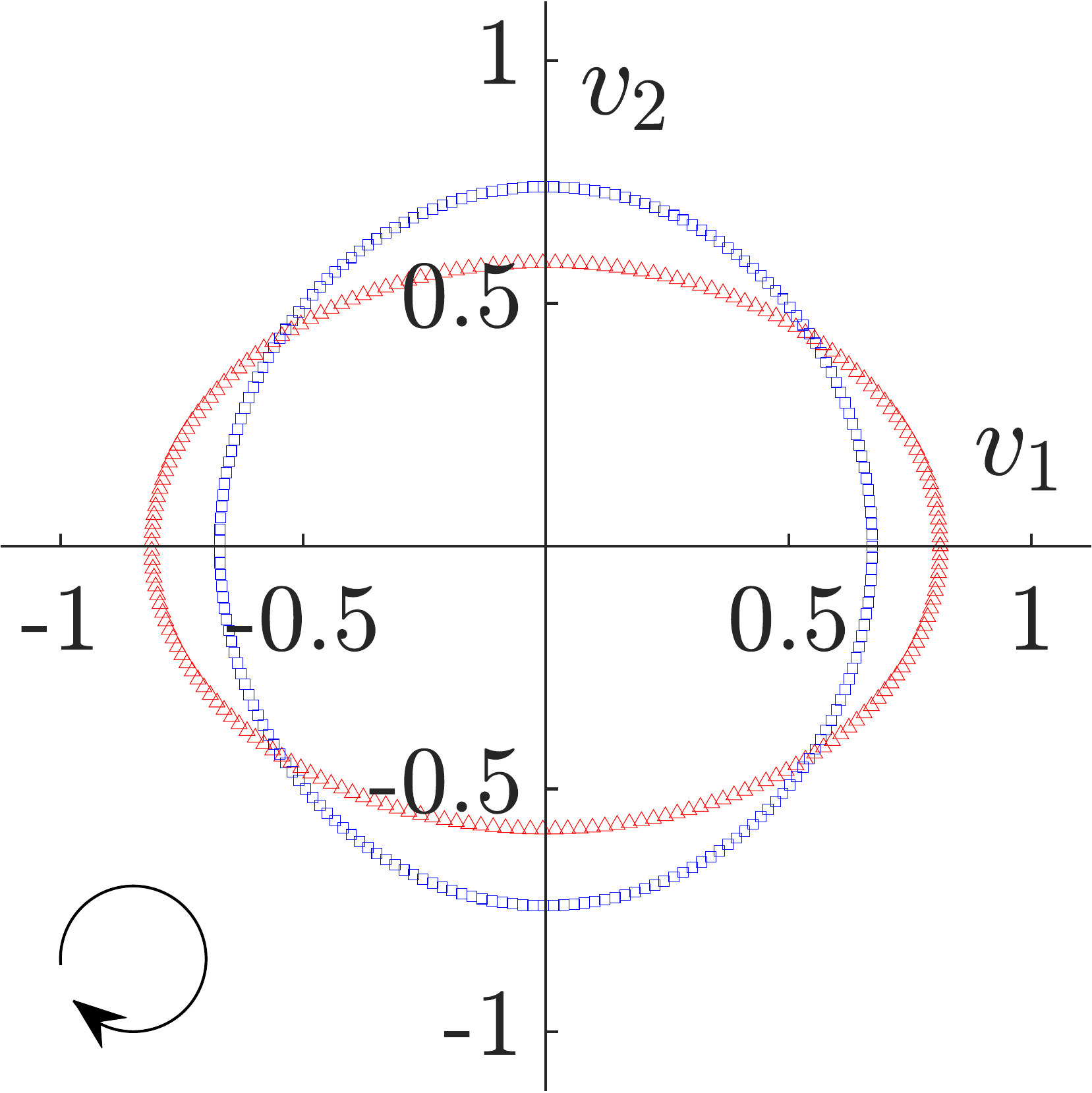}
        \caption{\label{fig:EP512_5}}
    \end{subfigure}
    \begin{subfigure}[b]{0.20\linewidth}
        \centering\includegraphics[height=90pt,center]{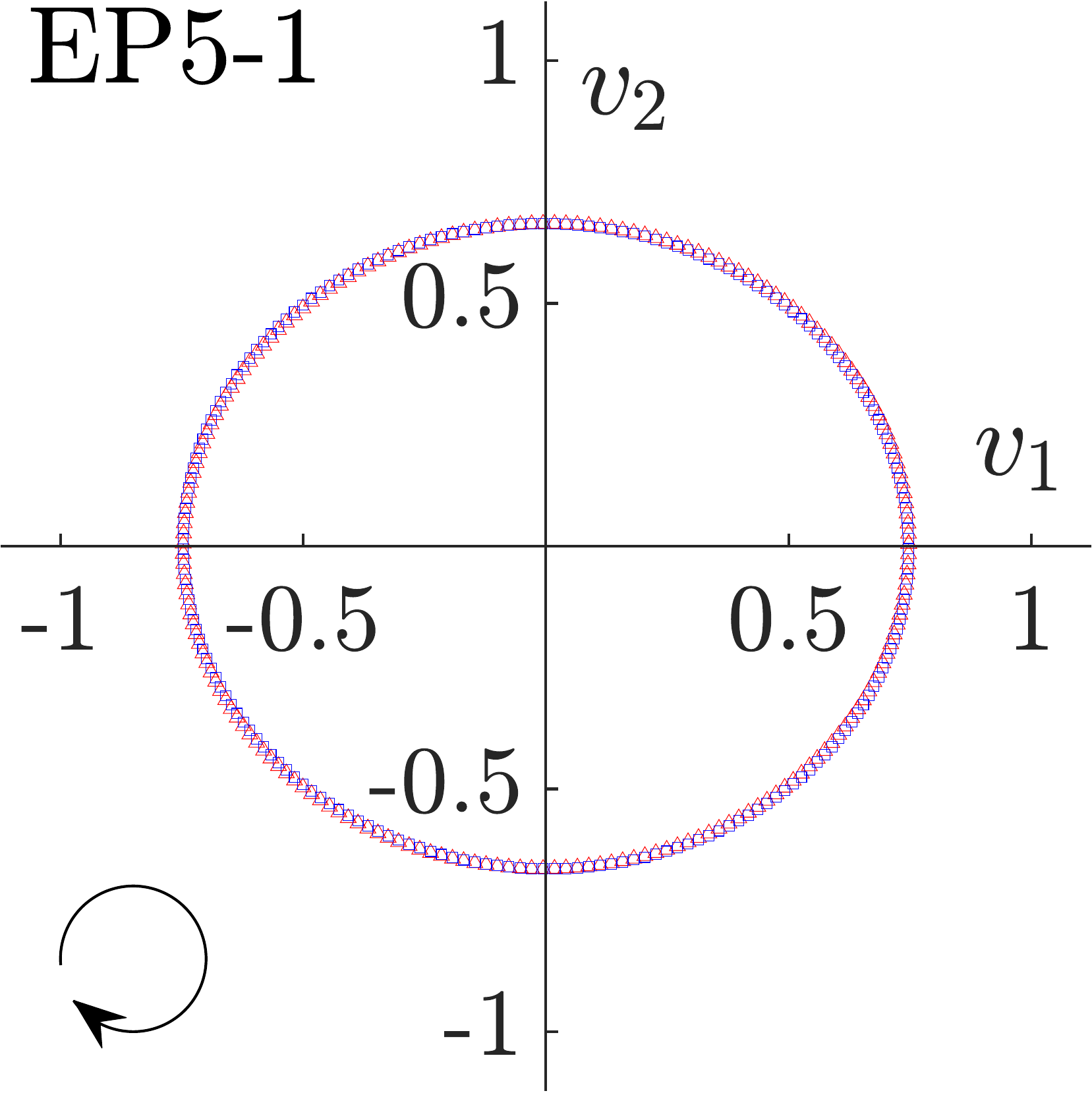}
        \caption{\label{fig:EP512_6}}
    \end{subfigure}%
    \begin{subfigure}[b]{0.20\linewidth}
        \centering\includegraphics[height=90pt,center]{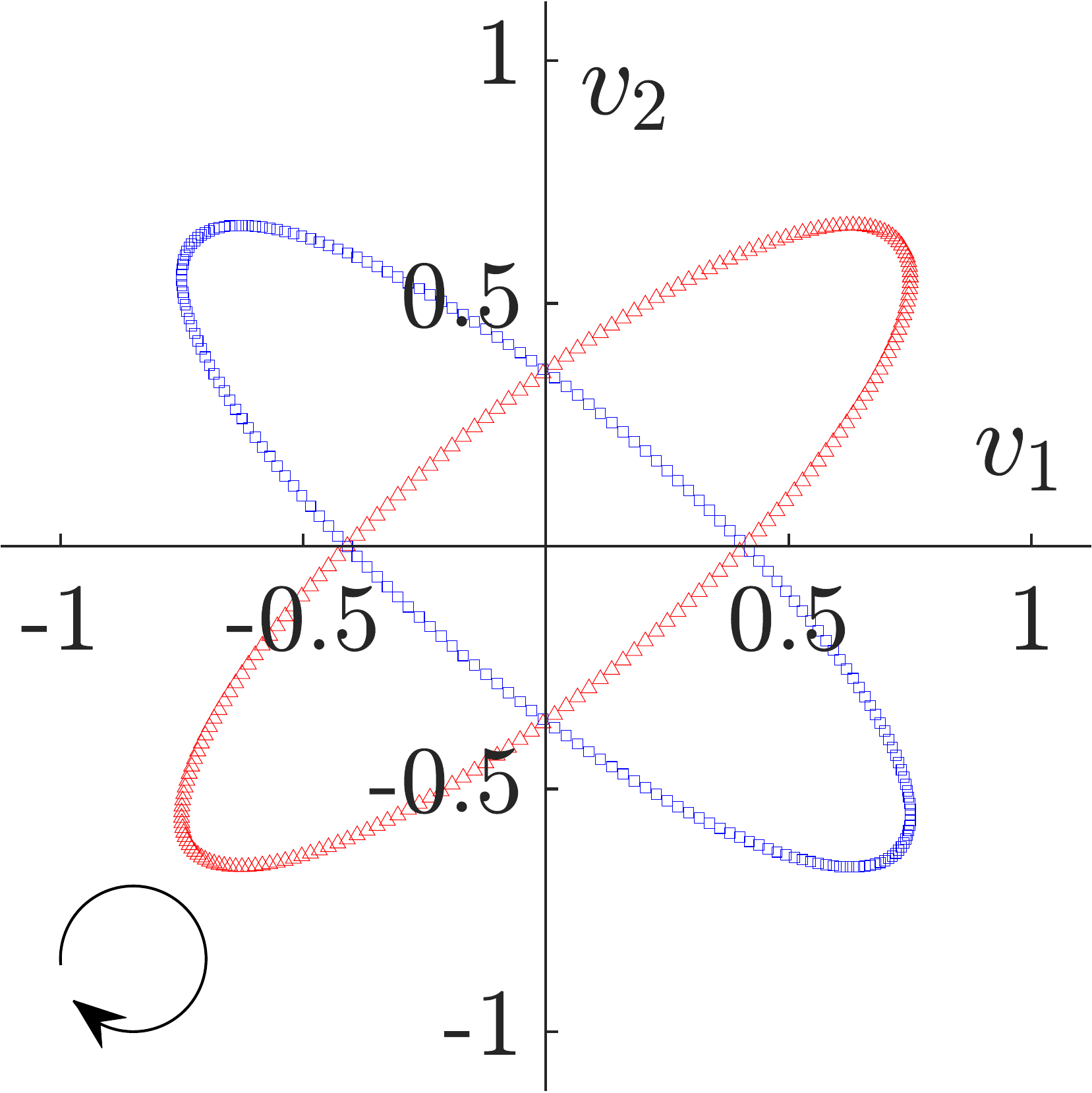}
        \caption{\label{fig:EP512_7}}
    \end{subfigure}%
    \begin{subfigure}[b]{0.20\linewidth}
        \centering\includegraphics[height=90pt,center]{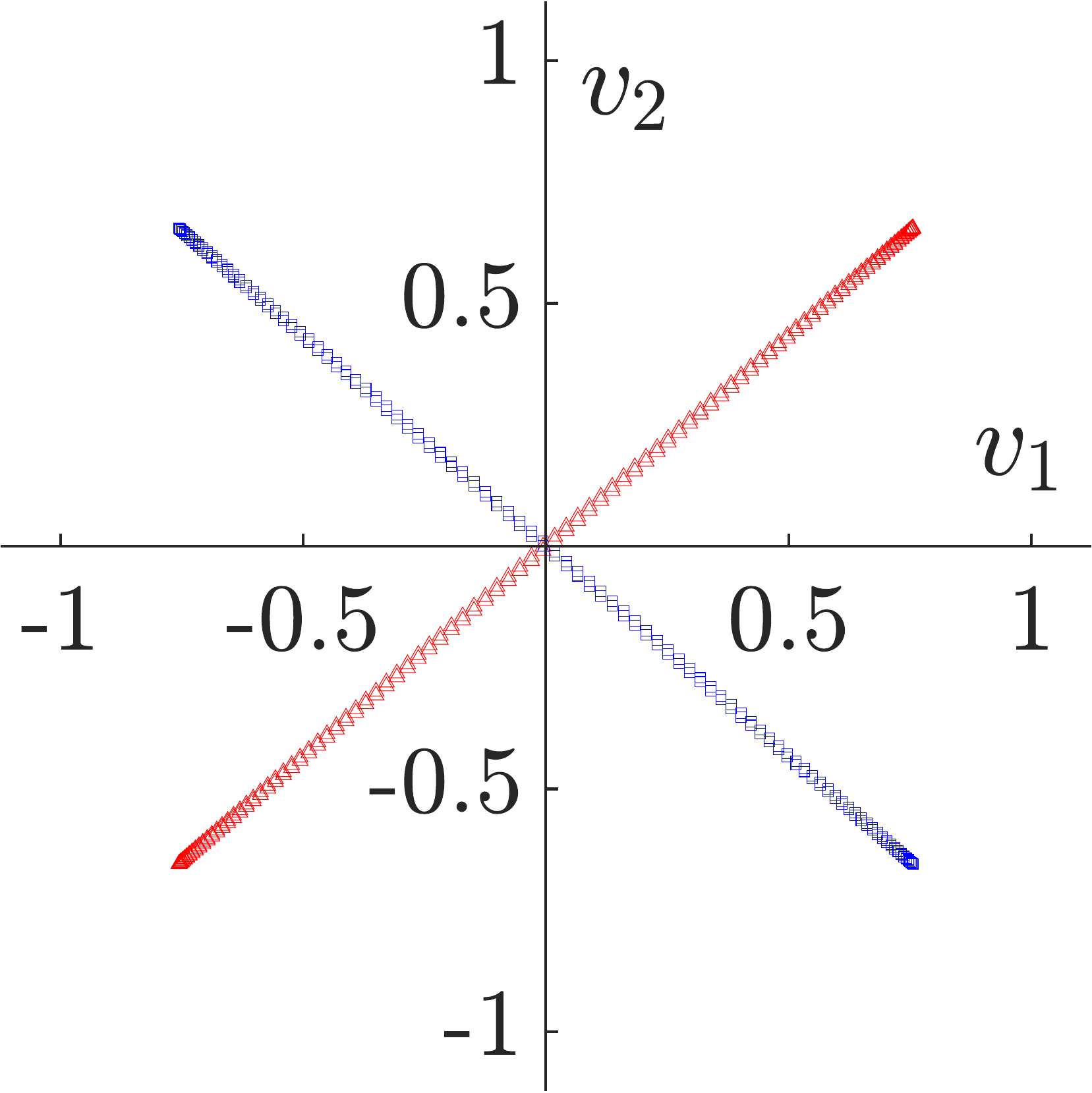}
        \caption{\label{fig:EP512_8}}
    \end{subfigure}%
    \begin{subfigure}[b]{0.20\linewidth}
        \centering\includegraphics[height=90pt,center]{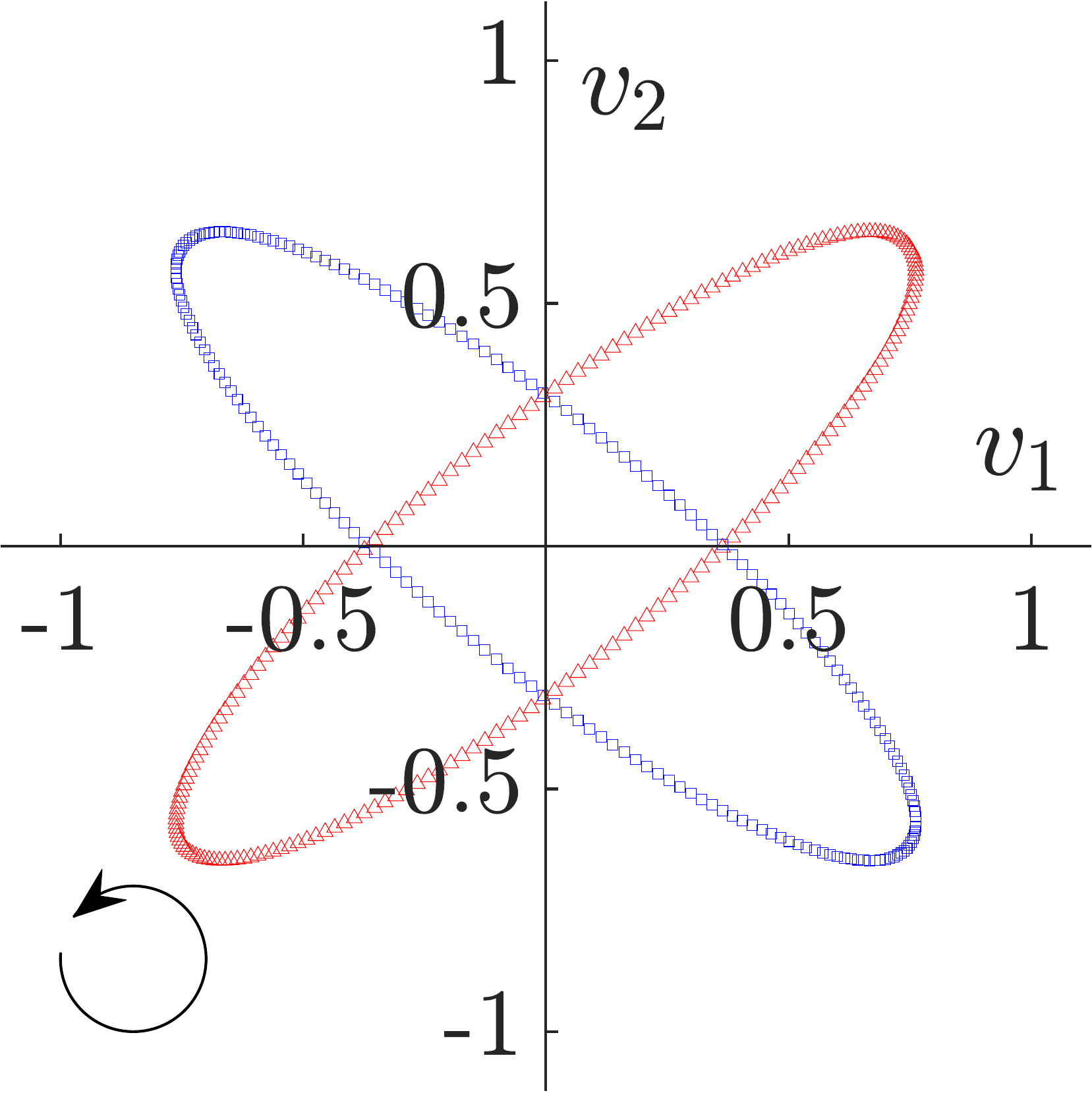}
        \caption{\label{fig:EP512_9}}
    \end{subfigure}%
    \begin{subfigure}[b]{0.20\linewidth}
        \centering\includegraphics[height=90pt,center]{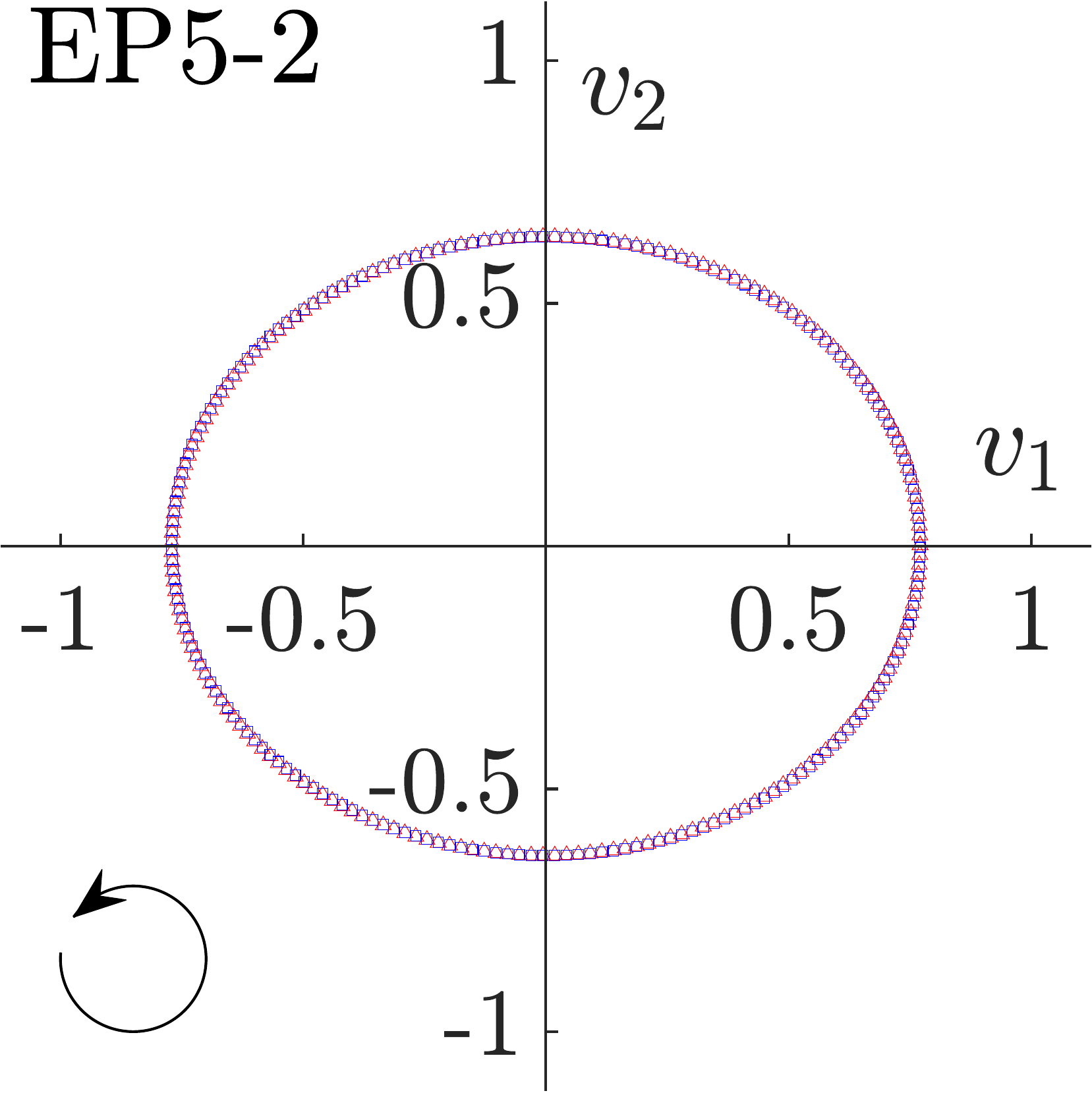}
        \caption{\label{fig:EP512_10}}
    \end{subfigure}
    \begin{subfigure}[b]{0.20\linewidth}
        \centering\includegraphics[height=90pt,center]{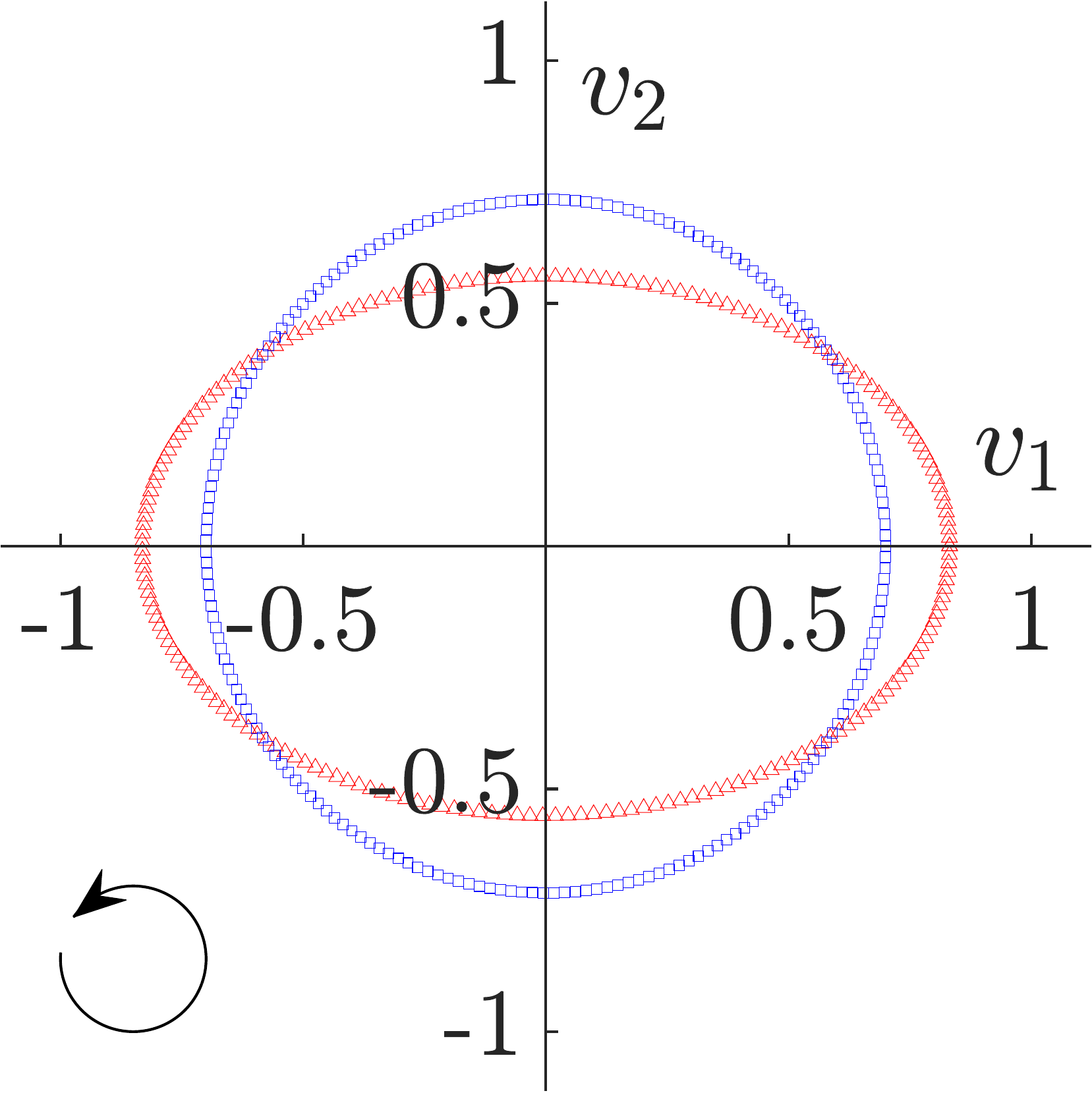}
        \caption{\label{fig:EP512_11}}
    \end{subfigure}%
    \begin{subfigure}[b]{0.20\linewidth}
        \centering\includegraphics[height=90pt,center]{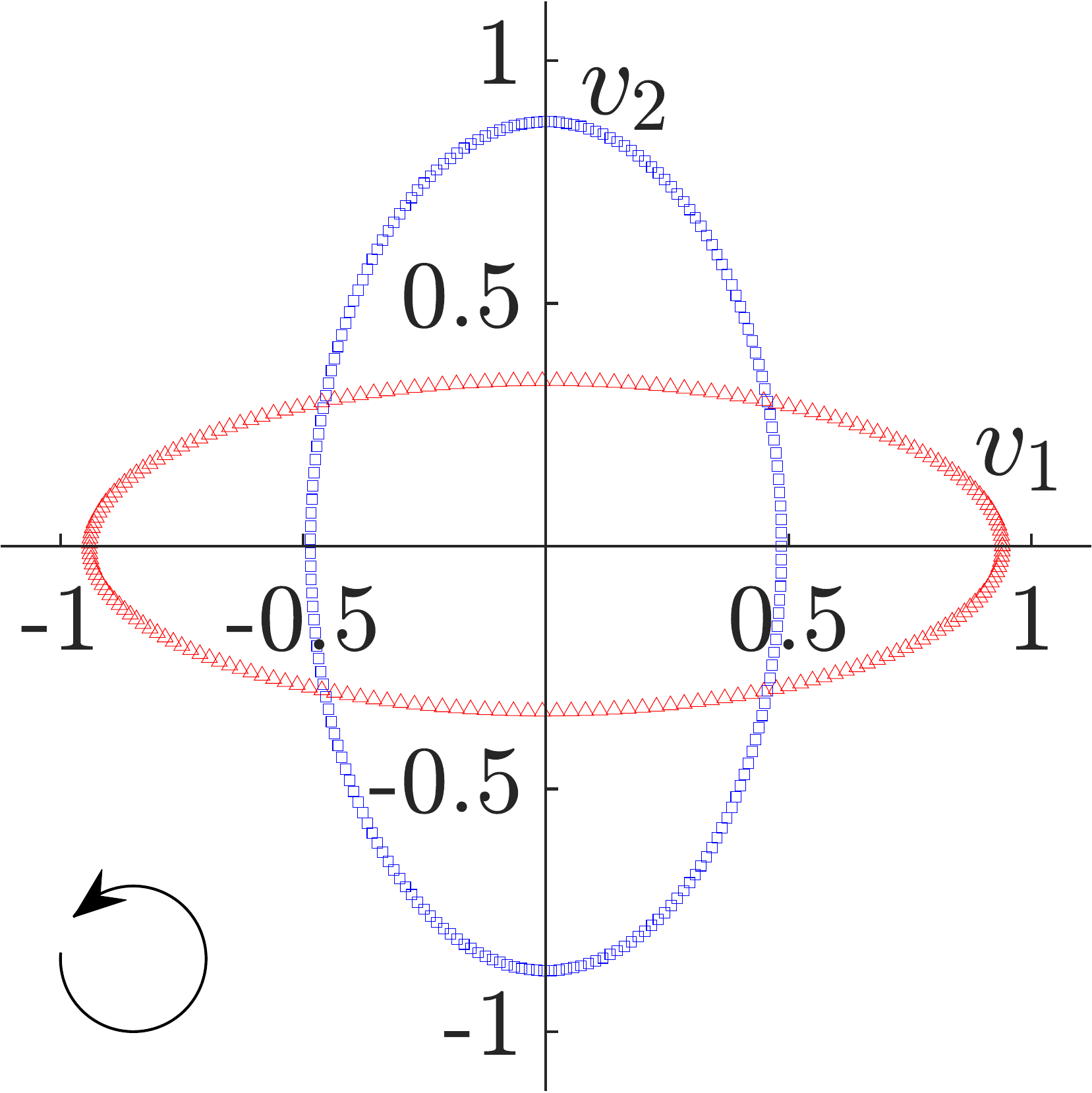}
        \caption{\label{fig:EP512_12}}
    \end{subfigure}%
    \begin{subfigure}[b]{0.20\linewidth}
        \centering\includegraphics[height=90pt,center]{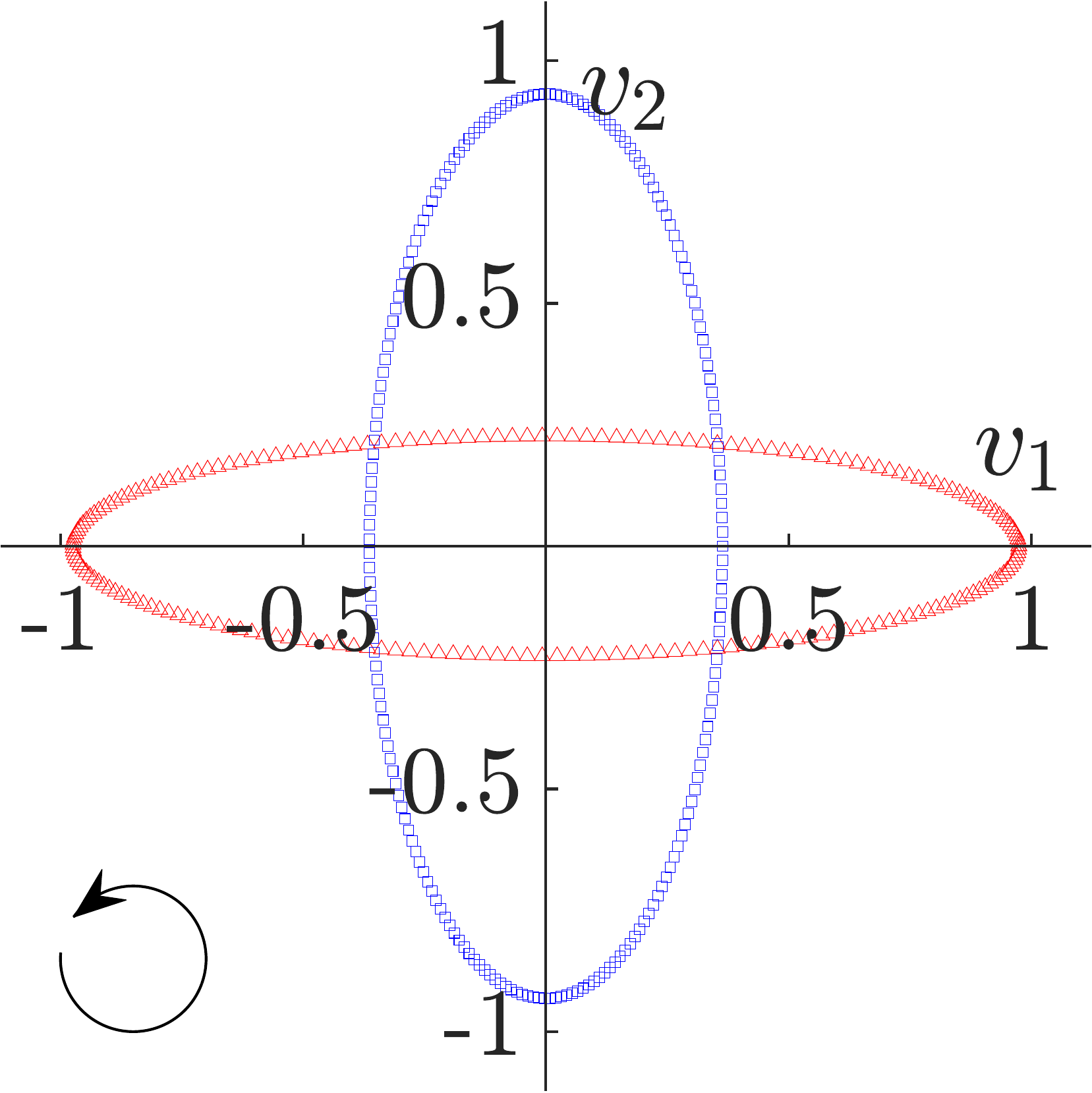}
        \caption{\label{fig:EP512_13}}
    \end{subfigure}%
    \begin{subfigure}[b]{0.20\linewidth}
        \centering\includegraphics[height=90pt,center]{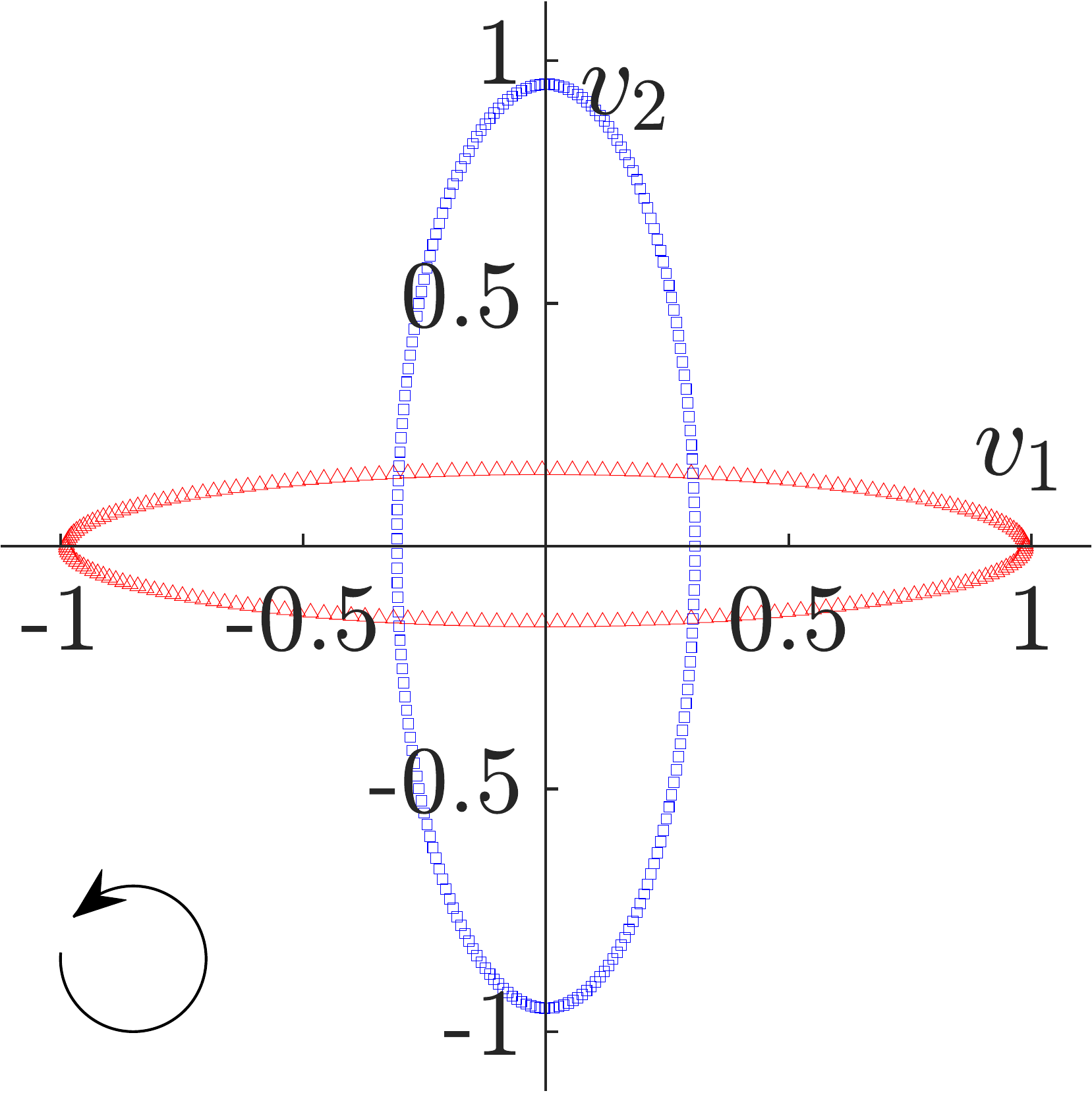}
        \caption{\label{fig:EP512_14}}
    \end{subfigure}%
    \begin{subfigure}[b]{0.20\linewidth}
        \centering\includegraphics[height=90pt,center]{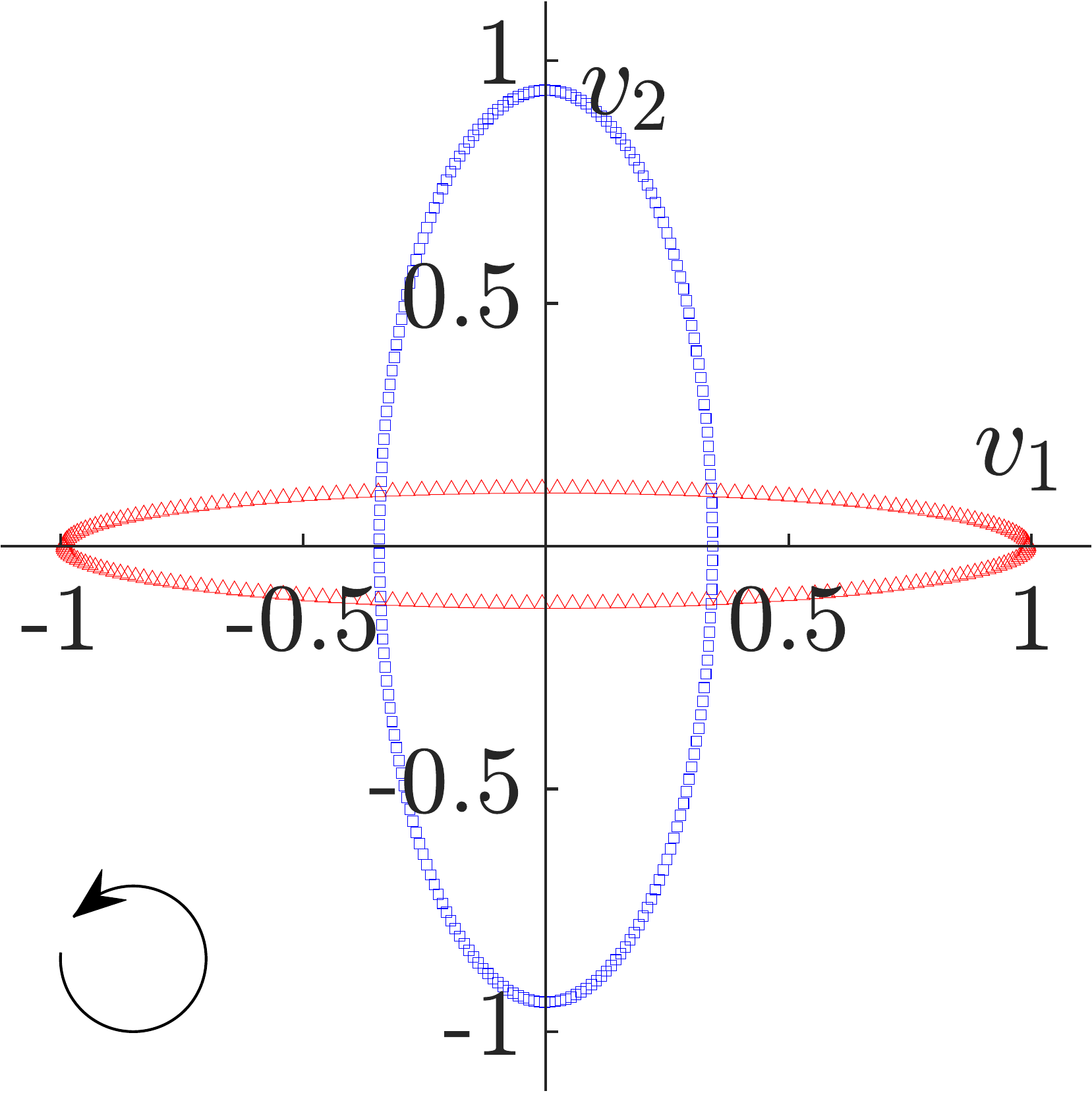}
        \caption{\label{fig:EP512_15}}
    \end{subfigure}%
\caption{\label{fig:pol_s2_015} The particle velocity trajectories at $x_2 = 0$ for modes P$^+$ and S$^+$ (negative imaginary parts) at a series of frequencies around exceptional points, EP5-1 and EP5-2 ((a) to (e) are at frequencies less than EP5-1, (f) to (j) are at frequencies between EP5-1 and EP5-2, with (f) and (j) being at the EPs, and (k) to (o) are at frequencies larger than that of EP5-2) The CW and CCW arrows represent the similar chirality of the two modes. Before and after the colliding region, there is a $\pi/2$ phase difference between $v_1$ and $v_2$, which results in vertical and horizontal elliptical trajectories. At the exceptional points the trajectories coalesce, though they are still not exactly circular; See Figure(\ref{fig:v1_s2}), (\ref{fig:v2_s2}). Within the ``collapse'' region the phase difference is no longer $\pi/2$ leading to oblique elliptical trajectories. At the center of this region the ellipses become two lines, after which chirality of the trajectories flip from CW to CCW. The reverse chirality evolution is observed for the P$^-$ and S$^-$ modes with positive imaginary parts. \gre{Quiver plot animations of selected modes (particle velocity vector) is provided in the supplementary materials.}
}
\end{figure}

It is also interesting to take a closer look at the modal behavior and investigate the polarizations more closely below, within, and above the intervals between the exceptional points. One can utilize the complex amplitudes shown in Figure~(\ref{fig:mode_s2}) to construct a cyclic trajectory with $e^{i(-\bm{k}.\bm{x} + \omega t)}$ phasor form. Figure~(\ref{fig:pol_s2_015}) shows the trajectory of real particle velocity vector, consisting of $v_1$ and $v_2$, at the center of unit cell, $x_2 = 0$. This is depicted at a number of frequency steps near EP5-1 and EP5-2 for the lossless 3-phase, 5 layer, unit cell at $s_2 = 0.15$, associated with the band structure shown in Figure~(\ref{fig:bnd_strc_s2_Q}). Outside the ``collapse'' region, the two modes are elliptically polarized, with axes in the tangential and normal directions (indicating $\pm \pi/2$ phase difference between the two components). At the exceptional points, the two trajectories coalesce. Between the two points the elliptical trajectories become oblique, turning essentially linear at the center of this region. Furthermore, at this point, the chirality of the trajectories flip from clockwise to counter-clockwise. 
Since in this case both exceptional points are located within the stop bands, one expects zero energy flux, which can be confirmed by writing the $x_1$ component of the flux averaged over a period:
\begin{equation}
	\mathcal{P}(f) = -\dfrac{1}{2}\Re\left(\tau_1v_1^* + \tau_6v_2^*\right),
\end{equation}
This gives zero values at EP5-1 and -2 and everywhere in between.

\begin{figure}[!ht]
    \begin{subfigure}[b]{0.20\linewidth}
        \centering\includegraphics[height=90pt,center]{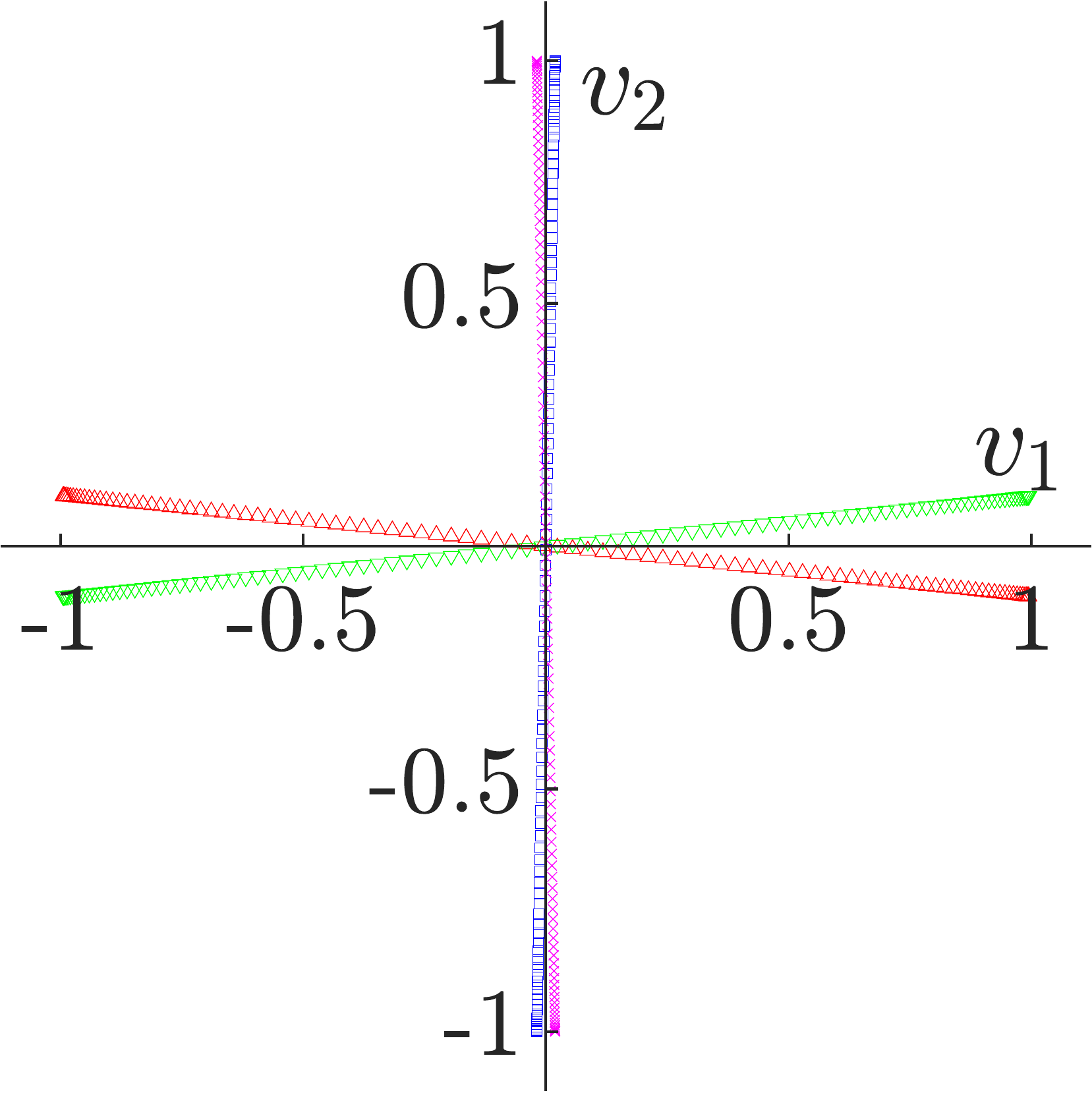}
        \caption{\label{fig:EP312_1}}
    \end{subfigure}%
    \begin{subfigure}[b]{0.20\linewidth}
        \centering\includegraphics[height=90pt,center]{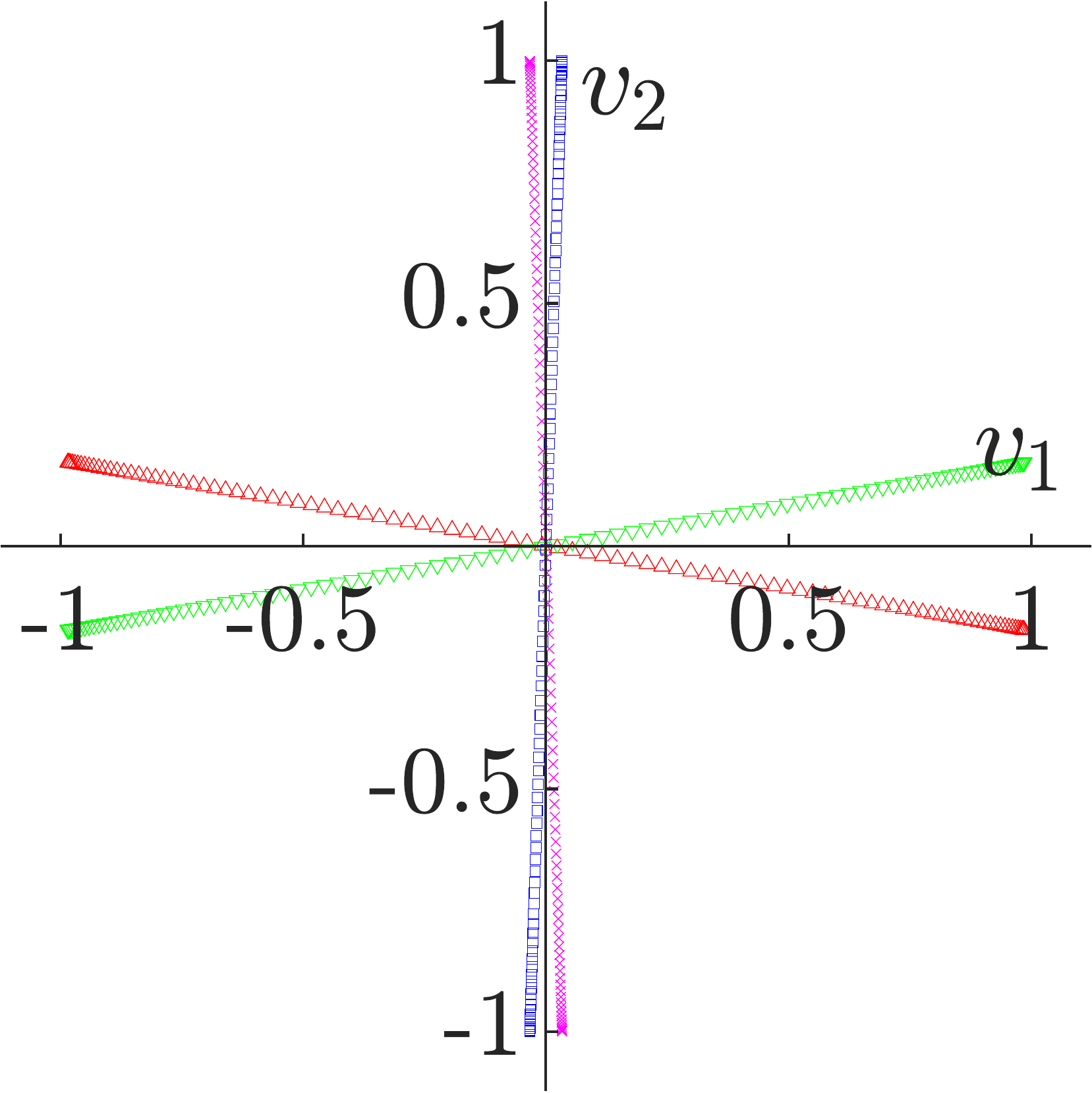}
        \caption{\label{fig:EP312_2}}
    \end{subfigure}%
    \begin{subfigure}[b]{0.20\linewidth}
        \centering\includegraphics[height=90pt,center]{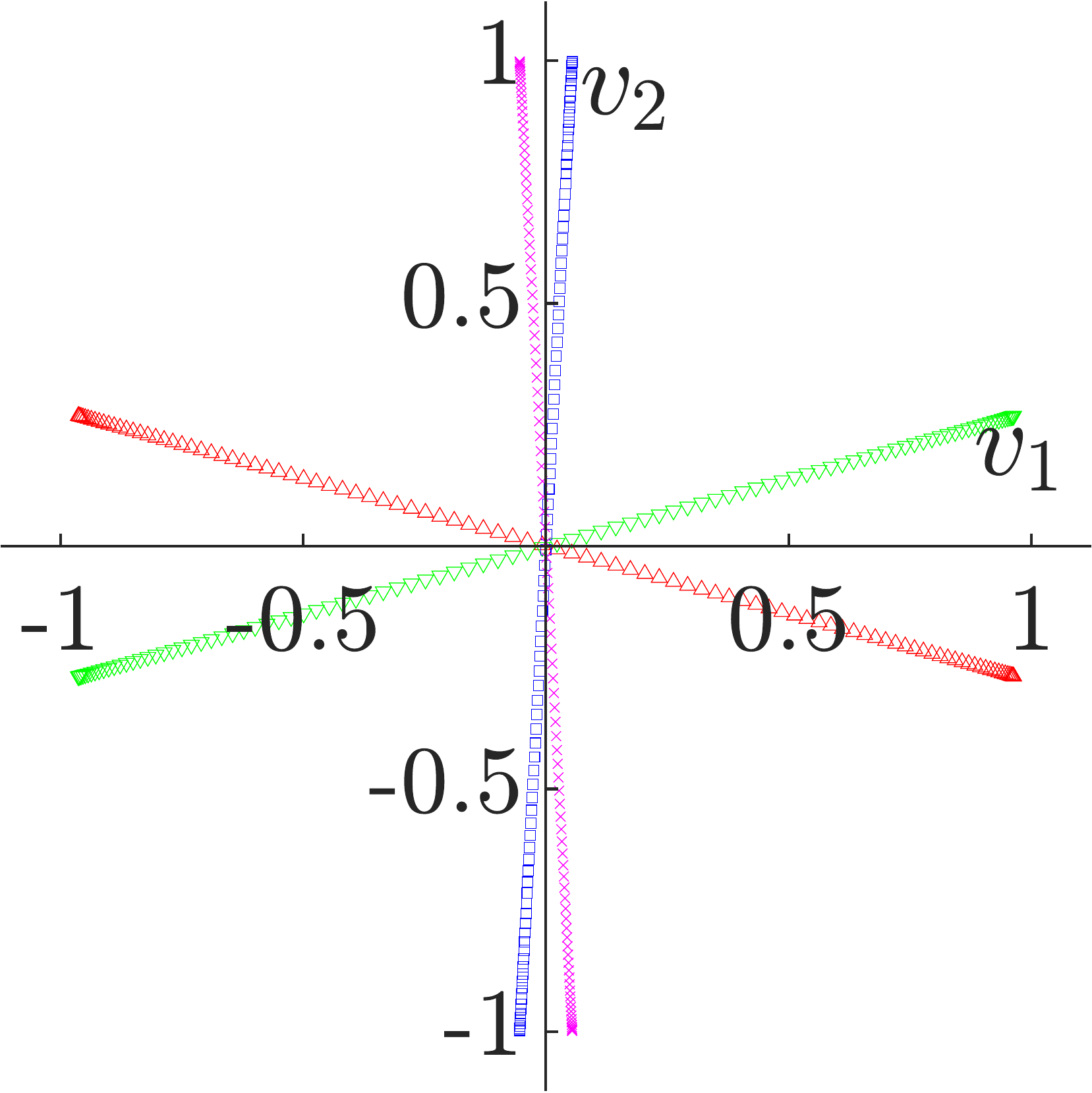}
        \caption{\label{fig:EP312_3}}
    \end{subfigure}%
    \begin{subfigure}[b]{0.20\linewidth}
        \centering\includegraphics[height=90pt,center]{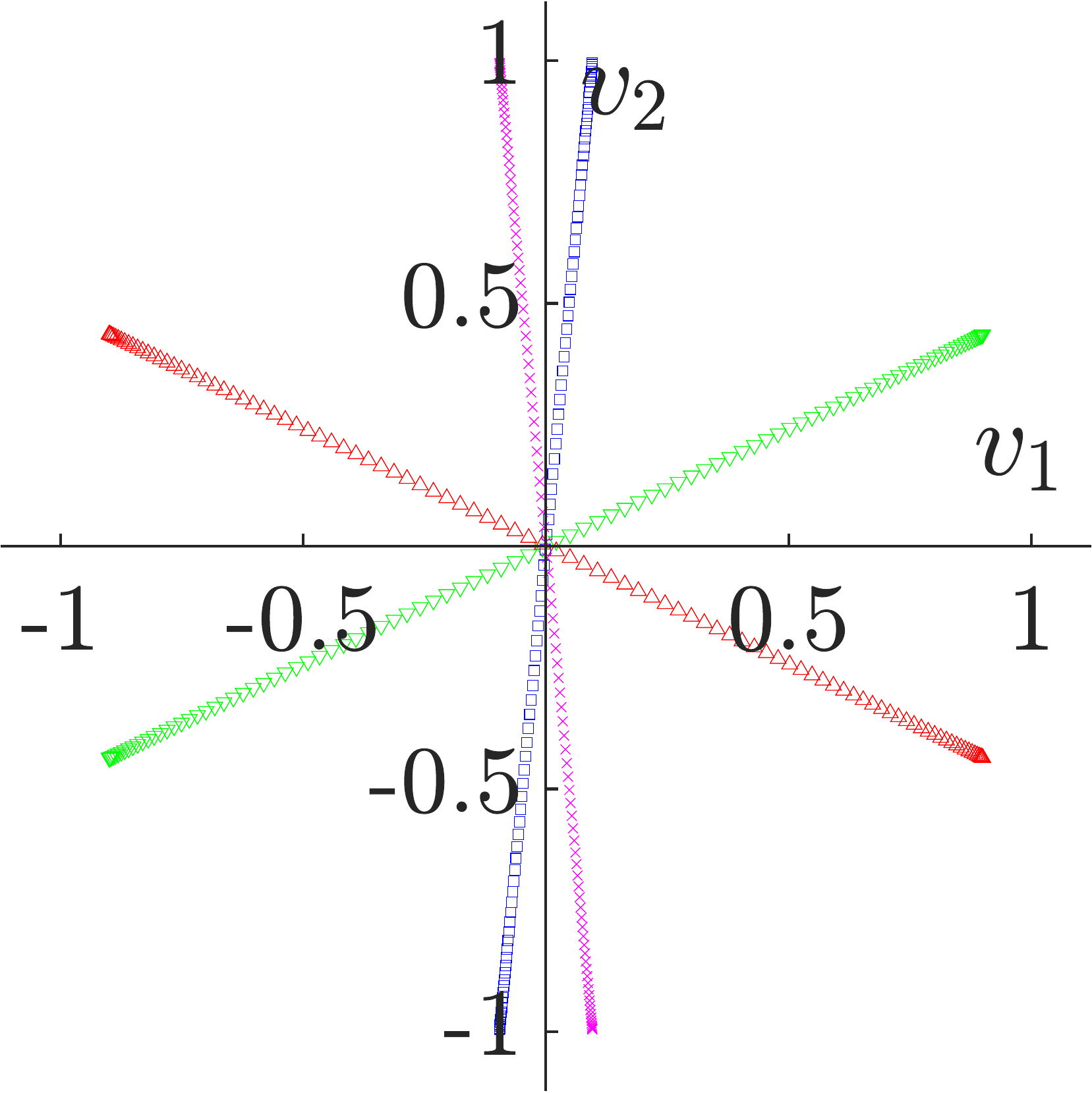}
        \caption{\label{fig:EP312_4}}
    \end{subfigure}%
    \begin{subfigure}[b]{0.20\linewidth}
        \centering\includegraphics[height=90pt,center]{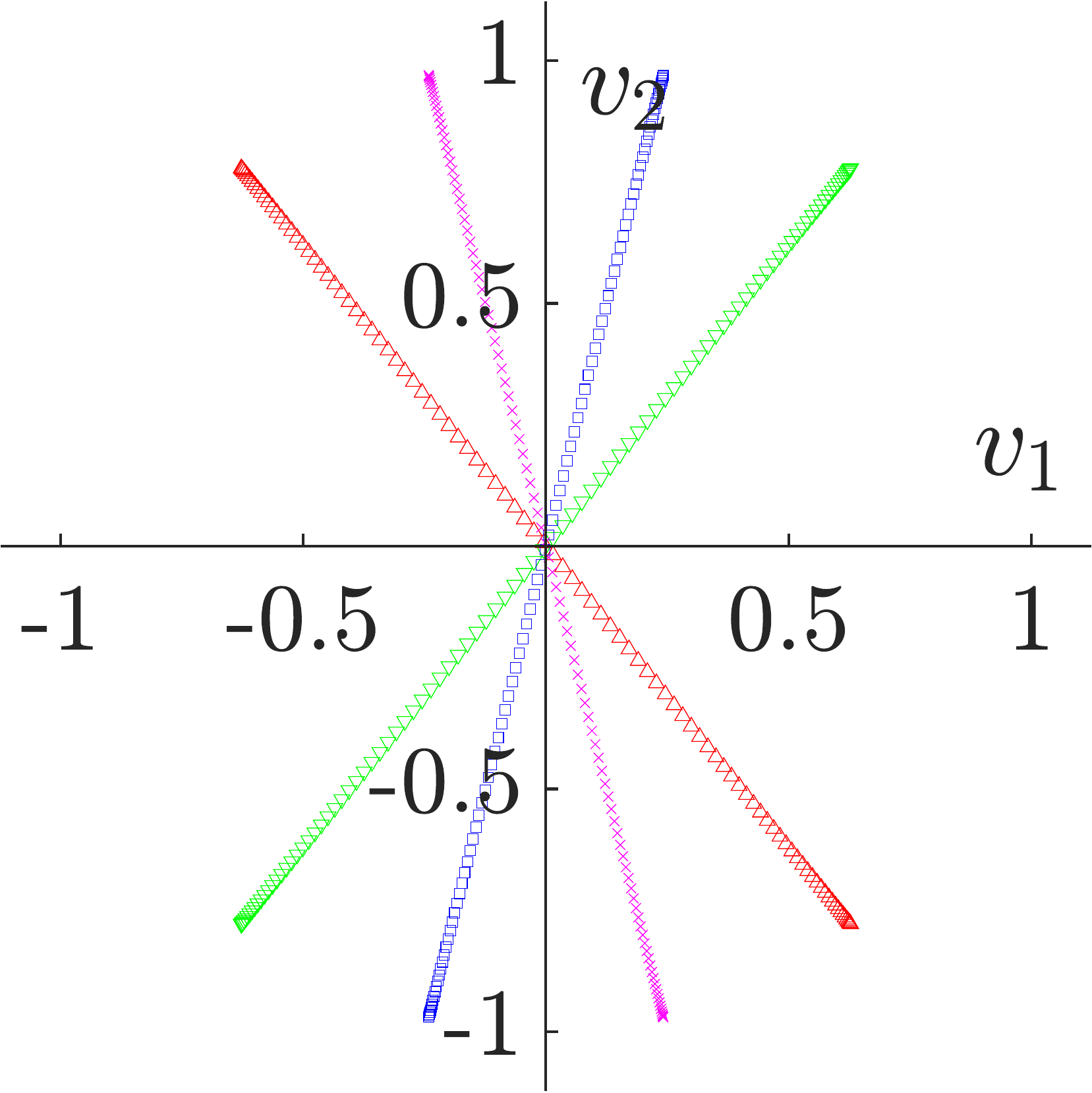}
        \caption{\label{fig:EP312_5}}
    \end{subfigure}
    \begin{subfigure}[b]{0.20\linewidth}
        \centering\includegraphics[height=90pt,center]{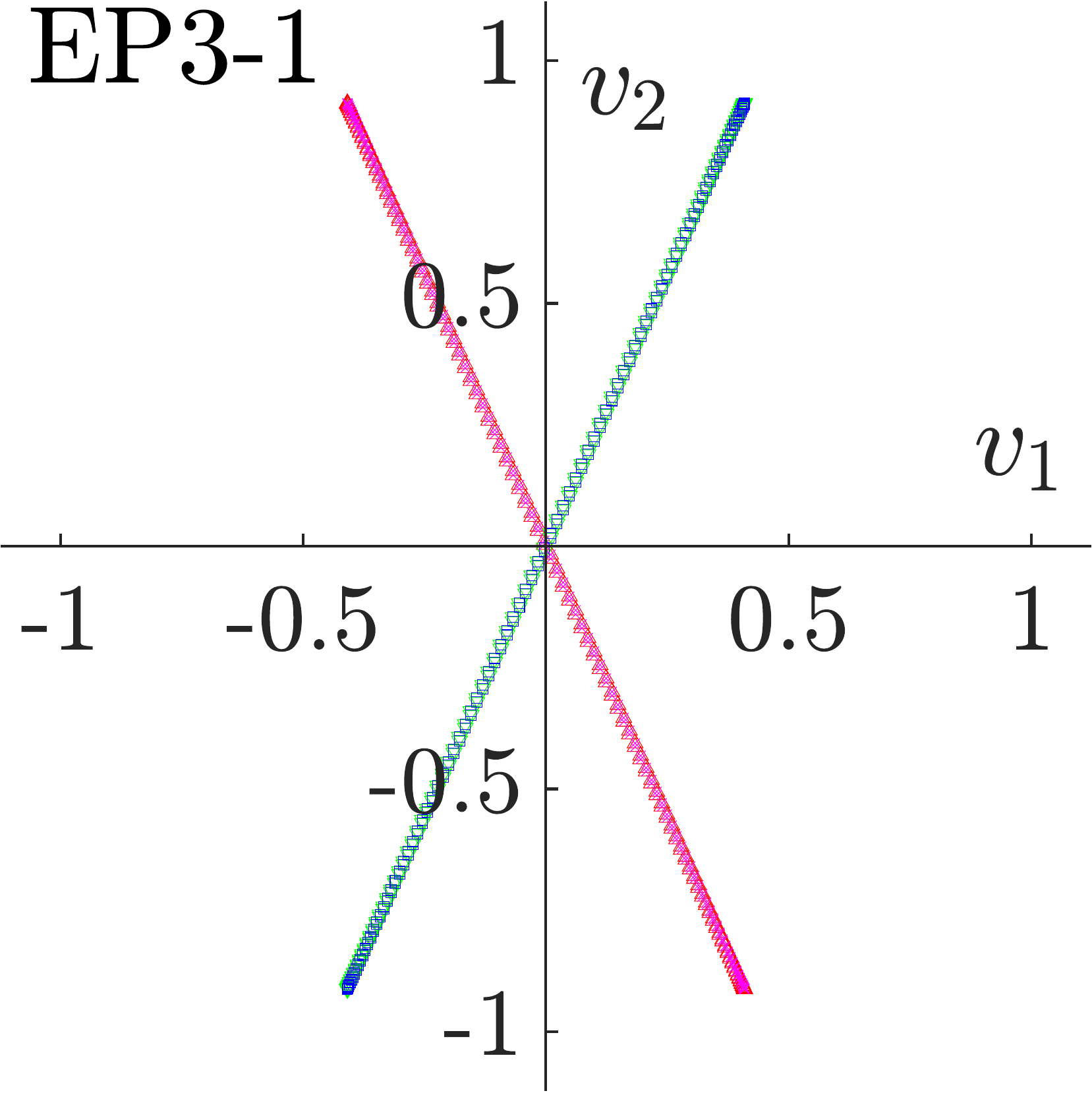}
        \caption{\label{fig:EP312_6}}
    \end{subfigure}%
    \begin{subfigure}[b]{0.20\linewidth}
        \centering\includegraphics[height=90pt,center]{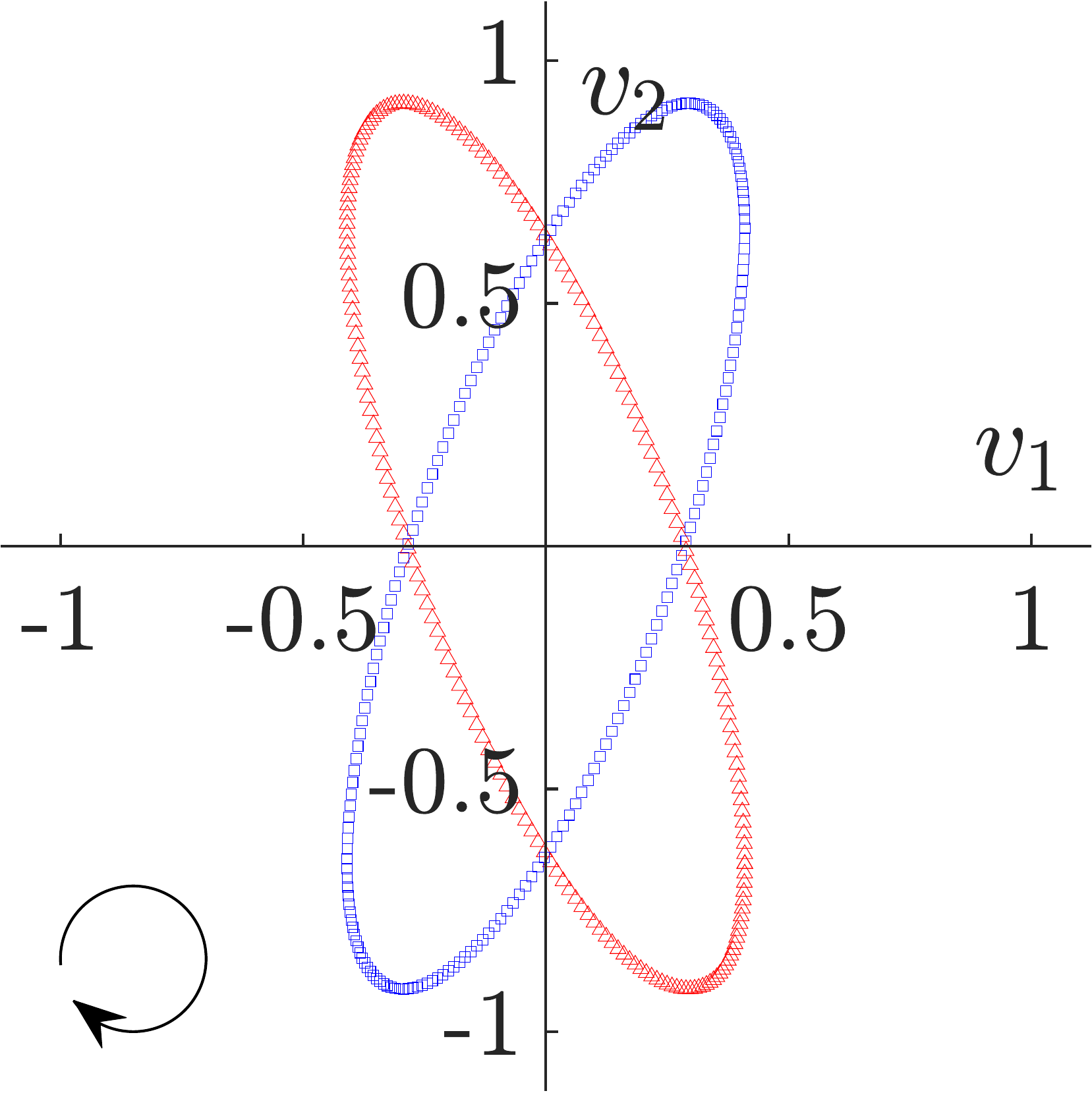}
        \caption{\label{fig:EP312_7}}
    \end{subfigure}%
    \begin{subfigure}[b]{0.20\linewidth}
        \centering\includegraphics[height=90pt,center]{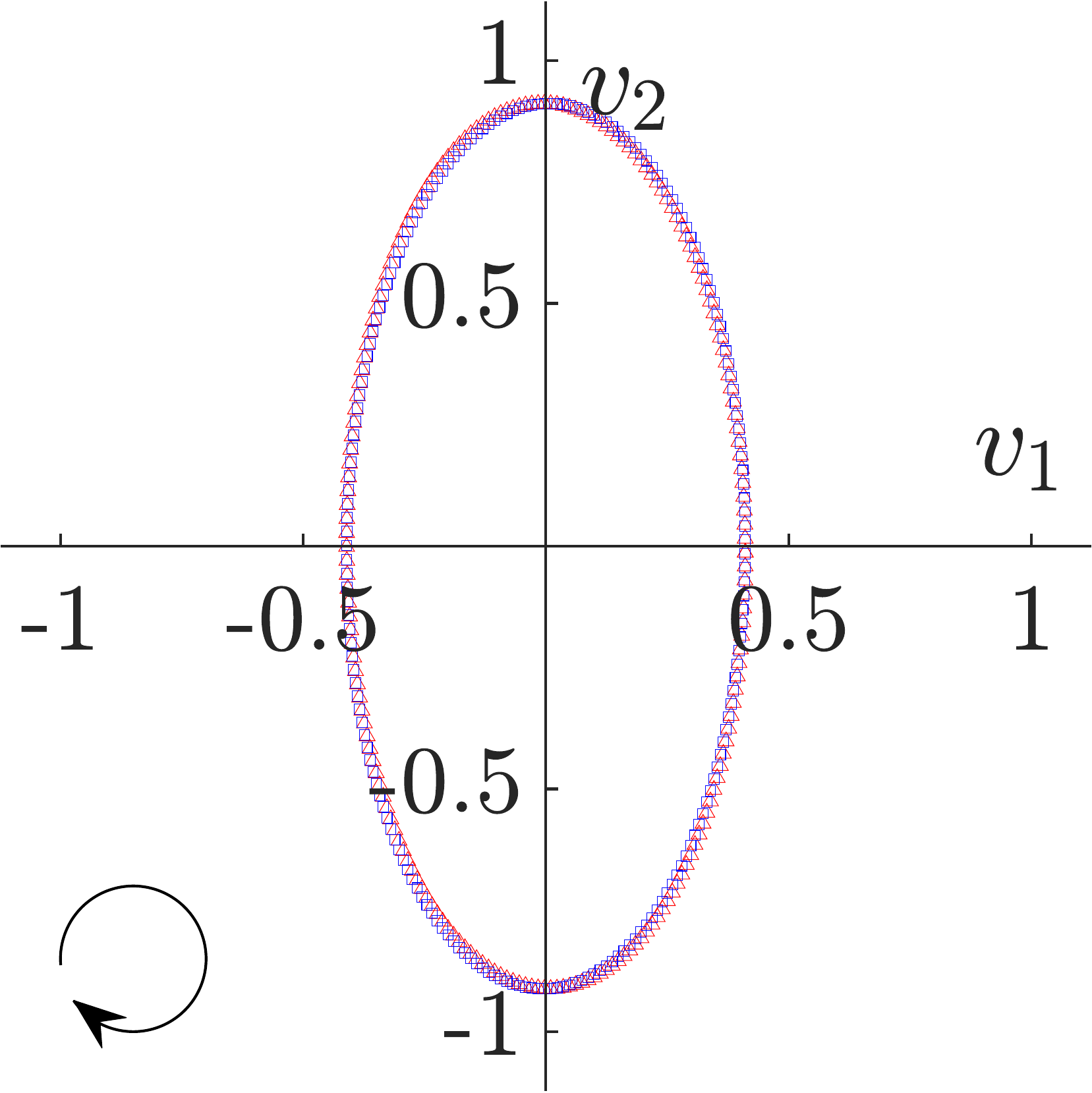}
        \caption{\label{fig:EP312_8}}
    \end{subfigure}%
    \begin{subfigure}[b]{0.20\linewidth}
        \centering\includegraphics[height=90pt,center]{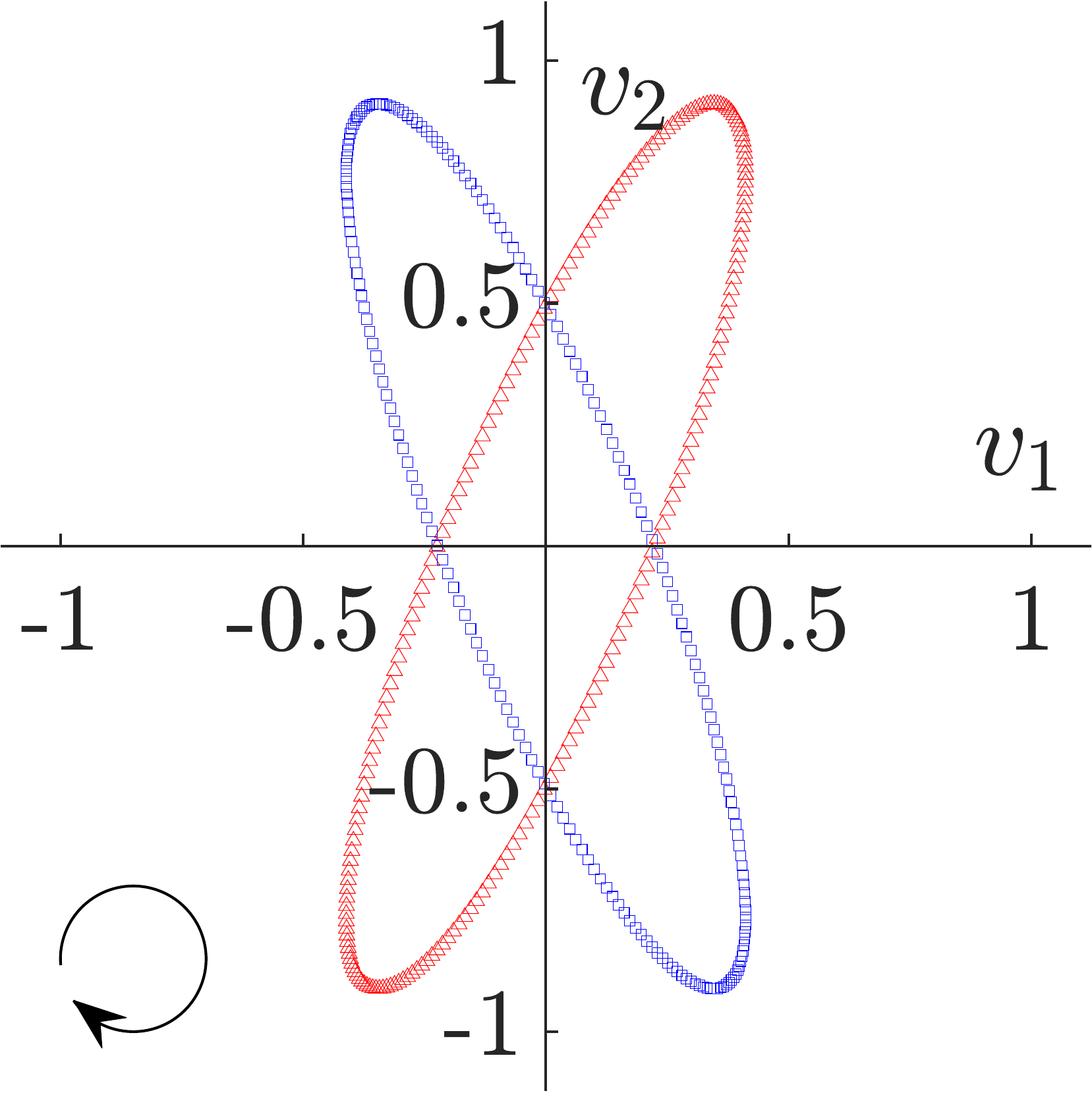}
        \caption{\label{fig:EP312_9}}
    \end{subfigure}%
    \begin{subfigure}[b]{0.20\linewidth}
        \centering\includegraphics[height=90pt,center]{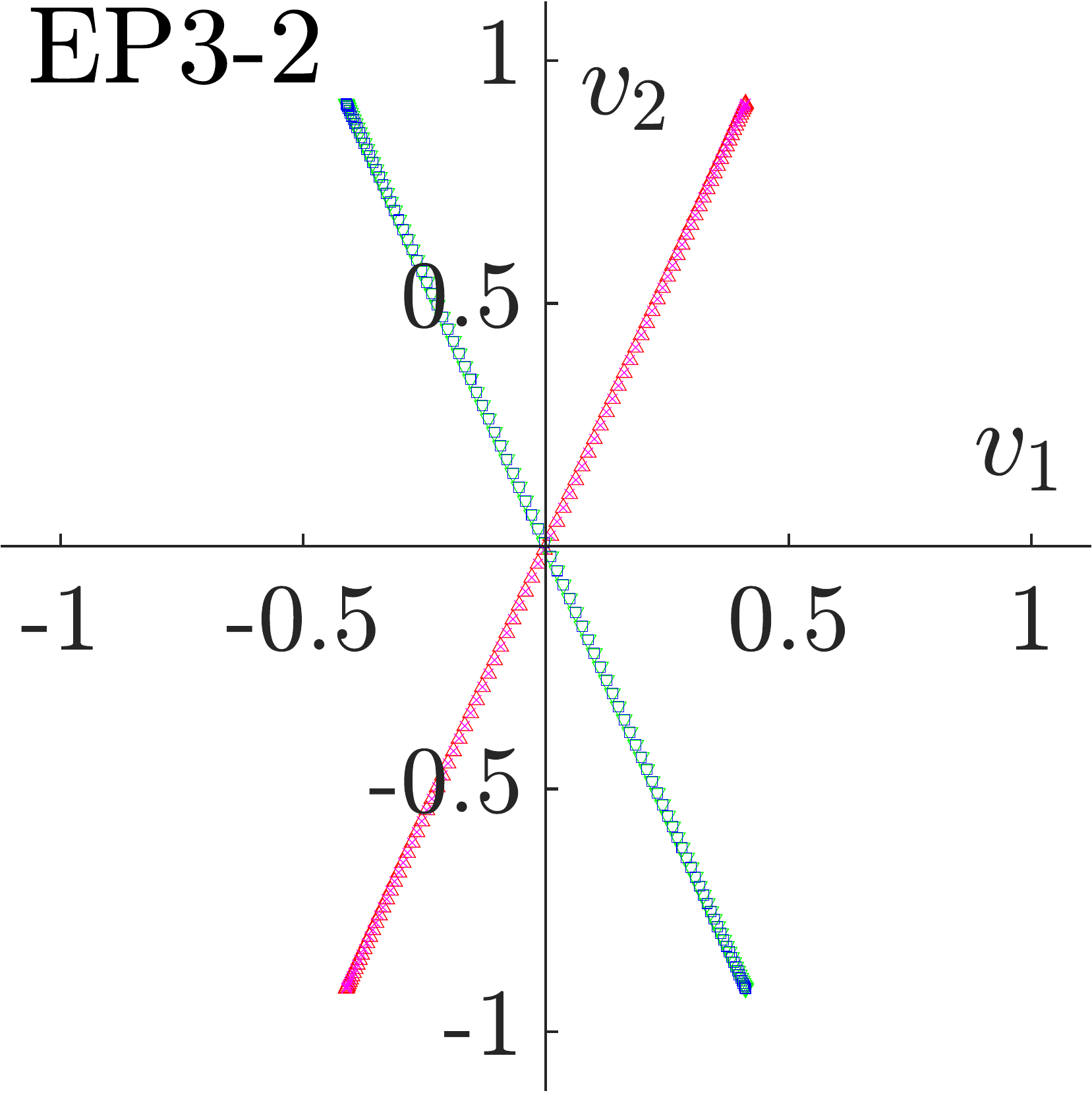}
        \caption{\label{fig:EP312_10}}
    \end{subfigure}
    \begin{subfigure}[b]{0.20\linewidth}
        \centering\includegraphics[height=90pt,center]{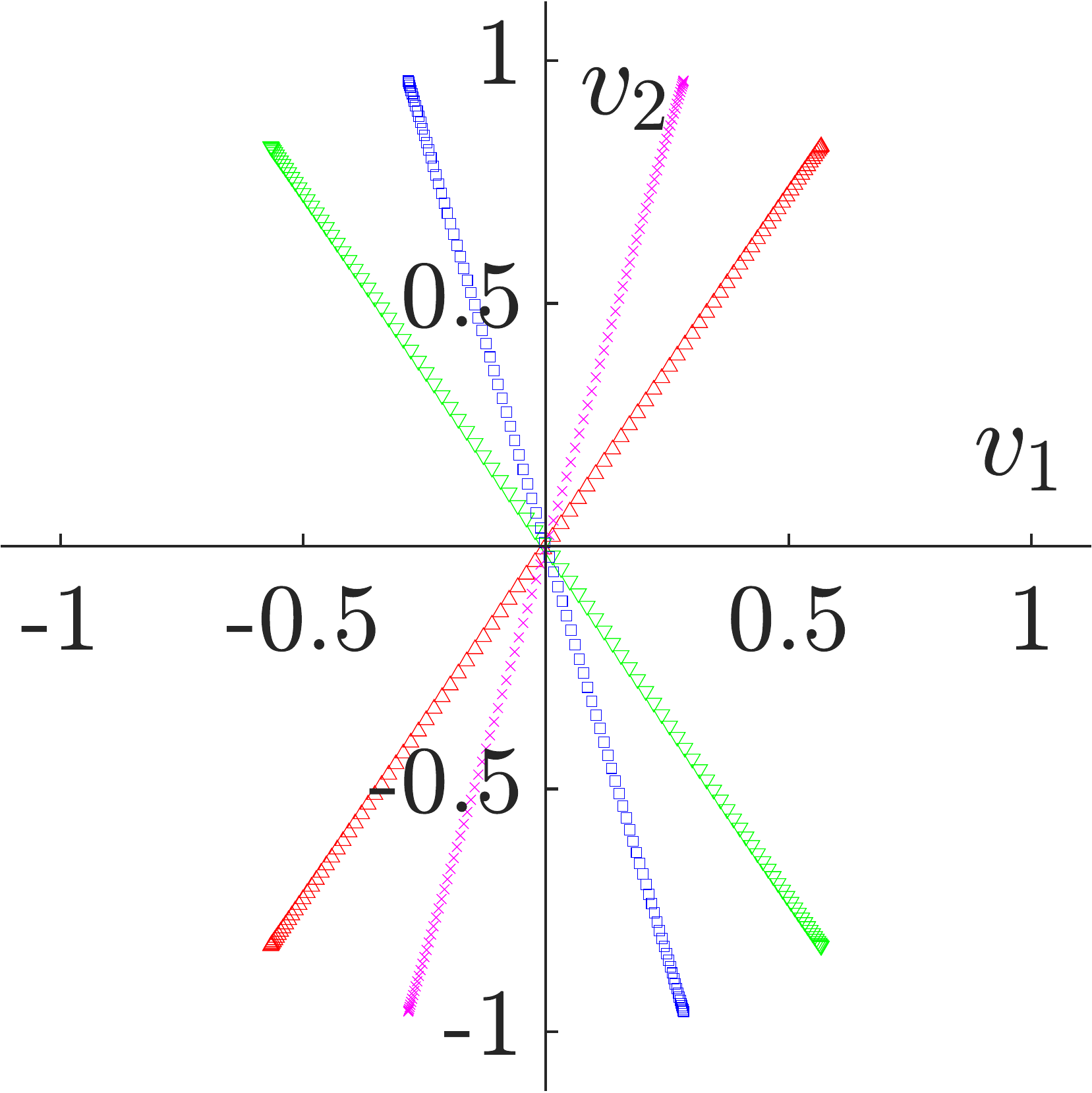}
        \caption{\label{fig:EP312_11}}
    \end{subfigure}%
    \begin{subfigure}[b]{0.20\linewidth}
        \centering\includegraphics[height=90pt,center]{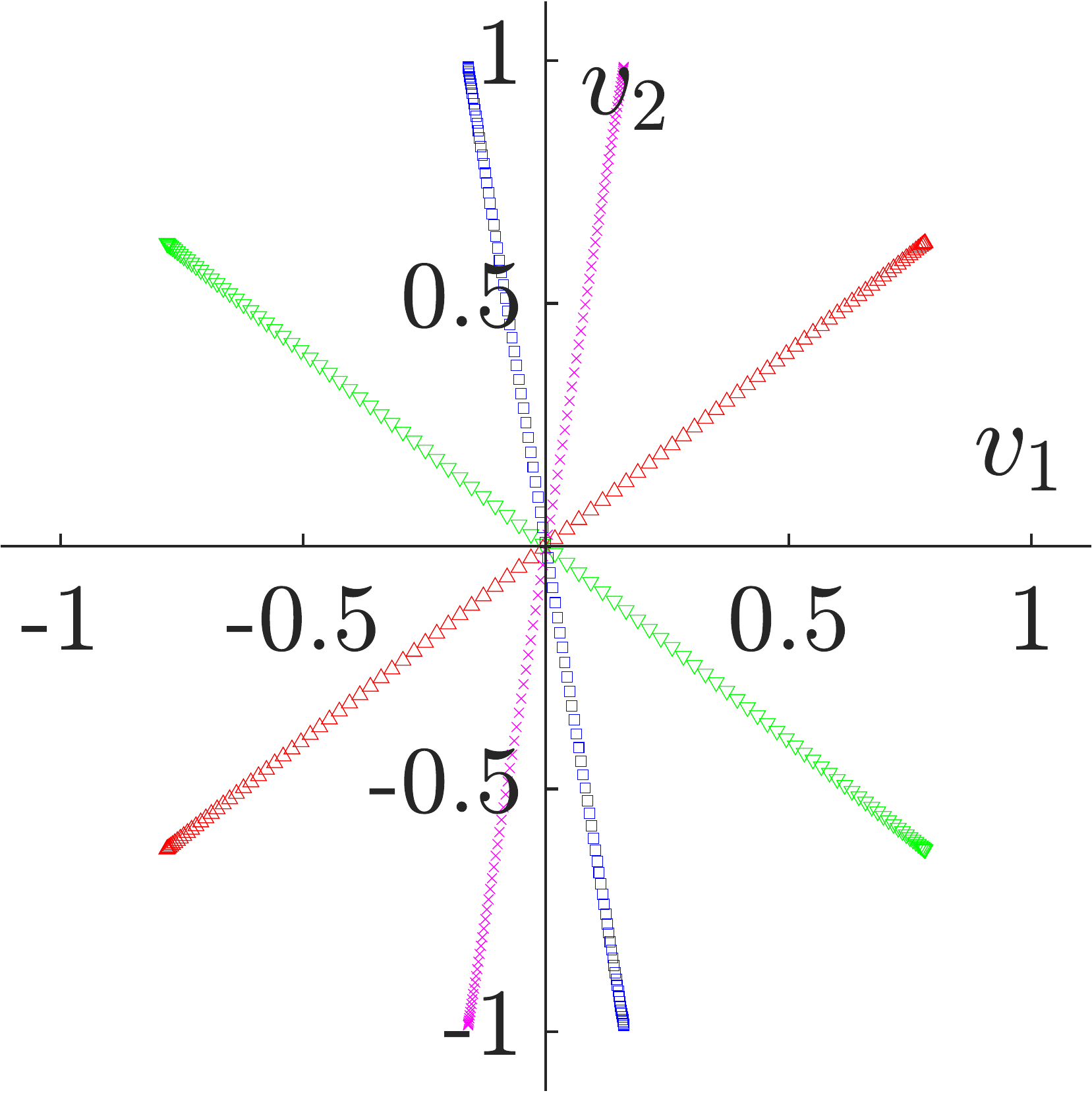}
        \caption{\label{fig:EP312_12}}
    \end{subfigure}%
    \begin{subfigure}[b]{0.20\linewidth}
        \centering\includegraphics[height=90pt,center]{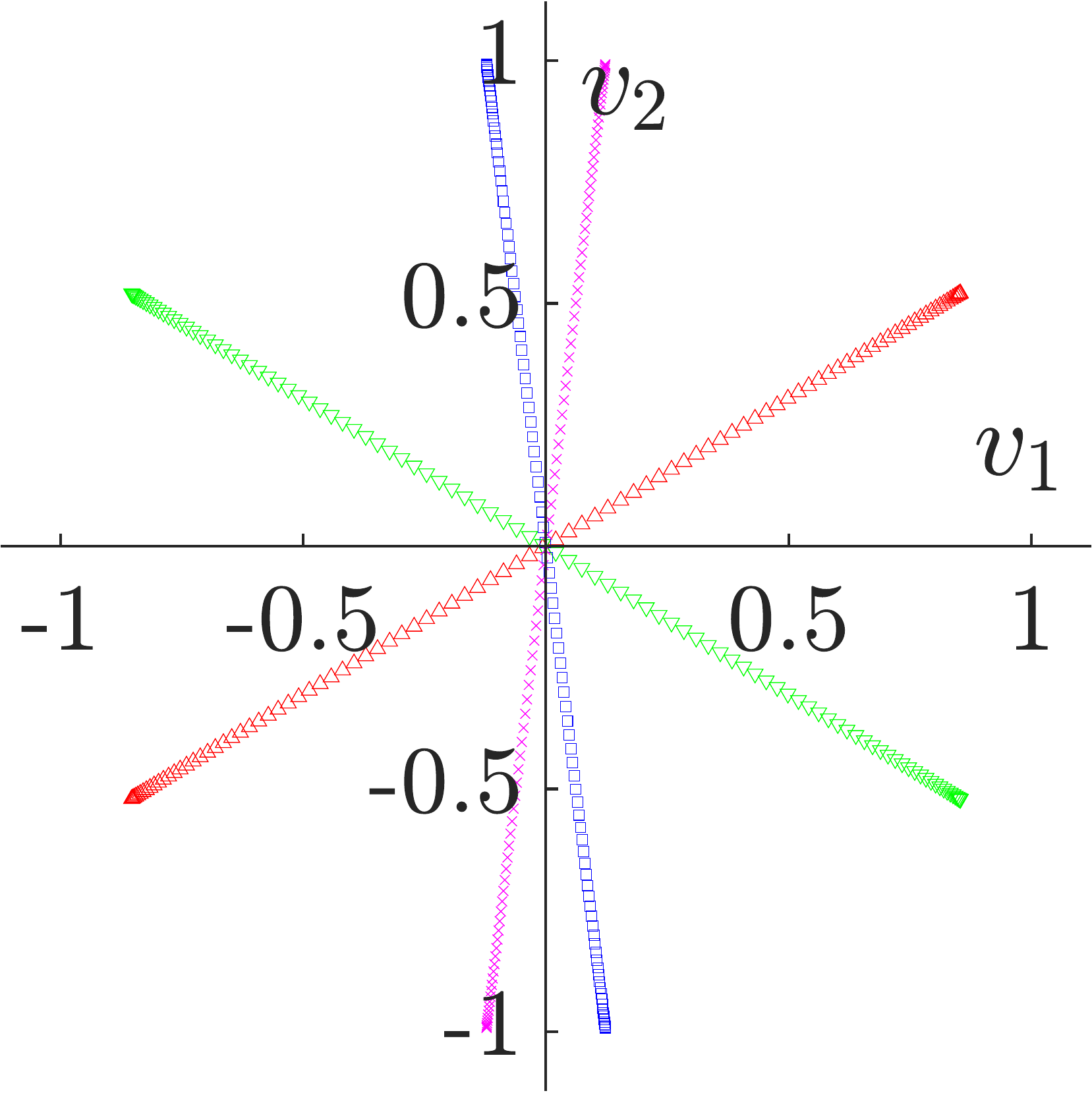}
        \caption{\label{fig:EP312_13}}
    \end{subfigure}%
    \begin{subfigure}[b]{0.20\linewidth}
        \centering\includegraphics[height=90pt,center]{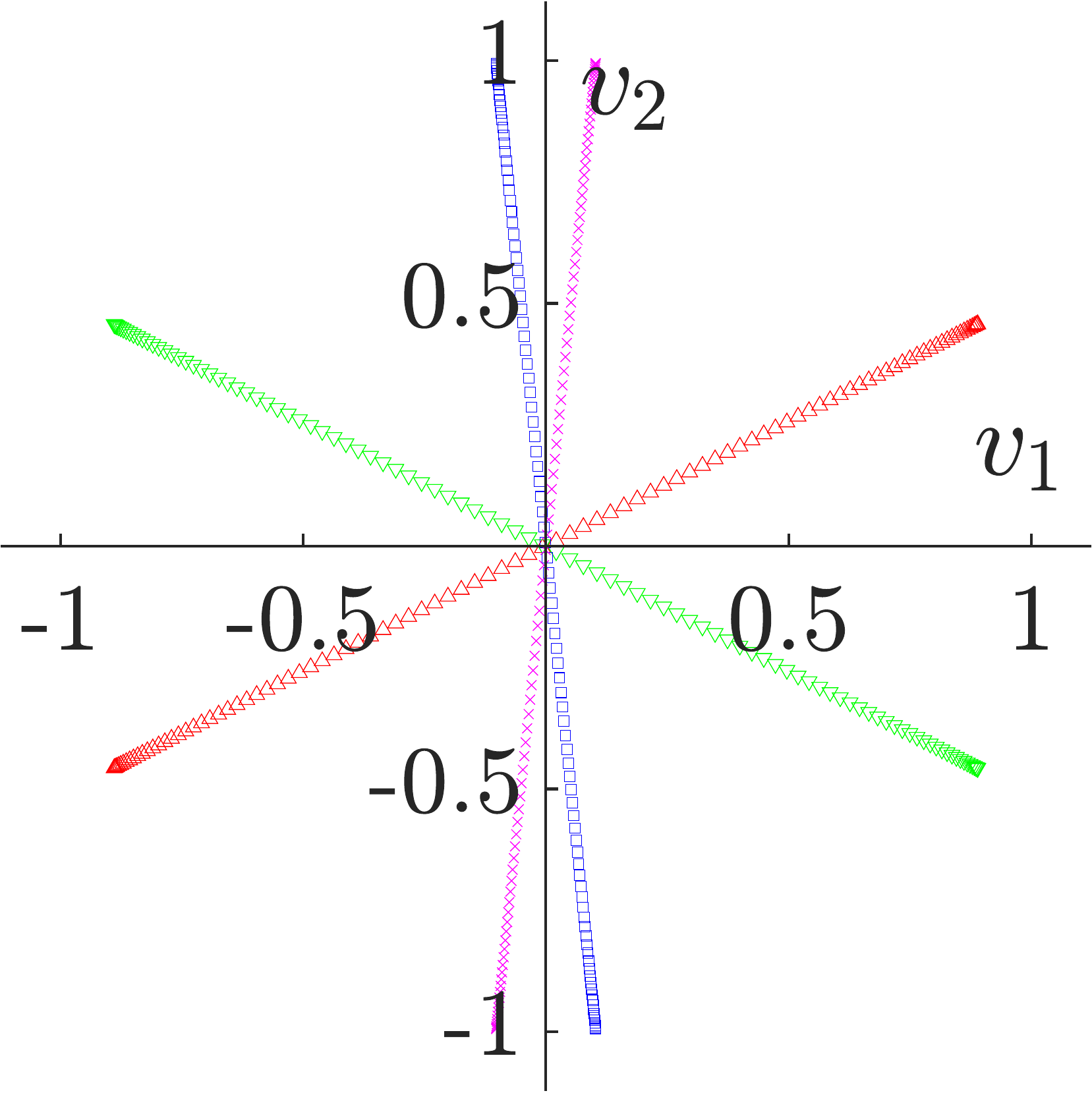}
        \caption{\label{fig:EP312_14}}
    \end{subfigure}%
    \begin{subfigure}[b]{0.20\linewidth}
        \centering\includegraphics[height=90pt,center]{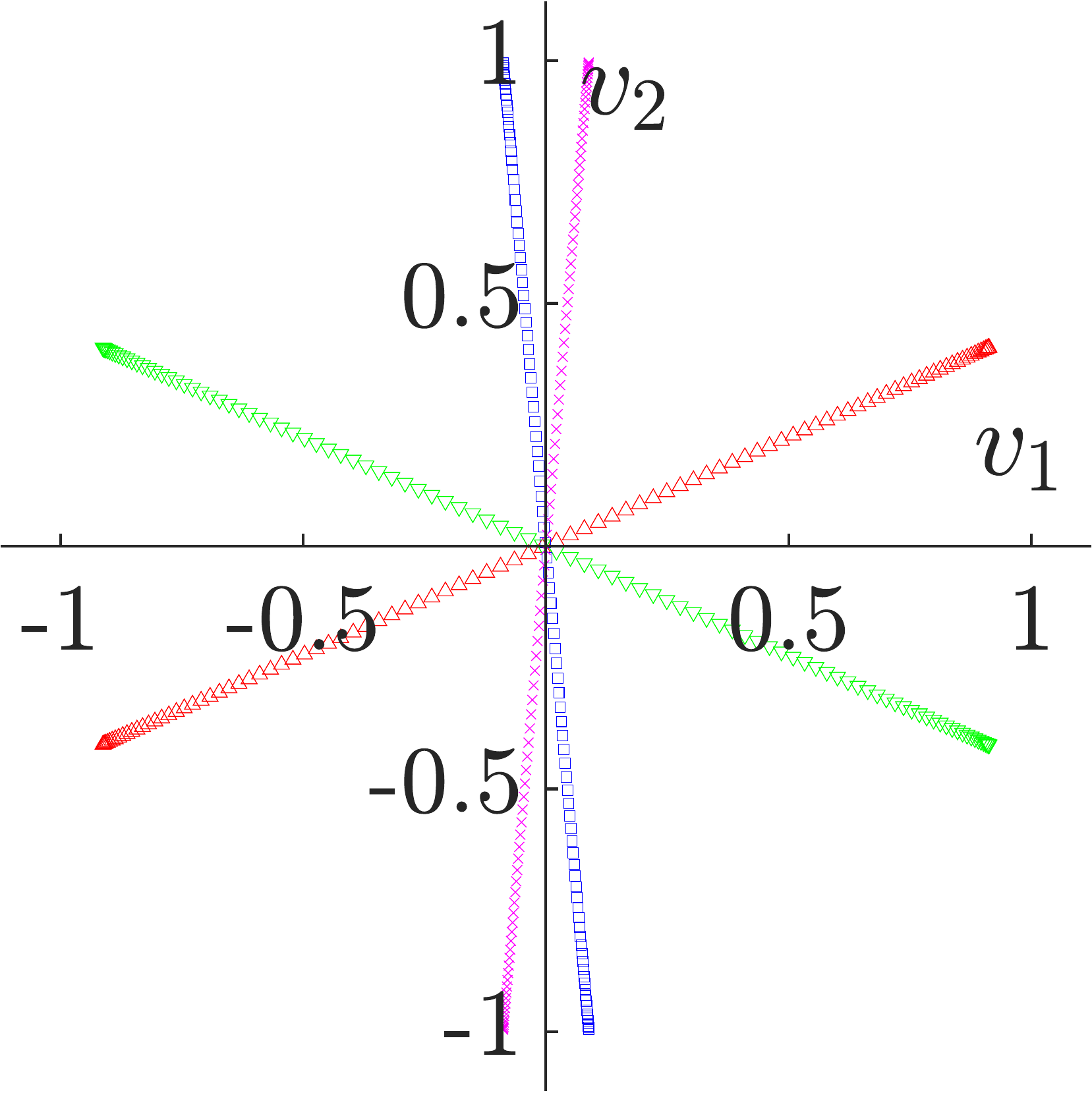}
        \caption{\label{fig:EP312_15}}
    \end{subfigure}%
\caption{\label{fig:pol_s2_004} The polarization trajectories near EP3-1 and -2 at the ``collapse'' region in Figure~(\ref{fig:3L_Q2_020_ri}). This region, unlike the previous case, is inside two pass bands. Mode degeneracy and collapse at EP3-1 and EP3-2 happens for waves of similar phase advance but opposing flux. Flux is zero at exceptional points and all the points in between with non-zero imaginary parts of the phase advance. For those frequency points in between ((g), (h), and (i)), only modes with negative imaginary parts are shown. Unlike the previous case, chirality does not flip in the middle of this stop band. At the center of the band, while it appears that the trajectories collapse, the eigenvalues are not the same. \gre{Quiver plot animations of selected modes (particle velocity vector) is provided in the supplementary materials.}
}
\end{figure}

The first two ``collapse'' regions in the case of 2-phase, 3-layer, unit cell occur in the middle of pass bands. Figure~(\ref{fig:pol_s2_004}) shows the $v_1-v_2$ trajectories at different frequency points near exceptional points, EP3-1 and EP3-2. The fluxes become zero, with elliptical trajectories, between the two exceptional points. The chiralities do not appear to flip, while the linear polarization angle swaps throughout the band. Note that the mode collapse in this case occurs for a pair of P and S modes with opposite flux directions. 

\subsection{Scattering} 

\begin{figure}[!ht]
    \begin{subfigure}[b]{\linewidth}
        \centering\includegraphics[height=215pt,center]{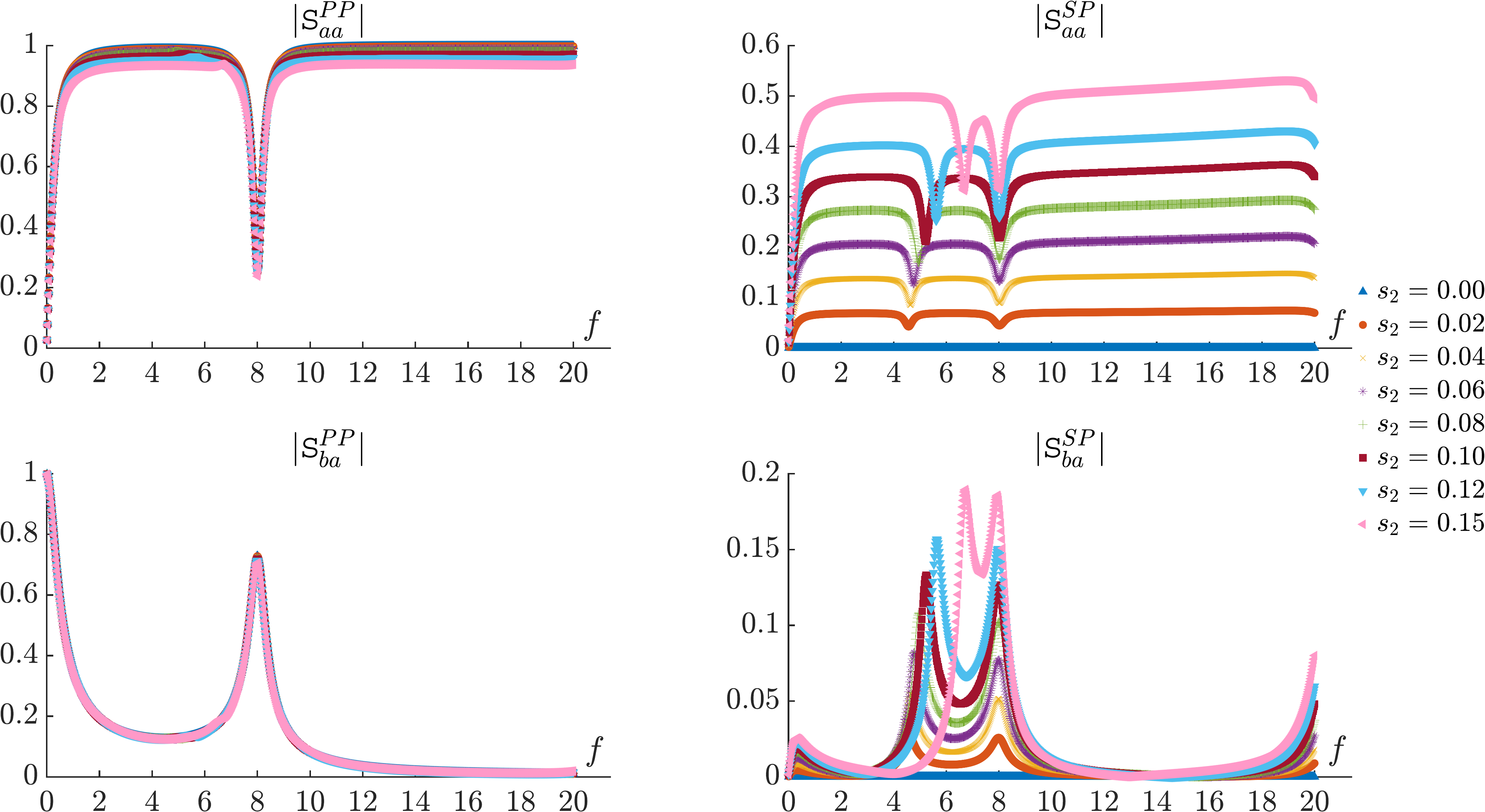}
        \caption{\label{fig:P_scat_mag}}
    \end{subfigure}
    \begin{subfigure}[b]{\linewidth}
        \centering\includegraphics[height=215pt,center]{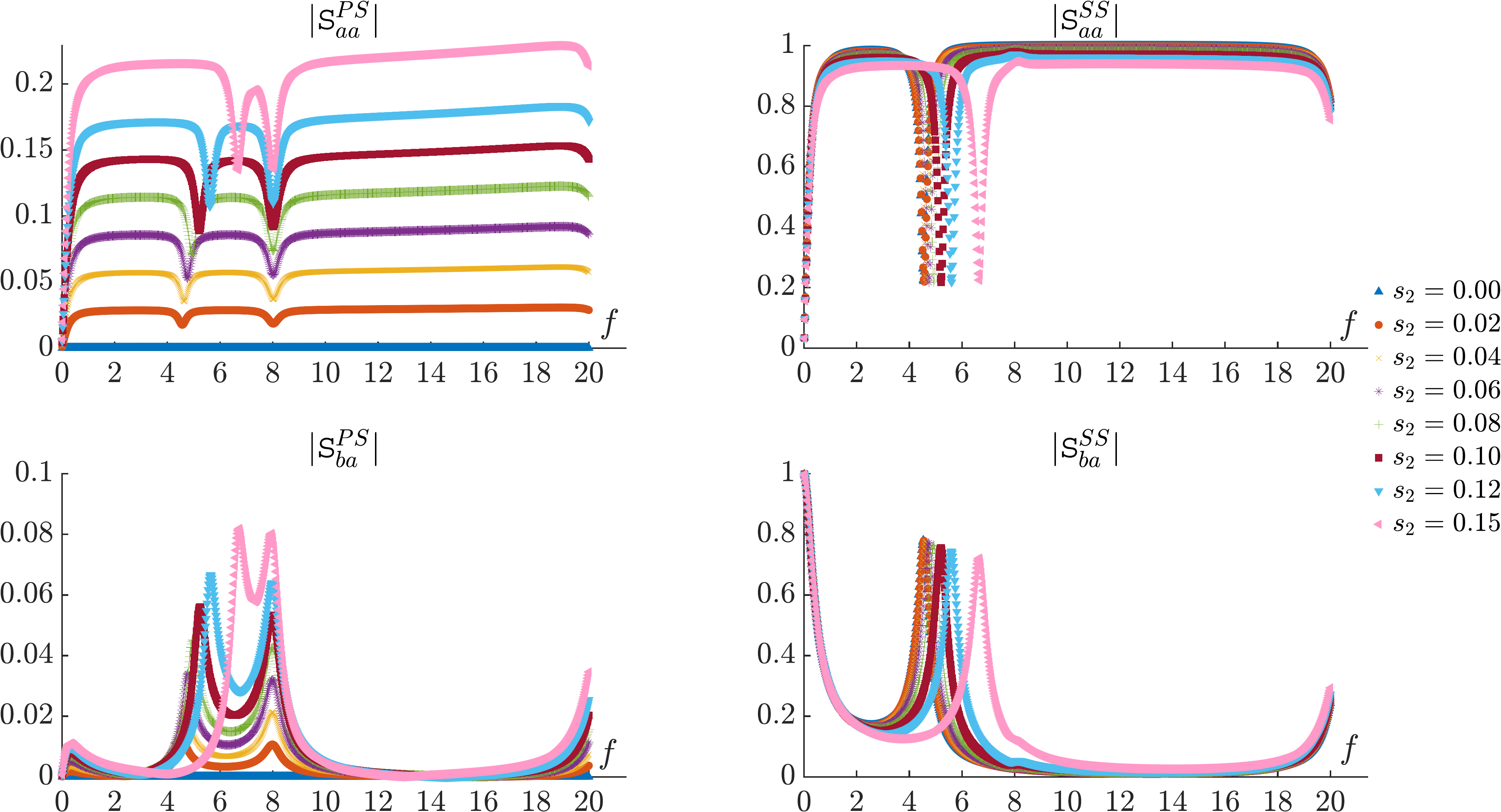}
        \caption{\label{fig:S_scat_mag}}
    \end{subfigure}
    \caption{\label{fig:scat_mag} Reflections and transmission amplitudes off of the 3-phase, 5-layer, medium with one unit cell ($n_c = 1$) for incident plane: (\subref{fig:P_scat_mag}) P-wave and (\subref{fig:S_scat_mag}) SV-wave. Note that 1\% loss in both wave speeds is added for layers 2 and 4 as the ratio of the imaginary to real parts of the wave speeds ($c''/c' = 0.01$). \gre{Frequency unit is kHz.}
    }
\end{figure}

\begin{figure}[!ht]
	\begin{subfigure}[b]{0.50\linewidth}
		\centering\includegraphics[height=130pt,center]{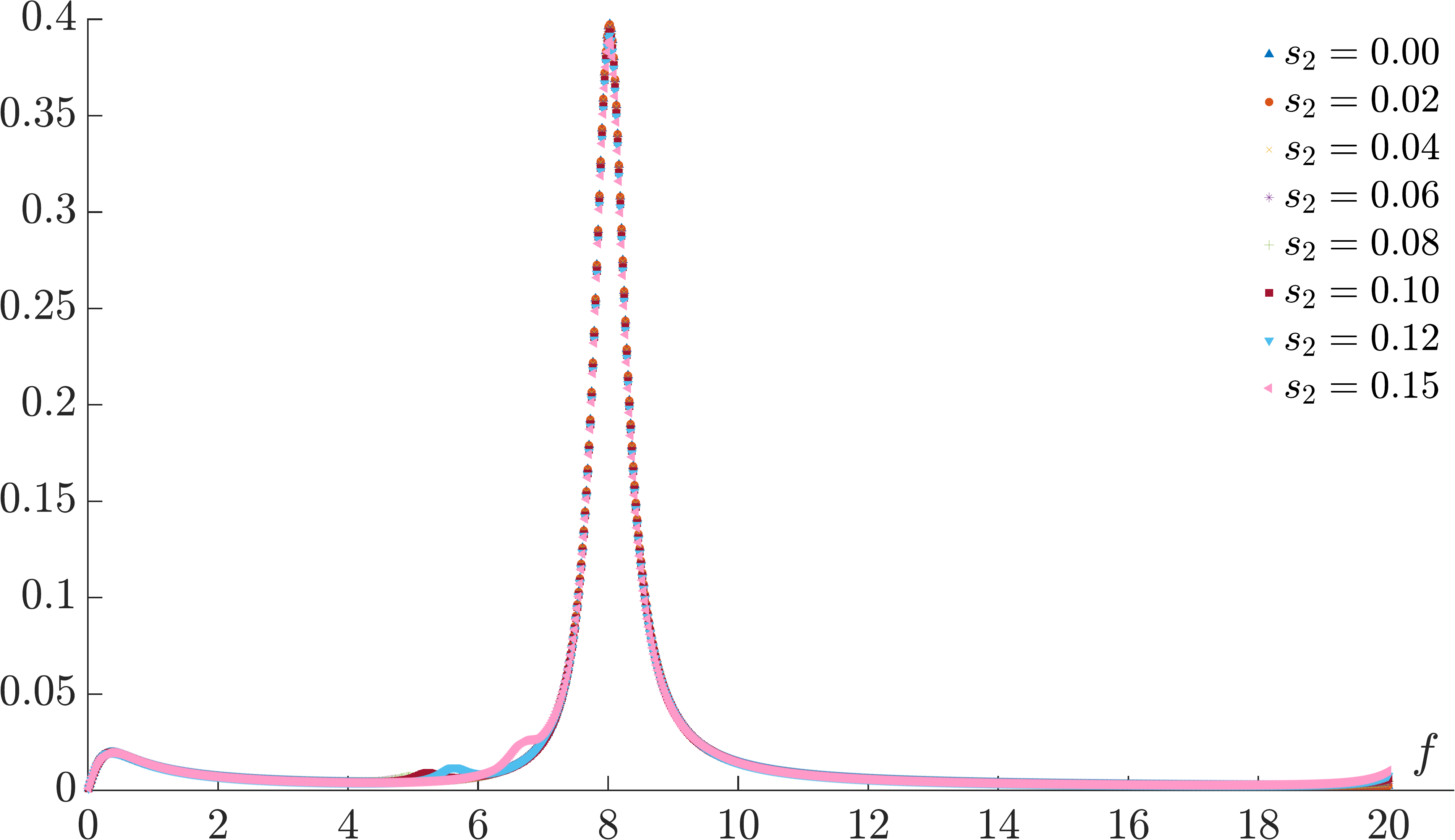}
		\caption{\label{fig:P_scat_absp}}
	\end{subfigure}%
	\begin{subfigure}[b]{0.50\linewidth}
		\centering\includegraphics[height=130pt,center]{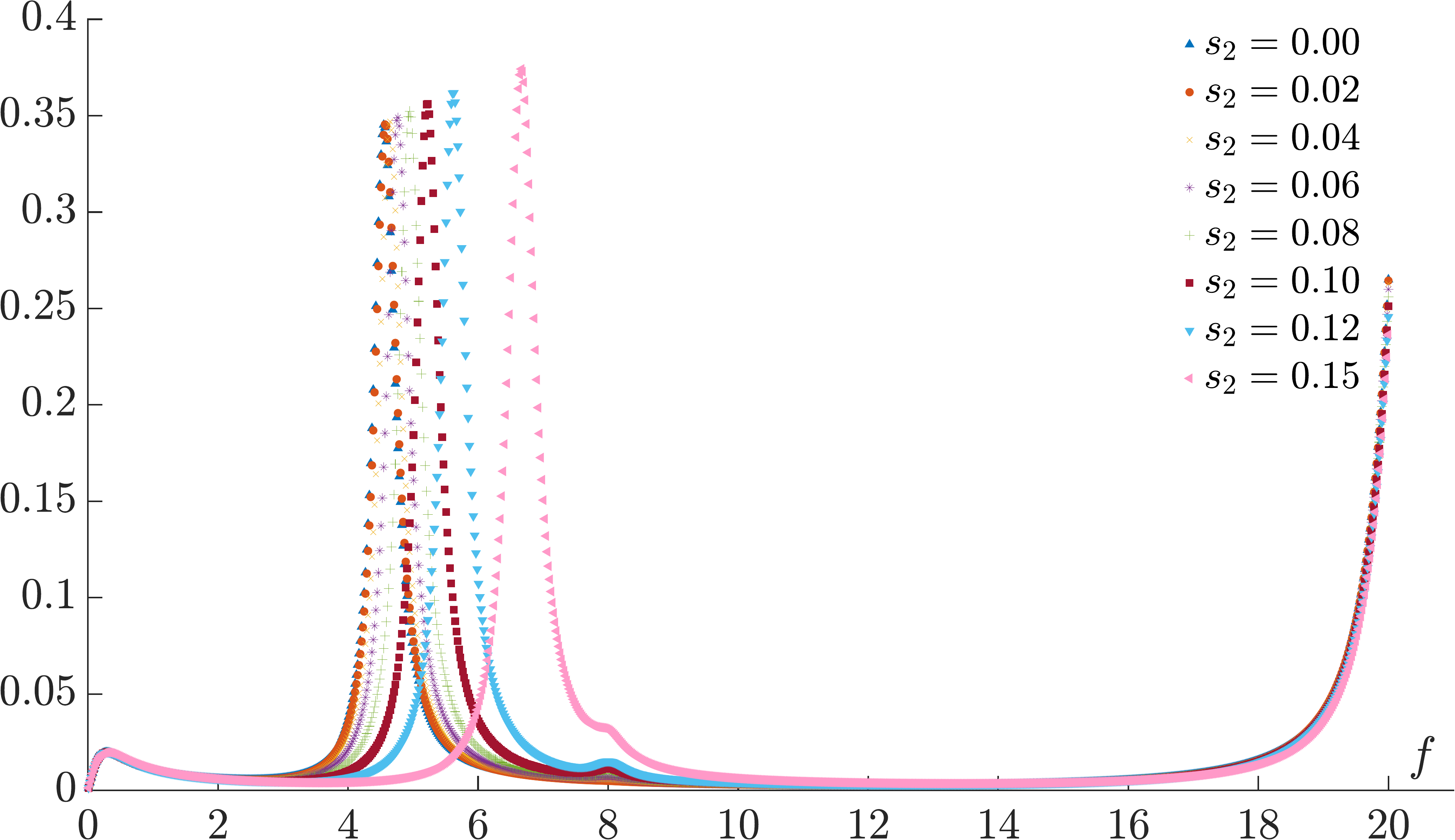}
		\caption{\label{fig:S_scat_absp}}
	\end{subfigure}
	\caption{\label{fig:scat_absp} Absorption in a single 3-phase, 5-layer, unit cell: (\subref{fig:P_scat_absp}) P-wave incidence, (\subref{fig:S_scat_absp}) SV-wave incidence. Layers 2 and 4 have 1\% loss in both wave speeds ($c''/c' = 0.01$). \gre{Frequency unit is kHz.}
    }
\end{figure}

Scattering coefficient amplitudes for a finite slab consisting of a single 3-phase, 5-layer, cell are shown in Figure(\ref{fig:scat_mag}) as functions of frequency for both P- and SV-wave incidence. The varying angles of incidence are indicated by the tangential component of the slowness vector $s_2$, with $s_2 = 0.48$ equivalent to about $\pi/3$ incidence angle. Figure~(\ref{fig:scat_absp}) shows plots of the absorption for these cases. 
Resonance peaks are varying with the nature of the waves and angle of incidence. The presence of exceptional points in the band structure do not seem to affect the scattering significantly as the main features are associated with the pass bands. The high transmission / low reflection in P-wave spectrum clearly matches with the beginning of the second longitudinal pass band (``optical branch''), which is relatively fixed around \SI{8}{kHz}, and the same can be observed for the second shear pass band as it changes with $s_2$; See Figure~(\ref{fig:bnd_strc_Q2}). Furthermore, this also corresponds with increased dissipation inside the system as indicated in Figure~(\ref{fig:scat_absp}). Finally for higher oblique angles (higher $s_2$), any incident wave would activate both shear and longitudinal waves inside the layered medium, which in turn lead to both longitudinal and shear waves transmission and reflection in the exterior domains, as indicated in the coupled scattering coefficients as seen in Figure~(\ref{fig:scat_mag}), with both effects becoming more prominent at more oblique angles. 

\section{Scattering behavior in $\mathcal{PT}$-symmetric media}

We now turn our attention from the topology of the band structure to the study of the scattering response with a focus on potential for lasing \gre{(requiring wave amplification and divergence of mode lifetime \cite{he2025laser})} and and highly asymmetric reflection. While the former requires active (gain) domains, the latter can be simply achieved through asymmetric structures, i.e. when $\mathcal{P}$-symmetry is broken. Here we show the occurrence of both wave amplification and asymmetric reflection in $\mathcal{PT}$-symmetric linear media. Such phenomena can occur when phase symmetry is broken, as has been shown earlier for simpler 2 DOF systems in \cite{wang2020exceptional} and elsewhere. In the cases studied below, a transmission coefficient with amplitude greater than 1 (through a single cell) is expected through the increase of the (balanced) gain. Furthermore, we show a difference of more than two orders of magnitude in directional reflection coefficients with very small and equal (i.e. balanced) gain and loss factors. 

The scattering matrix, $\SM$, is usually organized to calculate reflection and transmission (outgoing) coefficients at each ``port/mode'' through its operation on a vector representing the incidence (incoming) complex amplitudes of the same ordered set. Here ports are the two semi-infinite domains $a/b$, while modes are the P- or SV-waves in those domains, $P/S$. Ordering first by mode and then by port the scattering matrix can be written as
\begin{equation}
    \SM = \begin{pmatrix}
        \SM_m^{PP} & \SM_m^{PS} \\
			\\
        \SM_m^{SP} & \SM_m^{SS} \\
	\end{pmatrix},
\end{equation}
where 
\begin{equation}
    \SM_m = \begin{pmatrix}
        \SM_{aa} & \SM_{ab} \\ \SM_{ba} & \SM_{bb}
    \end{pmatrix}.
\end{equation}
It can be rearranged slightly by swapping rows $1\leftrightarrow2$ and $3\leftrightarrow4$ into 
\begin{equation}
    \TSM = \vect{P} \SM, 
\end{equation}
where the $\vect{P}$ matrix is constructed as
\begin{equation}
    \vect{P} = \begin{pmatrix}
        \vect{J}_2 & \vect{0} \\
        \vect{0}  & \vect{J}_2 
    \end{pmatrix}, 
\end{equation}
and
\begin{equation}
    \vect{J}_2 = \begin{pmatrix}
        0 & 1 \\ 1 & 0
    \end{pmatrix},
\end{equation}
is the 2-by-2 exchange matrix. In this form $\TSM$ transforms a vector consisting of the complex amplitudes of incoming waves from each domain and modality into a vector consisting of the complex amplitudes of outgoing waves of ordered with the same modality traveling in the same direction but in the other domain. Such a rearrangement lends itself very well to the focused study of this section. The eigenvectors of this operator represent combinations of incoming waves that will get transmitted with the same exact complex ratios (i.e. eigen-polarizations) after interacting with the specimen, while the eigenvalues represent the amplification/reduction factor or the phase advance. Therefore, the eigenvalues for a fully elastic or passive lossy specimen must have amplitude 1 (effectively only a phase advance) or less than 1, respectively. For a system with loss and gain in a $\mathcal{PT}$-symmetric arrangement, generally, the 4 eigenvalues still have amplitude 1, though exceptional points of the $\TSM$ operator can change this and lead to broken phase symmetry. 
In what follows, we present examples of the appearance of EP pairs in the spectrum of $\TSM$ along with the geometry and behavior of the ensuing branches. The scattering coefficients in these regions are calculated to evaluate transmission amplification factor above 1 as well as highly asymmetric reflection. 

\begin{figure}[!ht]
	\begin{subfigure}[c]{0.5\textwidth}
		\centering\includegraphics[height=160pt,center]{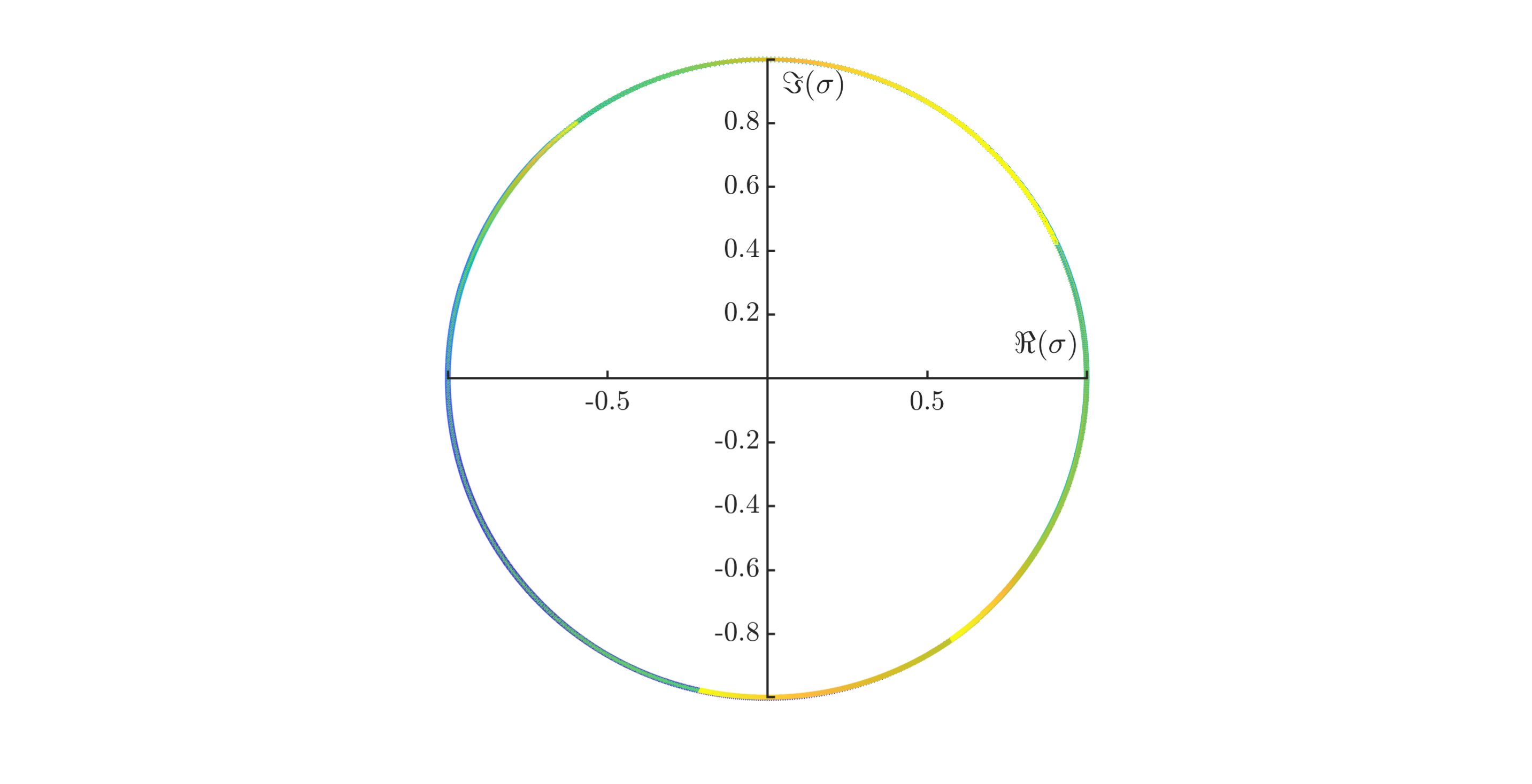}
		\caption{\label{fig:Stilde-sig-lossless}}
	\end{subfigure}%
	\begin{subfigure}[c]{0.5\textwidth}
		\centering\includegraphics[height=160pt,left]{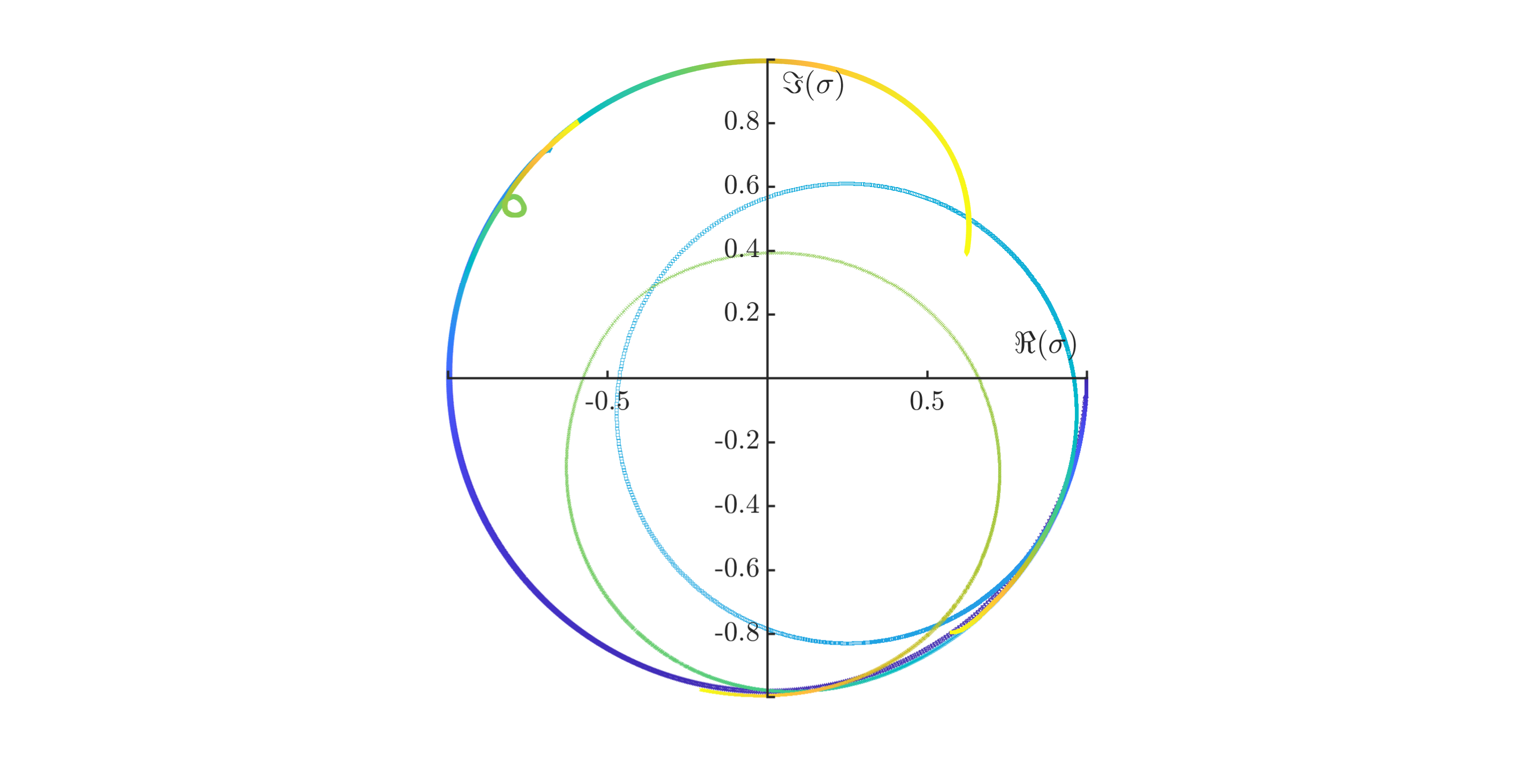}
		\caption{\label{fig:Stilde-sig-lossy1}}
	\end{subfigure}
    \begin{subfigure}[c]{0.5\textwidth}
		\centering\includegraphics[height=160pt,center]{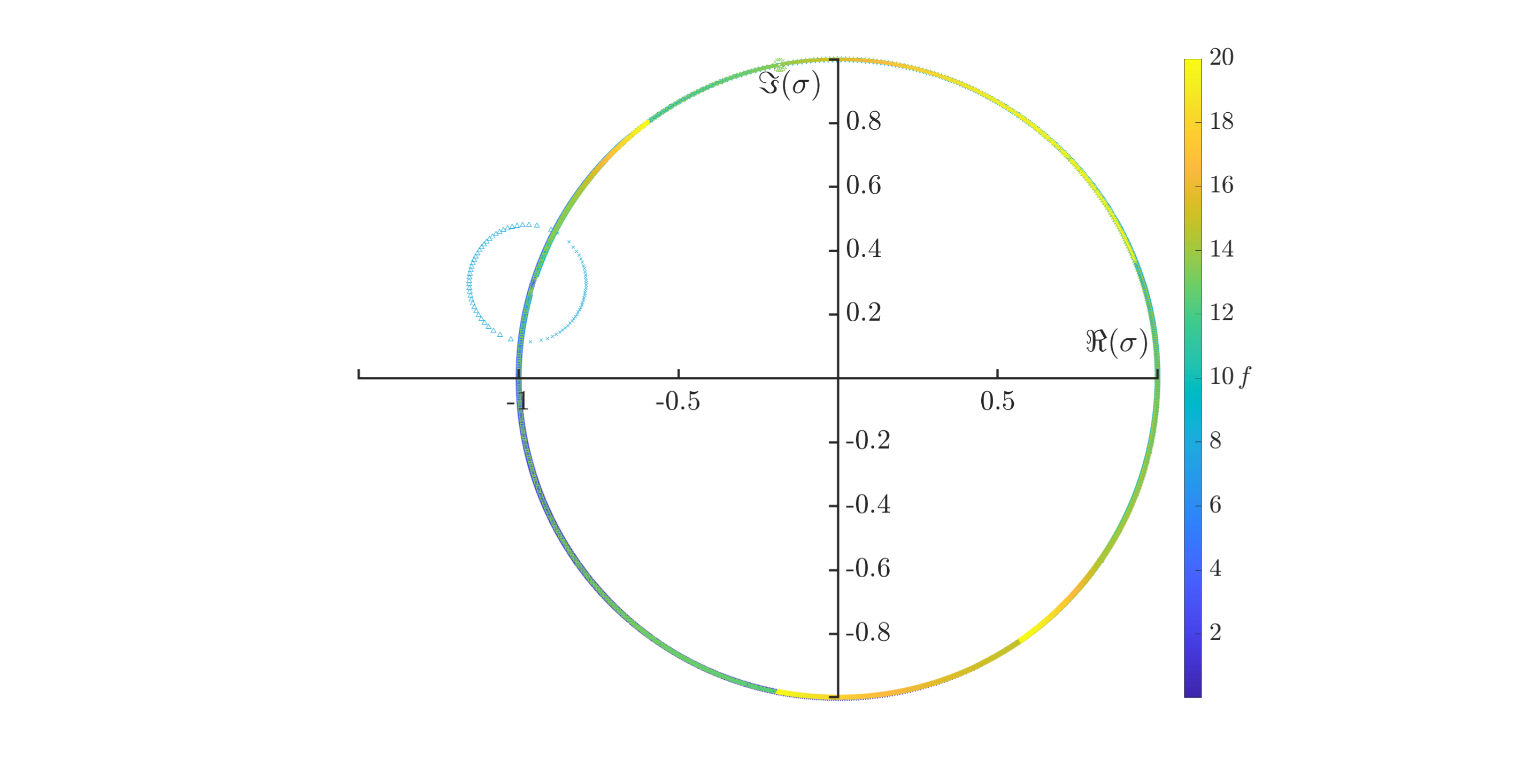}
		\caption{\label{fig:Stilde-sig-balg5}}
	\end{subfigure}
     \begin{subfigure}[c]{0.5\textwidth}
		\centering\includegraphics[height=160pt,left]{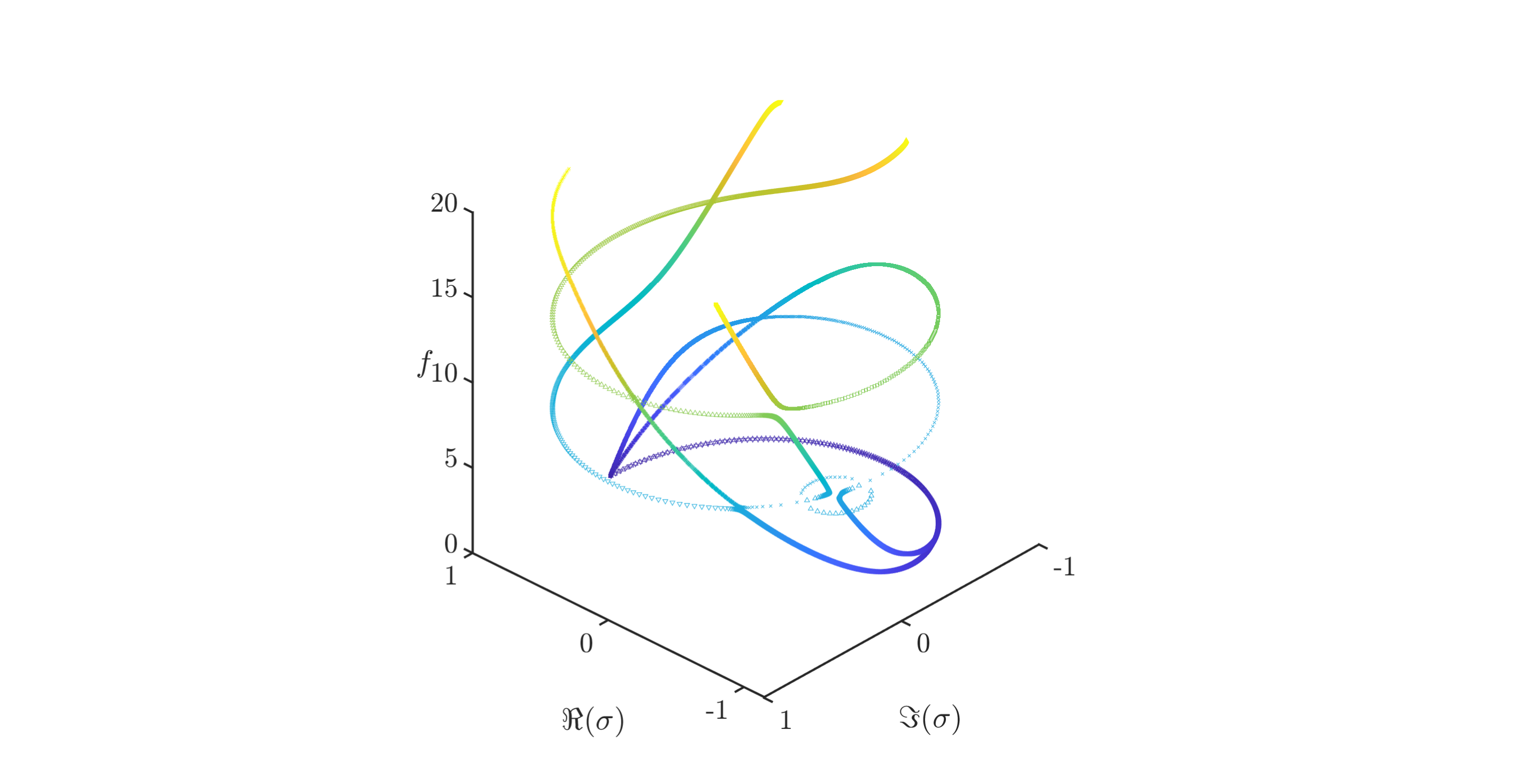}
		\caption{\label{fig:Stilde-sig-balg5-3d}}
	\end{subfigure}
	\caption{\label{fig:sig-simple} Location of the 4 eigenvalues of the modified scattering matrix $\TSM$ for the 3-phase, 5-layer, unit cell ($s_2 = 0.2$ in all cases). \gre{All subplots (a–d) correspond to the same frequency range, and the color bar shown (in kHz units) applies to all.} (\subref{fig:Stilde-sig-lossless}) Eigenvalues are always located on the unit circle for the lossless case and (\subref{fig:Stilde-sig-lossy1}) inside it for the lossy case ($c''/c' = 0.01$ in layers 2 and 4). The case of balanced gain and loss ($c''/c' = \mp 0.05$ in layers 2/4) is shown in (\subref{fig:Stilde-sig-balg5}) and (\subref{fig:Stilde-sig-balg5-3d}) for better view. Note the broken phase symmetry and the higher than 1 amplitude in the second quadrant.
    }
\end{figure}

Figure~\ref{fig:sig-simple} shows examples of the 4 complex eigenvalue of the $\TSM$ as functions of frequency for the 3-phase, 5-layer, unit cell shown in Figure~\ref{fig:mat} ($s_2 = 0.2$). Lossless and lossy case results behave as expected. For the case of $\mathcal{PT}$-symmetric specimen (with 2 layers providing balanced gain and loss), the amplitude is equal to 1, except between pairs of exceptional points, where the phase symmetry is broken. It is worth mentioning that while these frequencies are near ranges that EPs are also observed in the band structure, they do not match. In fact in the cases studied here they do not even have overlap, even though they are quite close to each other. 

\begin{figure}[!ht]
	\begin{subfigure}[c]{0.5\textwidth}
		\centering\includegraphics[height=180pt,center]{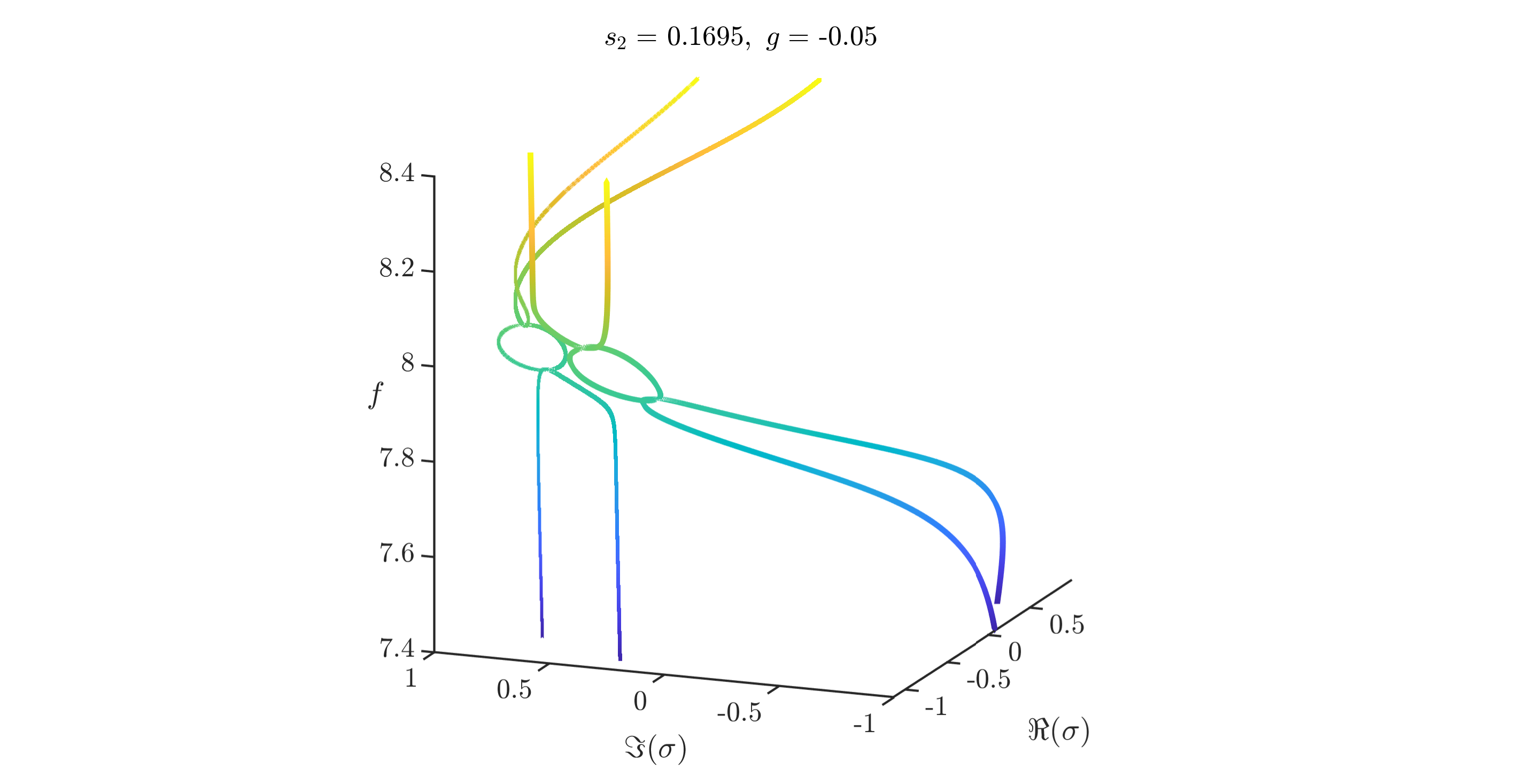}
		\caption{\label{fig:St-sig-s21695-g05}}
	\end{subfigure}%
	\begin{subfigure}[c]{0.5\textwidth}
		\centering\includegraphics[height=180pt,center]{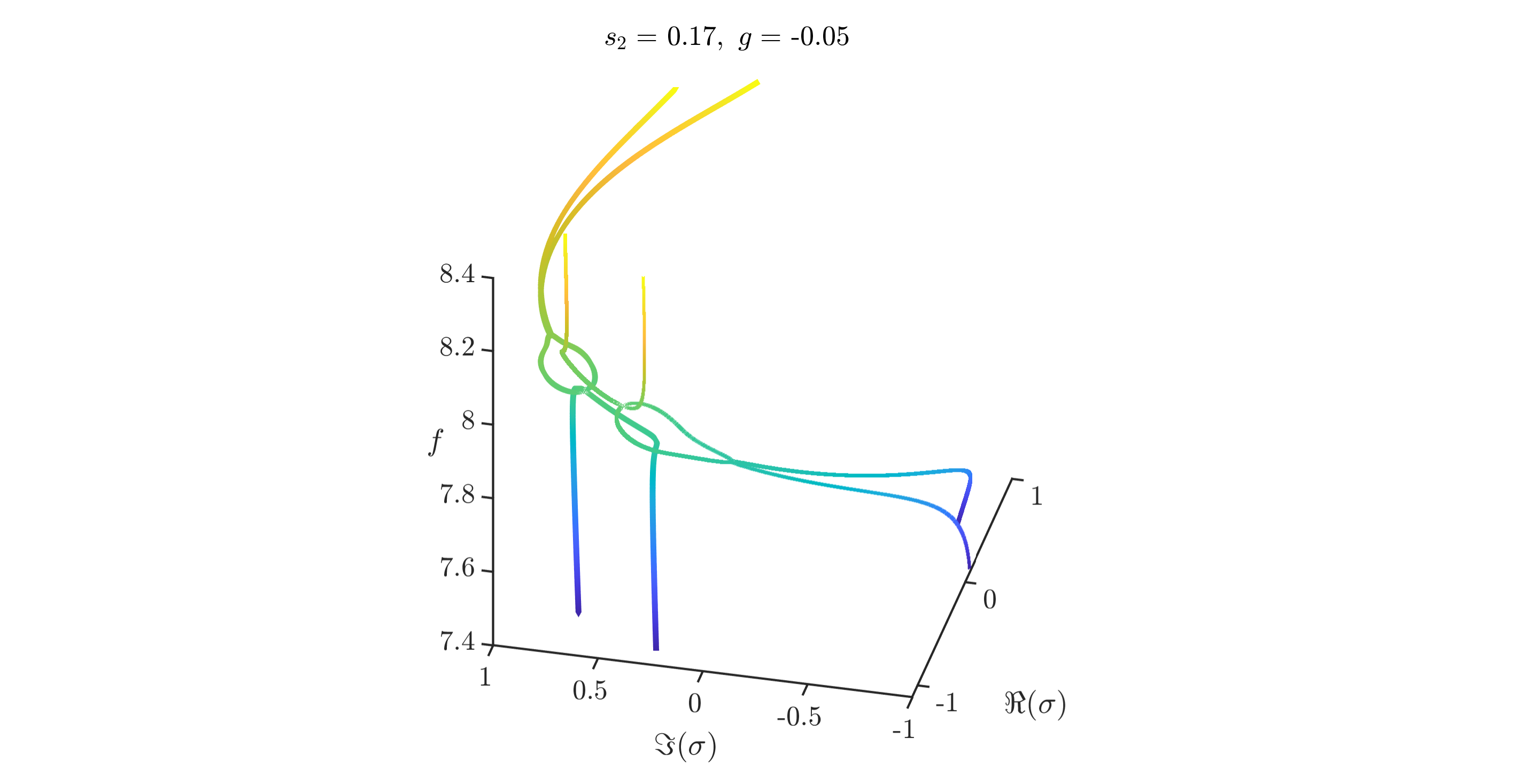}
		\caption{\label{fig:St-sig-s217-g05}}
	\end{subfigure}
    \begin{subfigure}[c]{0.5\textwidth}
		\centering\includegraphics[height=180pt,center]{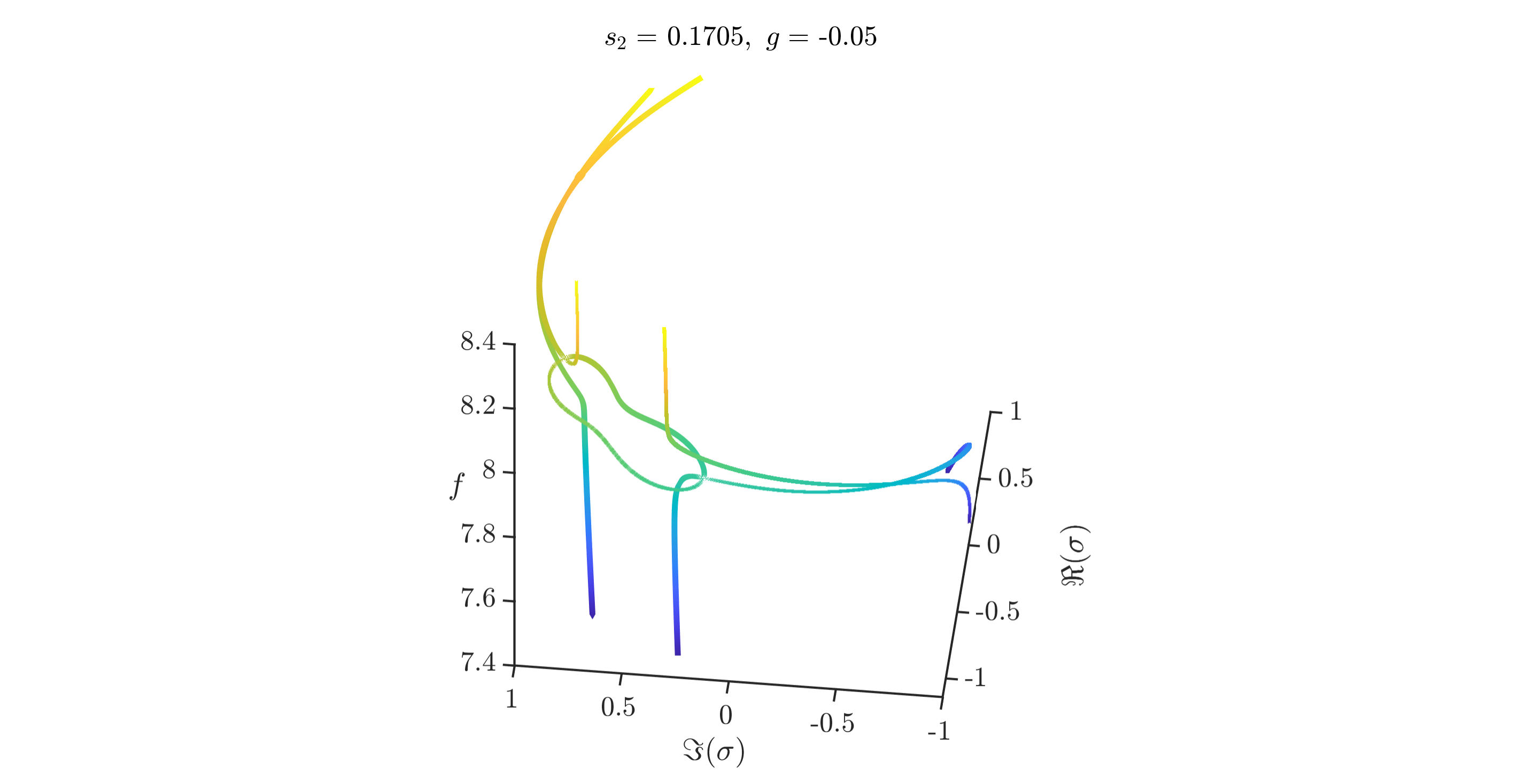}
		\caption{\label{fig:St-sig-s21705-g05}}
	\end{subfigure}
     \begin{subfigure}[c]{0.5\textwidth}
		\centering\includegraphics[height=180pt,center]{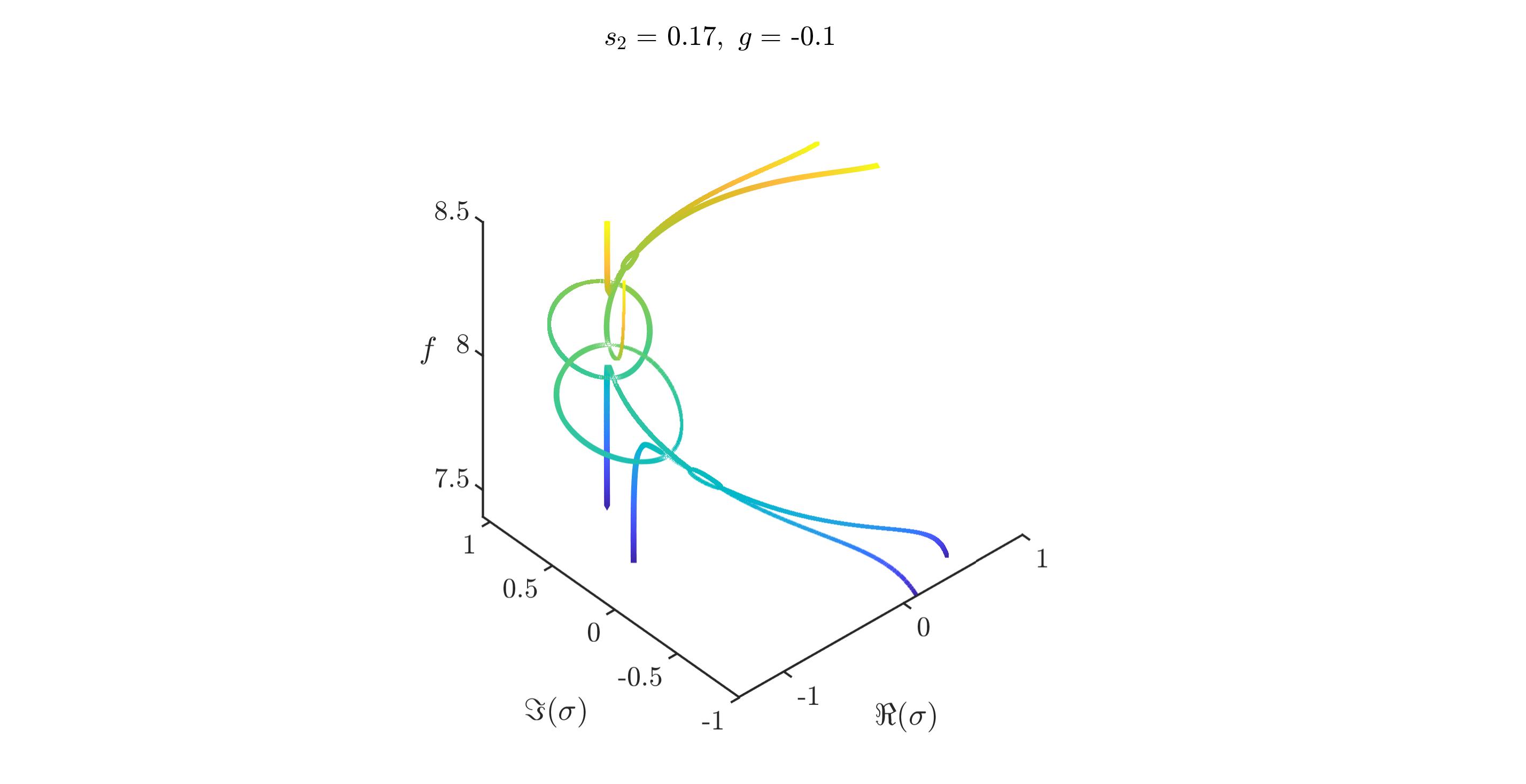}
		\caption{\label{fig:St-sig-s217-g10}}
	\end{subfigure}
	\caption{\label{fig:sig-variation} An example of the effect of varying the wave vector component $s_2$ and balanced gain parameter $g$ on the eigenvalues of the modified scattering matrix $\TSM$. \gre{Panels (a) and (c) share the same frequency range (in kHz units) and color scale as (b).} The same unit cell as previous figures is studied here. Each case is presented slightly differently for clarity. (\subref{fig:St-sig-s21695-g05}) through (\subref{fig:St-sig-s21705-g05}) show the significant geometrical changes associated with slight variation in wave vector. The latter appears to show a regular crossing, though this needs further high resolution or more detailed study. It also shows a double-lobed feature suggesting mixing of the EP pairs, or possibly a sign of existence of a quadruple EP in nearby complex domain of the parameter space. Comparison of (\subref{fig:St-sig-s217-g05}) and (\subref{fig:St-sig-s217-g10}) underlines the sensitivity to the gain parameter $g$, as a reduction in gain turns two elongated regions of broken symmetry into two.
    }
\end{figure}

Additionally, due to the higher dimensionality of this operator compared to those studied before (e.g. in \cite{wang2020exceptional}), multiple EP pairs may appear in the vicinity of each other. Since the $\mathcal{PT}$-symmetry requires the eigenvalues of $\TSM$ to always appear in pairs such that $\sigma \sigma' = 1$, triple branching is not possible. Although we have not observed any quadruple branching, we are nonetheless unable to readily reject their existence based on theoretical considerations. However, it may require perfect parametric conditions, unlike the prevalent double branching which appears to occur quite often and for broad ranges of balanced gain ratio. A more accessible study involves bringing multiple EP pairs into the vicinity of each other by exploring the $s_2$ space as well as the gain ratio $g = c''/c'$, denoted here as a negative value for the layers in which gain occurs (the loss value in the corresponding symmetric layers are positive and equal to $-g$). Such adjustments lead to exaggerated geometries of the eigenvalue traces, similar to those shown as examples in Figure~\ref{fig:sig-variation}. Additionally, one observes what appears to be regular crossing, which, given the apparent predominance of either level repulsion or exceptional points, is somewhat surprising. It must be emphasized that at very high magnification the presumed regular crossing may prove to be a very small level repulsion or an EP pair. This may be rigorously studied through a number of methods, such as encircling the crossing in the complex $s_2$ space \cite{Lu2018} or study of the eigenvectors or eigenvector subspaces \cite{kato_perturbation_1995, amirkhizi_reduced_2018}. Another interesting observation is the emergence of double-lobed features resembling the transition from one EP pair behavior to another neighboring one, or possibly a sign of existence of quadruple branching in the complex parameter domain. One can hypothesize that with minute adjustments to the geometry and/or material properties (including $g$ value), this quadruple branching may be brought onto the real parameter axis, though such a study is not carried out here. Other esoteric geometries show up as well, though their physical significance and/or potential features for application are unclear at the moment. 

\begin{figure}[!ht]
	\begin{subfigure}[c]{0.5\textwidth}
		\centering\includegraphics[height=140pt,center]{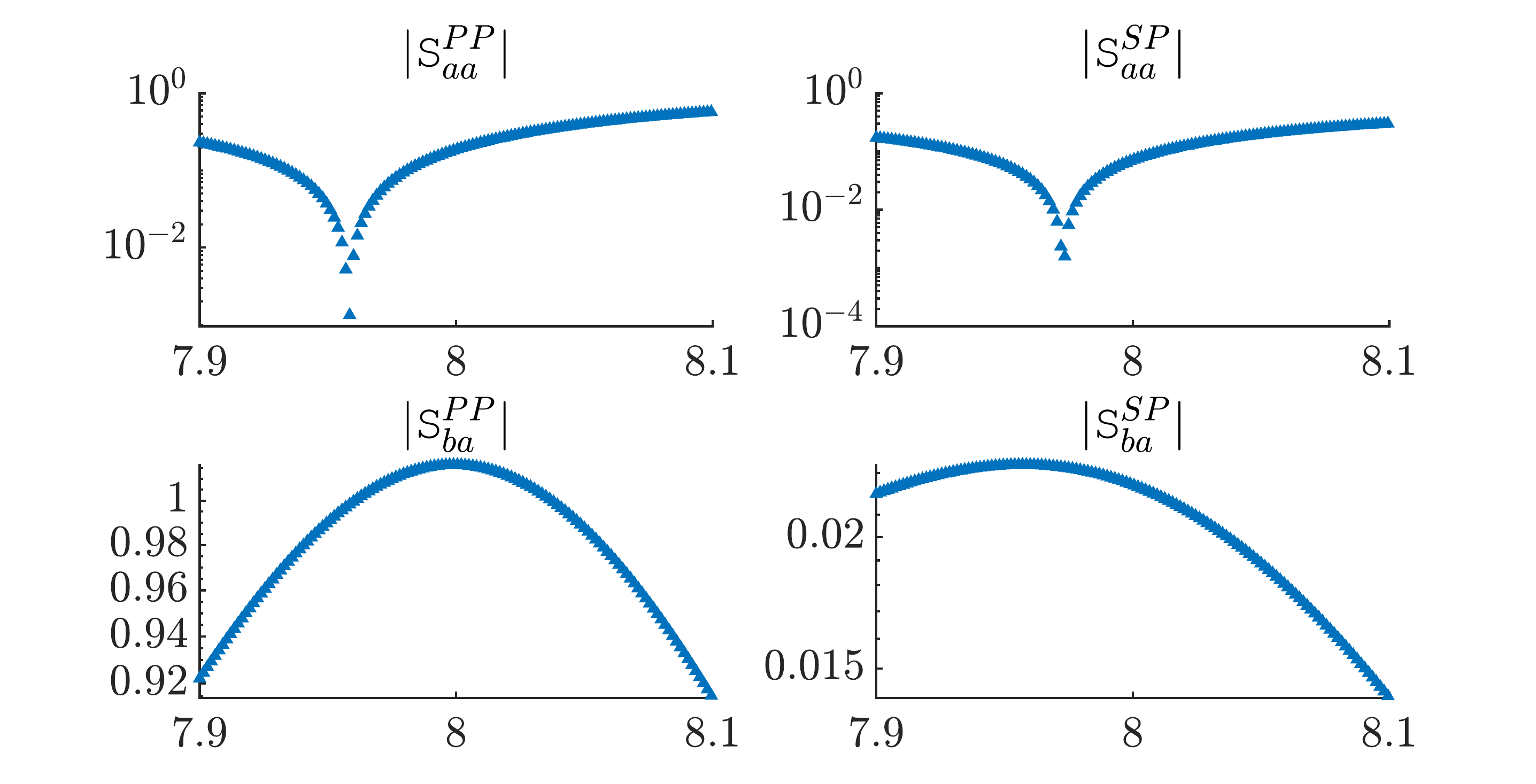}
		\caption{\label{fig:scgm05s21695P}}
	\end{subfigure}%
	\begin{subfigure}[c]{0.5\textwidth}
		\centering\includegraphics[height=140pt,center]{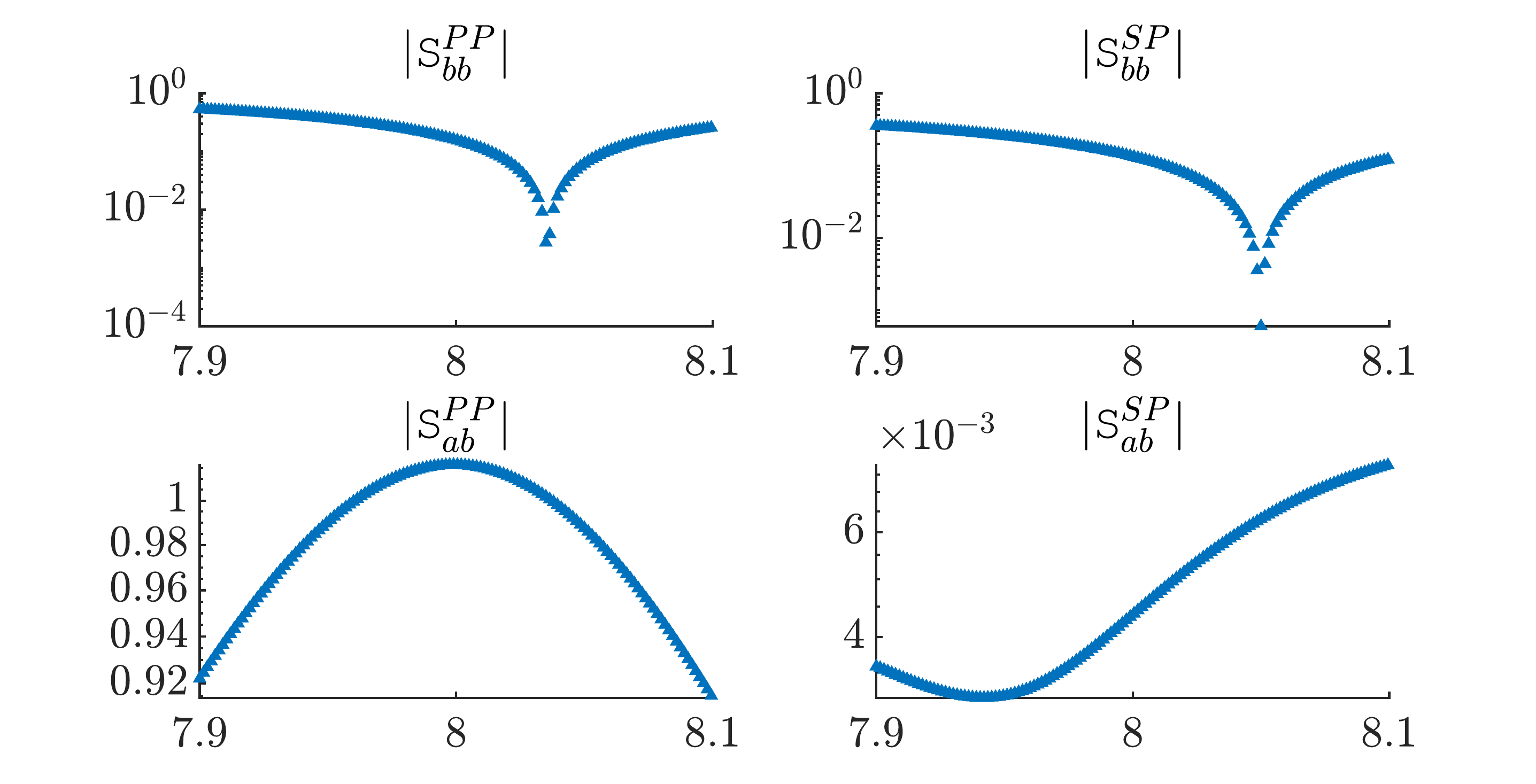}
		\caption{\label{fig:scgp05s21695P}}
	\end{subfigure}
    \begin{subfigure}[c]{0.5\textwidth}
		\centering\includegraphics[height=140pt,center]{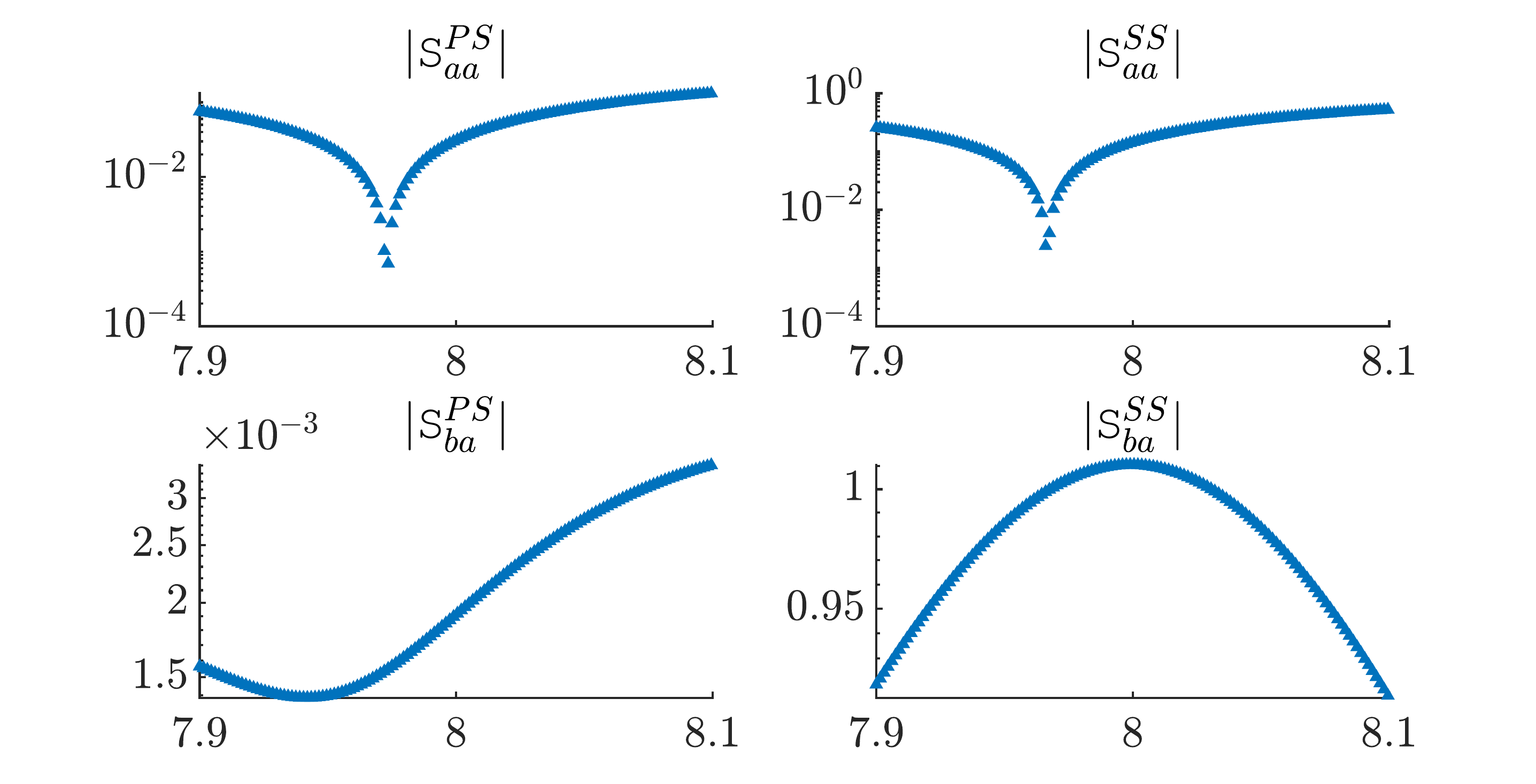}
		\caption{\label{fig:scgm05s21695S}}
	\end{subfigure}
     \begin{subfigure}[c]{0.5\textwidth}
		\centering\includegraphics[height=140pt,center]{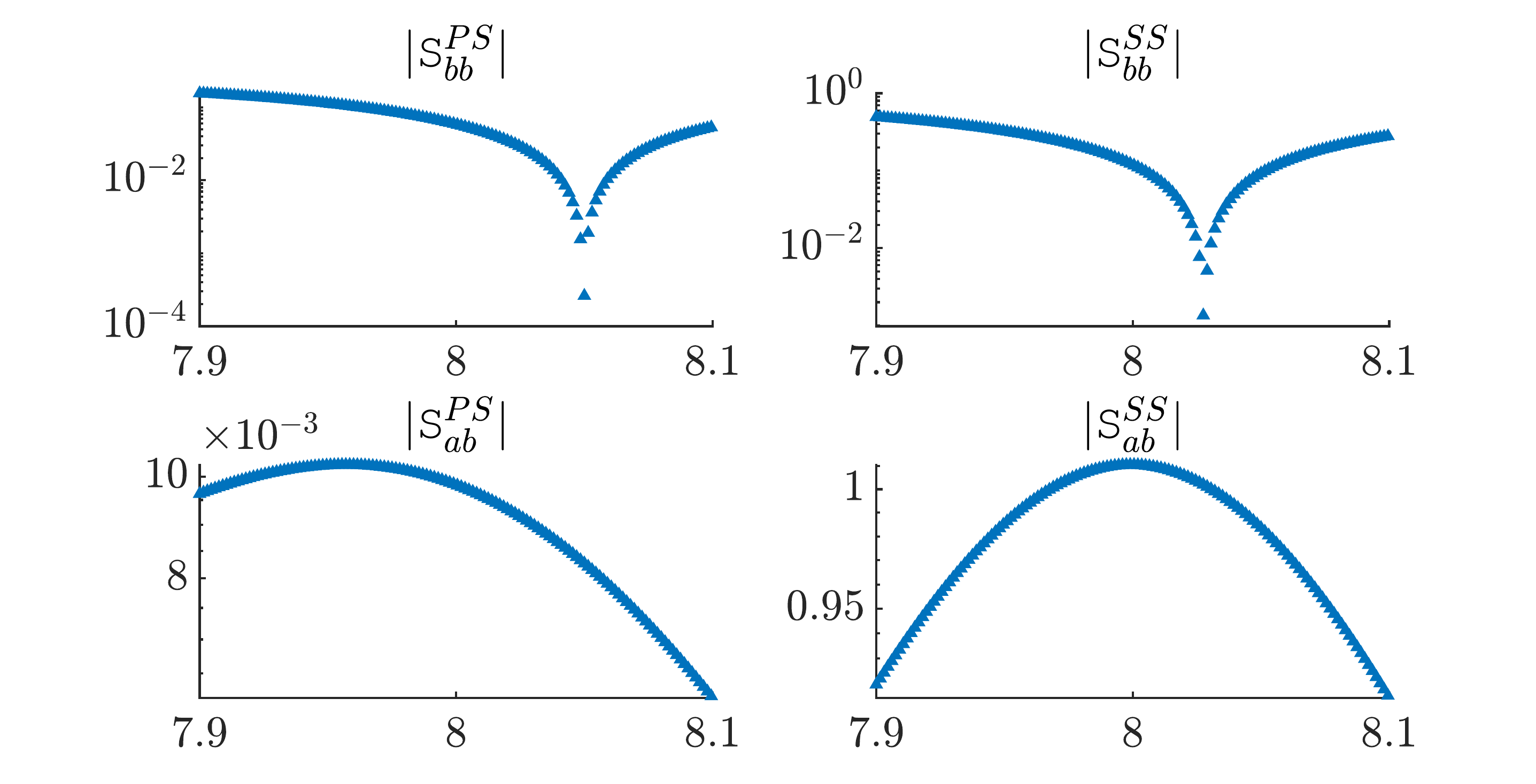}
		\caption{\label{fig:scgp05s21695S}}
	\end{subfigure}
	\caption{\label{fig:balg-scat-amp} An example of scattering around EP pairs. \gre{The x-axis in all subplots represents frequency (in kHz), and the y-axis shows the magnitude of the scattering coefficients.} The same unit cell as previous figures is studied here, with $s_2=0.1695$, $g = -0.05$. (\subref{fig:scgm05s21695P}) and (\subref{fig:scgp05s21695P}) show the scattering amplitudes for an incident longitudinal wave, while (\subref{fig:scgm05s21695S}) and (\subref{fig:scgp05s21695S}) show them for shear incidence. Note the difference in reflection coefficients in $ab$ vs. $ba$ directions, as it diminishes, depending on the direction at a few hundred Hz below and above \SI{8}{kHz}. Between these two points of single-sided reflection, the amplitude of transmission can be above 1 for $PP$ and $SS$ transmissions but relatively small for cross-polarization transmissions.
    }
\end{figure}

\begin{figure}[!ht]
	\begin{subfigure}[c]{0.5\textwidth}
		\centering\includegraphics[height=140pt,center]{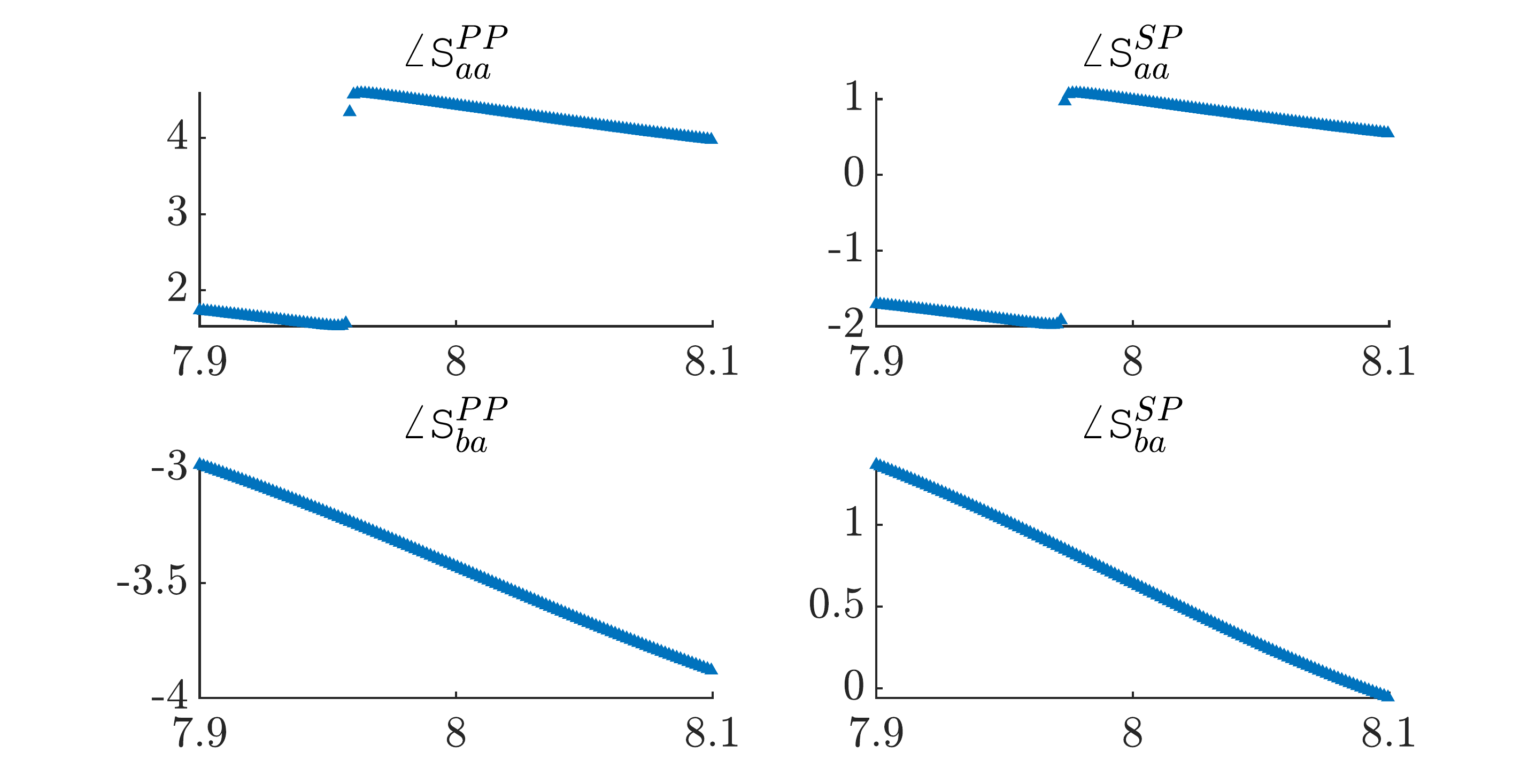}
		\caption{\label{fig:angscgm05s21695P}}
	\end{subfigure}%
	\begin{subfigure}[c]{0.5\textwidth}
		\centering\includegraphics[height=140pt,center]{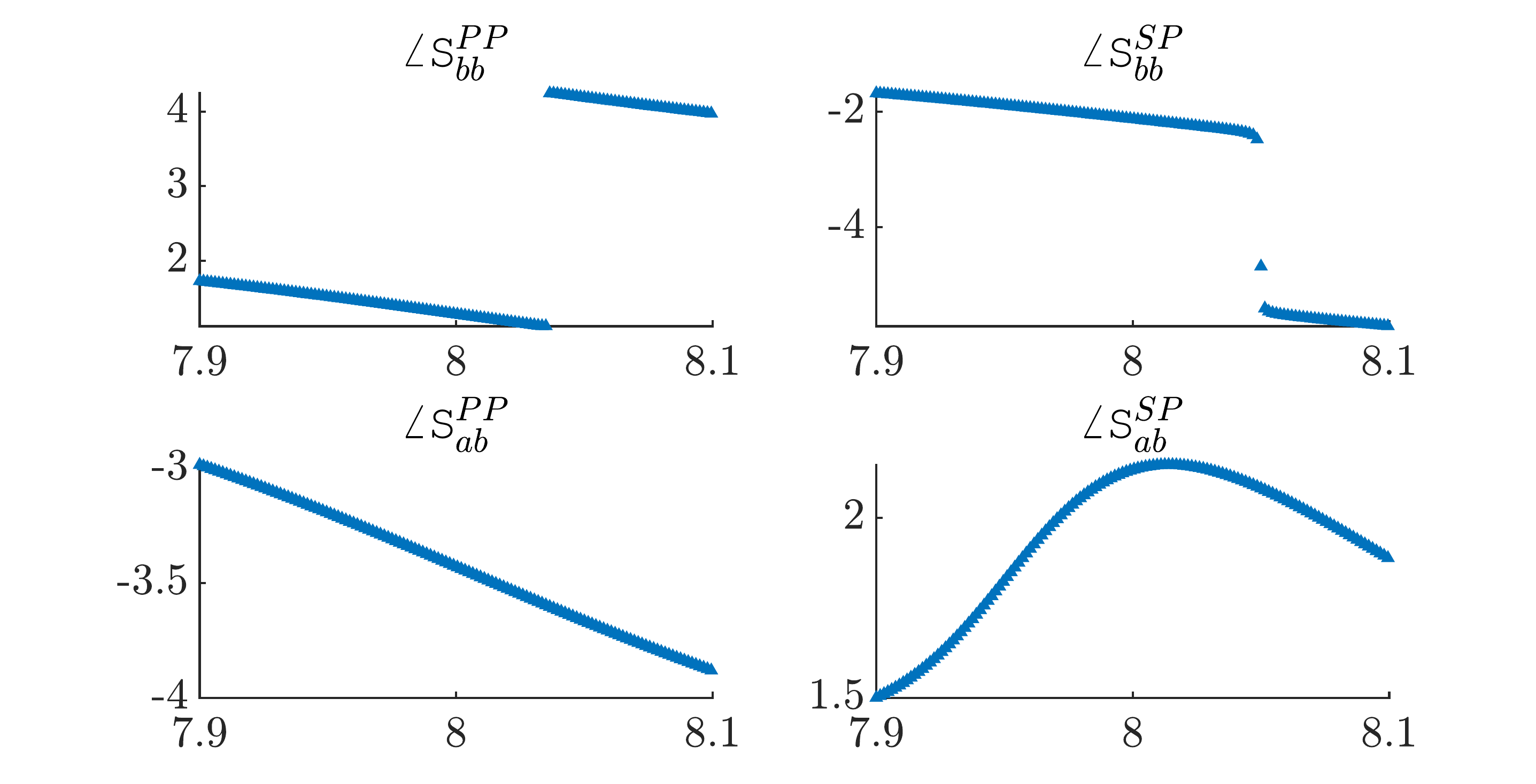}
		\caption{\label{fig:angscgp05s21695P}}
	\end{subfigure}
    \begin{subfigure}[c]{0.5\textwidth}
		\centering\includegraphics[height=140pt,center]{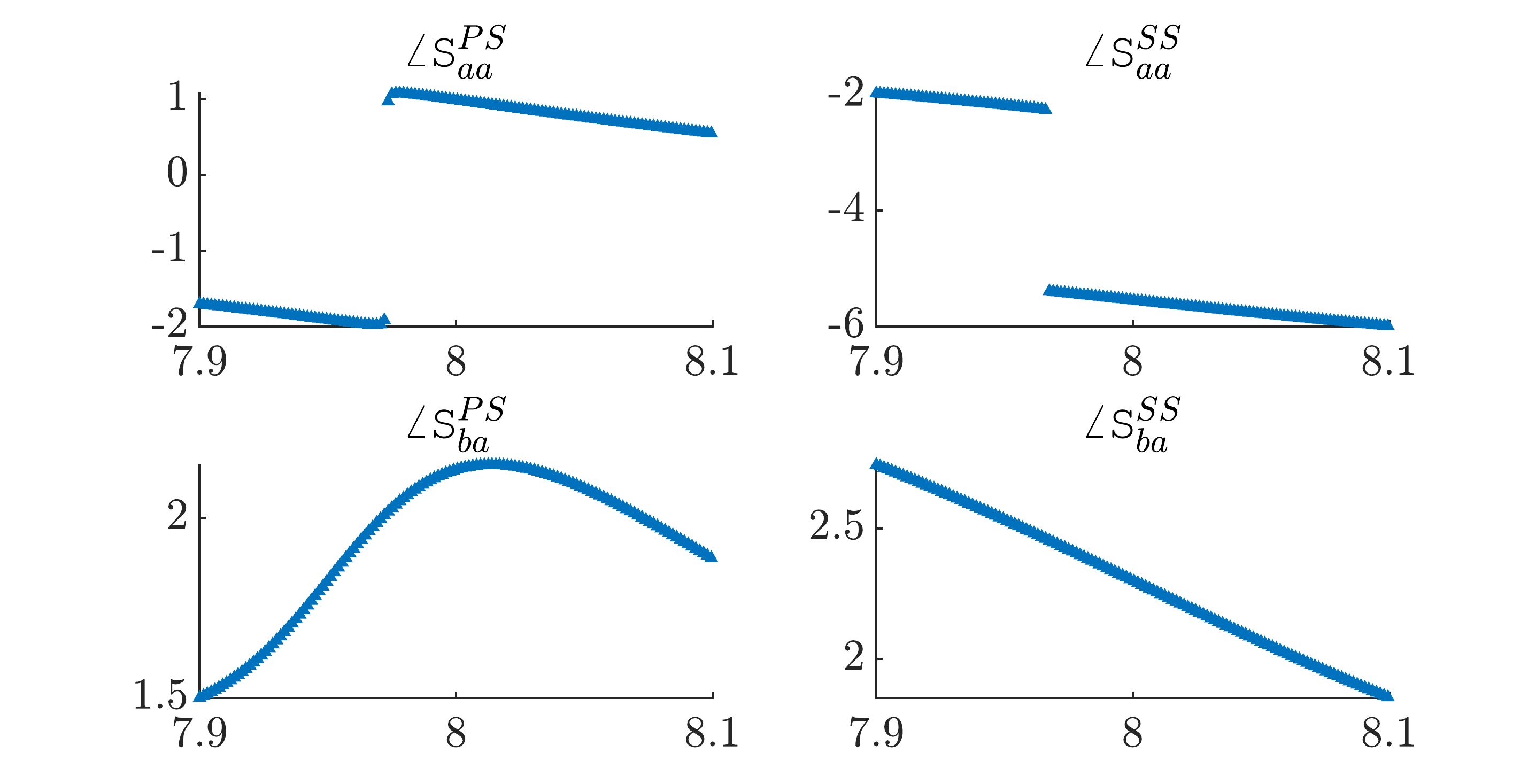}
		\caption{\label{fig:angscgm05s21695S}}
	\end{subfigure}
     \begin{subfigure}[c]{0.5\textwidth}
		\centering\includegraphics[height=140pt,center]{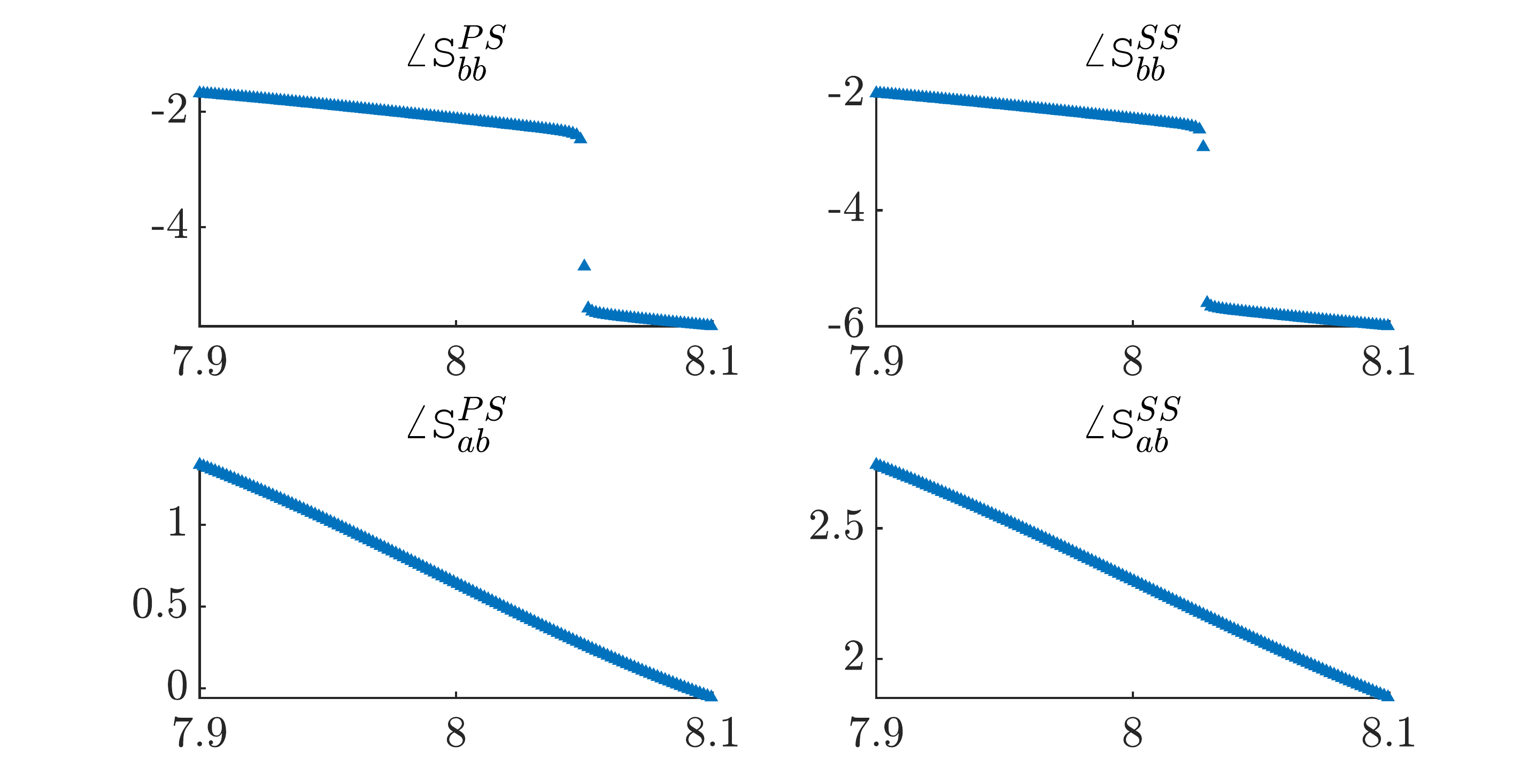}
		\caption{\label{fig:angscgp05s21695S}}
	\end{subfigure}
	\caption{\label{fig:balg-scat-ang}  An example of scattering around EP pairs. \gre{The x-axis in all subplots represents frequency (in kHz), and the y-axis shows the phase (in radians) of the scattering coefficients.} The same unit cell as previous figures is studied here, with  $s_2=0.1695$, $g = -0.05$. (\subref{fig:scgm05s21695P}) and (\subref{fig:scgp05s21695P}) show the scattering phase advances for an incident longitudinal wave, while (\subref{fig:scgm05s21695S}) and (\subref{fig:scgp05s21695S}) show them for shear incidence. The sharp changes are all around $\pi$ representing a change of sign as reflection amplitudes cross zero.
    }
\end{figure}
The application of such features may be found in single sided reflectivity and/or lasing potentials as discussed earlier. As such, the amplitude of eigenvalue pairs appears to be the dominant factor. Absent of increasing the absolute value of the gain, the interaction of EP pairs appears to also lead to higher deviations from unit amplitude for eigenvalue pairs. An example of such a study is shown in Figures~\ref{fig:balg-scat-amp} and \ref{fig:balg-scat-ang}. The wave amplification and asymmetric reflection is clearly observed. Additionally, while not shown here the jump in reflection phase as it diminishes near an EP is notable, especially as its direction is quite sensitive to $s_2$ (not shown here). This is equivalent to the reflection parameter encircling zero in clockwise or counter-clockwise directions. While at the EP due to the low value of amplitude, this might be a rather challenging measurement, it is possible to envision this asymmetry in phase at points slightly away from EP as the basis of an approach for determination of the wave vector. Source localization using micro-structured media has been discussed in \cite{wang_angle-dependent_2023}, and the use of EPs for enhancing their behavior is a potentially fruitful research deferred to later work. 

\section{Summary and conclusions}
In this work we studied the propagation of coupled in-plane longitudinal and shear stress waves in layered linear media. The machinery of the transfer matrix allows for exact calculations, though closed form results are prohibitive. The method allows for including linearly viscoelastic layers and even those with mechanical gain, presumably achieved through multi-physical mechanisms. Additionally, it readily allows for identification of modal features through calculation of the eigen-polarizations. The coupled physics immediately leads to the appearance of degeneracies in the spectrum, manifested as exceptional point pairs in the band structure. Though the pairs in band structure disappear from the real frequency axis with the addition of loss, for the lossless case they lead to fascinating effects in polarization (e.g. chirality associated with the direction of propagation). Scattering parameters for finite structures can also be calculated with the same tool sets, which lends itself quite readily to potential design and optimization. Furthermore, EP pairs appear also in the spectrum of the modified scattering matrix for $\mathcal{PT}$-symmetric finite structures composed of layers with balanced gain and loss. These EP pairs lead to unusual phenomena such as broken phase symmetry, transmission amplification (above 1), single-sided reflectivity, and abrupt changes in phase behavior of scattering parameters \gre{\cite{cai2023absorption, jin2022non}}. The exploration of the potential for use in ``lasing'' \gre{with transmission amplification and a rigorous analysis of mode lifetime \cite{he2025laser}} and wave vector measurement (source localization) is deferred to future studies.

\section*{Acknowledgments}
The authors acknowledge National Science Foundation support for this work through Grant No. 2219087 to University of Massachusetts, Lowell. The authors thank Dr. Weidi Wang for illuminating discussions throughout the course of this work.

\clearpage

\bibliographystyle{unsrtm}
\bibliography{P_SV}

\end{document}